\documentclass[10pt]{elsart}

\usepackage{epsf}
\usepackage{amssymb}
\usepackage{amsmath}

\begin{document}

\newcommand{\fig}[2]{\epsfxsize=#1\epsfbox{#2}}

\begin{frontmatter} 
\title{Topological transitions and freezing in 
XY models and Coulomb gases with quenched disorder:
renormalization via traveling waves 
}
\author{David Carpentier and Pierre Le Doussal} 
\address{ CNRS-Laboratoire de Physique Th{\'e}orique de l'Ecole Normale
Sup{\'e}rieure,\\ 
24 Rue Lhomond, Paris 75231 Cedex 05 France}
\begin{abstract}
We study the two dimensional XY model with quenched random phases and
its Coulomb gas formulation. A novel renormalization group (RG) method
is developed which allows to study perturbatively the glassy low
temperature XY phase and the transition at which frozen topological
defects (vortices) proliferate. This RG approach is constructed both
from the replicated Coulomb gas and, equivalently without the use of
replicas, using the probability distribution of the local disorder
(random defect core energy). By taking into account the fusion of
environments (i.e charge fusion in the replicated Coulomb gas) this
distribution is shown to obey a Kolmogorov's type (KPP) non linear RG
equation which admits travelling wave solutions and exhibits a
freezing phenomenon analogous to glassy freezing in Derrida's random
energy models. The resulting physical picture is that the distribution
of local disorder becomes broad below a freezing temperature and that the
transition is controlled by rare favorable regions for the defects,
the density of which can be used as the new perturbative parameter.
The determination of marginal directions at the disorder induced
transition is shown to be related to the well studied front velocity
selection problem in the KPP equation and the universality of the
novel critical behaviour obtained here to the known universality of
the corrections 
to the front velocity. Applications to other
two dimensional problems are mentionned at the end.
\end{abstract}
\end{frontmatter}

\section{Introduction}\label{part:intro}

\subsection{Overview}

Topological phase transitions in two dimensions, which are induced by
the proliferation of topological defects, are naturally described via
Coulomb gas formulations. In the description of the
Kosterlitz-Thouless transition of the XY model, the corresponding
Coulomb gas (CG) with integer charges is obtained as an effective
theory for the topological defects of the model : the vortices
\cite{kosterlitz73,kosterlitz74}. Similarly, a collection of
dislocations whose proliferation induce the melting transition in a
two dimensional elastic lattice can be described by a Coulomb gas with
vector charges which belong to the reciprocal
lattice\cite{nelson79,young79}. Finally, most of two dimensional
statistical models, such as the Ising, Potts and Askin-Teller models,
can be transformed into Coulomb gases \cite{kadanoff78} (see also the
review \cite{nienhuis87}). In all these cases, the transitions can be
studied by renormalization in considering the screened interaction
between two test Coulomb charges. This interaction is logarithmic at
large distance ${\bf r}$, and the renormalization procedure consists
in neglecting the higher terms in an expansion in $1/r$ as being
irrelevant \cite{nienhuis87,minnhagen89}. This renormalization is thus
valid for a dilute gas of charges, and is usually implemented by an
expansion of the Coulomb gas partition function in powers of the {\it
charge fugacity} $y$. This fugacity $y=\exp (-E_{c}/T)$ is related to
the {\it core energy} $E_{c}$ of the charges (i.e the local energy to create a
defect or their chemical potential). Thus this Coulomb gas
renormalization procedure is perturbatively controlled in the limit of
large $E_{c}$, or equivalently at finite temperature in the limit of
small fugacity. Several topological transitions in two dimensions have
been successfully described using this CG technique
\cite{kosterlitz74,nelson79,nienhuis87}.

Soon afterwards, several authors attempted to extend these techniques
to models with quenched disorder \cite{cardy82,rubinstein83,nelson83}.
The randomness in the original statistical models translates into
random fields in the Coulomb gas formulation. The averaged free energy
(instead of the partition function) of this Coulomb gas is then
expanded, as in the pure case, in powers of the fugacity $y$ of the
charges. Usual scaling techniques then allow to study the topological
transitions in these disordered systems.

It does not seem to have been realized at that time that these
approaches rely on the crucial assumption of a {\it uniform fugacity} 
for the charges in the sample, or equivalently of 
a core energy {\it spatially uniform over the sample}.
Although this assumption is natural for pure models,
it is at least questionable in the presence of a random potential
\cite{giamarchi95}. The
randomness may favor the appearance of topological defects (the
charges of the CG) in some sites where the core energy will be effectively lower
than on other sites. At high temperature, the randomness of the core
energy $E_{c}$ is irrelevant as it is averaged out by
thermal phase fluctuations. In that case the large scale behaviour 
of the model can indeed be
described by simply considering the averaged fugacity over the sample
$\overline{y}$, as in \cite{cardy82,rubinstein83,nelson83}. However,
as we show here, when the temperature is lowered, the spatial
inhomogeneities in the local core energy become more and more
relevant. As a result, approaches based on a single uniform fugacity 
are doomed to fail. A correct detailed description of the scaling behaviour 
of the site-dependent fugacities (or core energies) becomes 
then necessary to determine the
phase diagram and describe quantitatively the transitions in the
random model. It is the purpose of this work to define a novel
renormalization method which allows this description in a consistent
and perturbatively controlled manner. We present here a
detailed analysis, a shorter account of the method and
results has appeared in \cite{carpentier98bis}.
Although we focus here on the random phase XY model,
these techniques can be applied to
a much wider class of systems which can be formulated as disordered
Coulomb gases (e.g. see Section \ref{part:sinegordon} for a random
Sine-Gordon formulation). In particular, extensions to the melting of
two dimensional crystals in presence of disorder are studied in
\cite{carpentier99bis,carpentier98,giamarchi98} 
and applications to the statistics of localized
wave functions of electrons in random magnetic fields
as well as entropic transitions in Liouville theory and other 
models are studied in \cite{carpentier99ter}. 

\subsection{Random phase XY model}

In this paper, we focus on the 2d XY model with randomness in the
phase (see eq. (\ref{xy})), originally studied in \cite{rubinstein83}
in the context of XY magnets with random Dzyaloshinskii-Moriya
interactions. The topological defects (vortices) of this model are
represented by integer charges $n_{\bf r}$ at sites ${\bf r}$. Two
charges $n_{\bf r},n_{\bf r'}$ interact for a large separation via the
usual Coulomb interaction of strength $J$, with the corresponding
energy $ - 2 J~n_{\bf r}n_{\bf r'} \ln (|{\bf r-r}'|/a)$ where $a$ is the
typical size of the core region of the vortices, {\it i.e} the short
distance cut-off (as discussed in section \ref{part:model}).
These charges also couple to a random potential $V_{{\bf r}}$, which
arises from the randomness of the XY model, via a term $n_{{\bf
r}}V_{{\bf r}}$ which depends on the sign of the charge. The crucial
property of this random potential is that it has long range logarithmic
correlations:
\[
\overline{(V_{\bf r} - V_{{\bf r}'})^2} = 4 \sigma J^2 
\ln|{\bf r}-{\bf r}'| + O(1)
\]
the on-site variance being $\overline{V^{2}_{{\bf r}}}=\Delta \sim
2\sigma J^{2}\ln L $ where $L$ is the system size. 

\begin{figure}
\centerline{\fig{8cm}{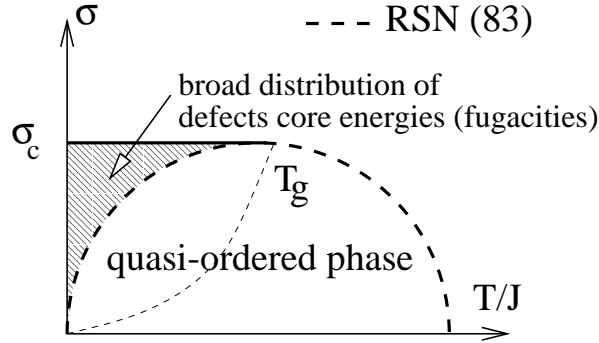}}
\caption{Phase diagram of the two dimensional  XY model with random
phase fluctuations. $J$ corresponds to the stiffness, $T$ to the
temperature and $\sigma $ to the strentgh of the randomness. The
dashed line represents the transition line of Rubinstein, Schraiman and
Nelson  \cite{rubinstein83} between the quasi-ordered (XY) phase and a
disordered phase.}
\label{fig:phasediag-intro}
\end{figure}

Rubinstein, Schraiman and Nelson \cite{rubinstein83} identified the
most relevant vortices as the $n=\pm 1$ charges. They derived scaling
equations for the stiffness $J (l)$, the disorder
strength $\sigma (l)$ and a {\it single fugacity} for these charges 
$y (l)$. As in the pure case, the nature of the phase is determined by
the behaviour of the fugacity : in a quasi-ordered phase, 
$y(l)$ decreases with the scale while in a disordered phase it
increases. In this last case, vortices appear at the
scale $ae^{l^{*}}$ at which their renormalized core energy $E_{c} (l)$ 
is about zero, which corresponds to a fugacity of order $1$ : $y (l^{*})\simeq 1$ 
(see {\it e.g} \cite{nelson78}). Interestingly the divergence of
the correlation length at criticality found in \cite{rubinstein83},
$\xi \sim \exp (cste/ |T-T_{c}|^{\frac{1}{2}})$,
is identical to the result of Kosterlitz-Thouless for the pure system.
The phase diagram was found to be reentrant at low temperature,
and is shown as a dashed line in fig. \ref{fig:phasediag-intro}
as a function of the renormalized value of $\sigma$ and $J$.

 Recently, several authors \cite{cha95,nattermann95} have 
proposed a modified phase diagram for the same model, where the
reentrance of the phase transition of fig. \ref{fig:phasediag-intro}
disappears. The main point, further developed in
\cite{korshunov96c}, is an energy argument for a single vortex at
zero temperature in a finite size sample, in the spirit of the
Kosterlitz-Thouless argument 
for the transition in the pure model \cite{kosterlitz74}. 
The energy to create a single defect has two main
components (neglecting the bare core energy) : the elastic energy
$E_{el}\simeq J \ln L$ and the disorder energy. Indeed the defect can
take advantage of the spatial 
variations of $V_{{\bf r}}$ and choose the minimal value  of 
$V_{{\bf r}}$ in the $N= (L/a)^{2}$ sites of a sample of size $L$. 
To estimate this minimum $V_{min}$ of $V_{{\bf r}}$ over $N$ sites,
the above authors
neglected the spatial correlations of $V_{{\bf r}}$.
Under that hypothesis, the problem of this single particle in the
random potential $V_{{\bf r}}$ becomes {\it identical} to the Random Energy
Model (REM) defined and solved in the seminal work of Derrida
\cite{derrida81}. In particular, the averaged  
$\overline{V}_{min}$ of the minimum behaves for
large $N$ as 
$V_{min}\simeq -\sqrt{2 \Delta}  (\ln N)^{\frac{1}{2}}
\simeq -\sqrt{8\sigma }J\ln L$, where $\Delta $ is the onsite variance
(see above). Adding the elastic
contribution $E_{el}$ yields a creation energy  
$E\simeq J (1-\sqrt{8\sigma })\ln L$. 
Hence this simple single vortex argument points towards the 
existence of a topological phase transition at 
$\sigma =\sigma_{c}=\frac{1}{8}, T=0$
where vortices proliferate. The modified phase diagram is shown as a
solid line on figure \ref{fig:phasediag-intro}. 

Although appealing, this single vortex argument is not sufficient by itself
to prove the existence of a phase transition in the system of many interacting
charges, and even less so to describe the critical behaviour.
Screening by mutual interactions usually plays an important role 
in Coulomb systems, and screening of the disorder could also in principle
modify the results. The single vortex argument can only become
valid asymptotically (at infinite scaling length) 
if there indeed exists a phase with finite renormalized values
for $J$ and $\sigma$ and vanishing fugacity at zero temperature.
The very existence of the XY phase in the present problem 
has been questionned in recent works \cite{mudry99} and a further
analysis is thus necessary.
Thus, we cannot avoid the use of a renormalization approach to
determine the true phase diagram. Such a controlled RG description
should correctly take into account the correlations of the random potential,
neglected in the above argument. It should allow to  
precisely caracterize the universal features of the critical behaviour 
at the disorder driven transition.    

 Several attempts to construct a RG method were 
proposed \cite{nattermann95,korshunov96,scheidl97,tang96}
prior to this work, but no agreement was found between the results 
of these different approaches. As we discuss below, this is
largely because none of these approaches was internally consistent.
In particular, they were all based on expansions in {\it dipole} fugacities
while considering single charge fugacities not only appears more
natural but is essential in constructing a consistent renormalization 
procedure. As a result these approaches missed the very important contribution 
of the fusion of environments (see below). Korshunov and
Nattermann started by approximating the free energy of a given
disordered Coulomb gas by the sum of the free energies of {\it
independent dipoles}\footnote{We will come back to this expansion in
section \ref{part:other-approaches}} of charges \cite{korshunov96}.
They found no renormalization of the disorder strength $\sigma$ and a
modified 
phase diagram in agreement with the
figure \ref{fig:phasediag-intro}. An interesting replica approach 
was developed by Scheidl \cite{scheidl97}, with the use of a Coulomb gas of
$m$-component charges ${\bf n}$, where $m$ is the number of replica.
Scheidl noticed that, in the renormalization procedure, one must a priori
take into account charges with a number $p>1$ of non zero
components, contrarily to \cite{rubinstein83} where only single
component charges were considered. His RG equations are not compatible
with the one of \cite{korshunov96}: since the disorder
strength $\sigma$ is renormalized, it leads to a different phase
diagram when expressed in the bare constants. Upon closer examination
in Section \ref{part:other-approaches} and interpretation within our
formalism, the assumption implicitly underlying Scheidl's work \cite{scheidl97}
appears to be the gaussian nature of the renormalized local disorder,
while we are able to prove that the local disorder does not remain gaussian
upon coarse-graining.
Although we are also able to show a posteriori that in the XY phase some
results are compatible between the two approaches, the non gaussian nature 
of the local disorder becomes crucial to determine the critical
behaviour at the transition.
Finally Tang \cite{tang96} developed an appealing physical picture
for the single vortex freezing phenomena in this model
beyond the REM approximation mentionned above, and correctly foresaw 
the existence of a connection \cite{tang96,tang97}
with the problem of the Directed Polymer 
on the Cayley Tree (DPCT). However, no precise Coulomb gas renormalization 
procedure was developed following these ideas.

 Finding a controlled RG procedure which allows to describe this
physics of freezing in a collection of interacting Coulomb charges in a random
environments is thus a non trivial and challenging problem.
As we will see (and as can be expected from the above single charge argument),
it requires a quantitative description of the scale dependence
of the probability distribution of the fugacities associated with the rare
sites were the charges are frozen. 

Before embarking in the (sometimes technical) remainder of the
paper, it is useful to depict the spirit of the method and
summarize the main results.

\subsection{Description of the methods}

 To capture the physics of freezing in a gas of Coulomb charges
we find it crucial to start by  
realizing that upon increasing the size of the charges (cutoff
$a$), their core energy aquires a random component from the 
potential $V_{{\bf r}}$. Increasing $a$ to
$\tilde{a}=a (1+dl)$, this potential  
$V_{{\bf r}}$ with logarithmic correlation splits into the rescaled
logarithmically correlated potential $V^{>}_{{\bf r}}$ at scale $\tilde{a}$, and a
random part $v_{{\bf r}}$ uncorrelated at scales larger than
$\tilde{a}$.
This local potential is naturally incorporated in the (renormalized) core energy
$E_{c}$, which thus becomes random and site dependent : 
$E_{c}^{(a)}\rightarrow
E_{c}^{(\tilde{a})}=E_{c}^{(a)}\pm v_{{\bf r}}$. 
 Thus, upon coarse graining, the fugacities of the $\pm 1$ charges of
the Coulomb gas become random (and dependent on the sign of the charge) :
$z_{\pm}=ye^{\pm \beta v_{{\bf r}}}$. Let us stress that 
the definition of $v_{{\bf r}}$ (and thus of $z_{\pm}$) depends on 
details of the cutoff procedure. However the RG
procedure developed here depends only on the logarithmic behaviour
of the correlator of $V^{>}_{{\bf r}}$ at large distances, which is cut-off independent.
As a result, a remarkable universality will emerge at the end of the procedure,
which happens to be related to universality of front solutions in non-linear
equations (see below). 

\begin{figure}
\centerline{\fig{14cm}{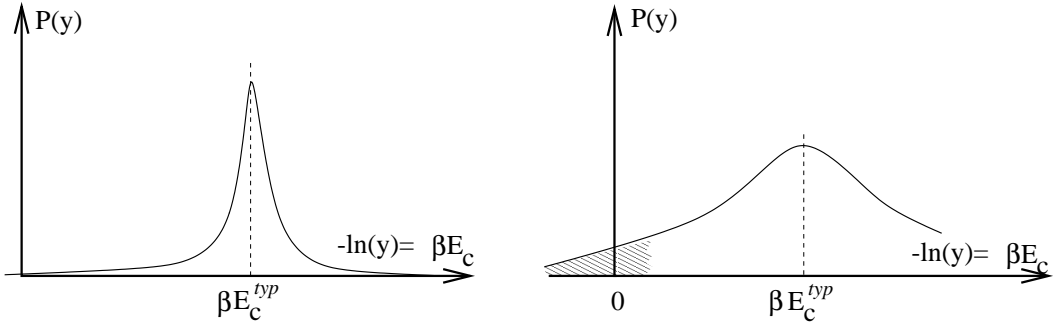}}
\caption{Schematic representation of the probability distribution $P(y)$ of
the local fugacities $y$, plotted as a function of 
$- \ln y = \beta E_c$ (where $E_c$ is the renormalized core energy).
At high temperature (left) it is narrow and a description in terms of the
averaged fugacity is correct. At low temperature
the distribution of $\beta E_c$ is broad and non gaussian
and information about the whole distribution is a priori needed}
\label{fig:distribution}
\end{figure}

 We are now faced with the description of the
scale-dependence of a Coulomb gas with random fugacities, and thus
{\it a priori} of the full corresponding fugacity distribution. 
Since we find that, at low temperature, the distribution of the local core energy 
does not remain gaussian under coarse graining we cannot restrict a priori the 
renormalization study to follow a small number of variables (such as
the mean and the variance) as illustrated in figure \ref{fig:distribution}.
Instead, we must study the scale dependence
of the full distribution $\tilde{P}(E_c)$ and especially the precise
form of its tails around $E_{c} \lesssim 0$ (i.e $y\sim 1$) which
control the low temperature physics (at $T=0$ defects appear where 
the local core energy is negative). This requires new techniques 
in two dimensions.

 Before developing a new RG procedure, we have to
find a correct perturbative parameter to study the disorder driven
transitions at low temperature. Around the pure topological
transition, this parameter corresponds to the charge fugacity $y$ (for the
most relevant charges $n=\pm 1$).
However, we expect that, upon lowering the temperature, a freezing of the
defects occurs. In that case, most sites have a large core energy
($y \sim 0$) while only a few sites with $E_{c} \lesssim 0$ are favorable to the
defects.
A natural choice for a perturbative parameter at low temperature 
in this highly non-homogeneous Coulomb gas is thus the {\it density of favorable
sites}\footnote{Actually, the random fugacities $y_{{\bf r}}$
depends on the sign of the charge : $z^{{\bf r}}_{\pm}$, and the
perturbative parameter is the density $P (z_{+}\sim 1)=P (z_{-}\sim
1)$ of sites favorable either to $+1$ charges or to $-1$ charges.
One must thus follow the RG flow of the full probability distribution
$P_{l}(z_{+},z_{-})$.} which we note
$P (y\sim 1)$. $P_l(y\sim 1)$ plays an analogous role here than 
$y(l)$ in the pure case: the
quasi-ordered phase is caracterized by a density $P_l(y\sim 1)$
decreasing with the scale $l$, while a disordered phase, where disorder
induced topological defects proliferate, corresponds to an increasing
$P_l(y\sim 1)$. This new small parameter allows us to study in 
a controlled perturbative expansion the physics of
the vortices at low temperature and around the transitions.

\begin{figure}
\centerline{\fig{8cm}{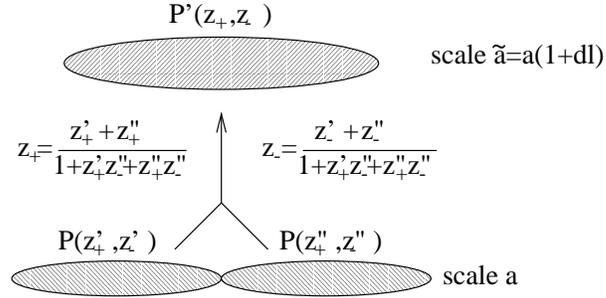}}
\caption{Schematic representation of the fusion of two local
environments upon coarse-graining. Two regions, distant from less than
the new cut-off $\tilde{a}$, are merged when the cut-off is increased.
The corresponding fusion rule for the distribution function of the
local variables $z_{\pm}^{{\bf r}}$ is shown on the figure.}
\label{fig:fusion-intro}
\end{figure}

 An essential contribution to the renormalization of $P_{l}(z_{+},z_{-})$,
absent in previous approaches, originates from 
what we call the {\it fusion of random environments}. The main
idea is the followoing. Each random fugacities $z_{\pm}^{{\bf r}}$ is
associated with a region of size $a$ around ${\bf r}$.
Upon increasing the cut-off, the size of these regions increases, and we
thus have to merge two regions distant from less than the new cutoff
$\tilde{a}$ (see fig. \ref{fig:fusion-intro}). The probability distribution
of the fugacities in the new region can be deduced from the
one in the two merged regions by a {\it fusion rule}
(see fig. \ref{fig:fusion-intro}). 
 The final renormalization of $P_{l} (z_{+},z_{-})$ can be formulated
into a single differential equation for the distribution function
$P_{l}(z_{+},z_{-})$  (given in Section \ref{part:replica}). 
Most interestingly we find that the universal features at
the transition of the model ($\sigma^{R} =\frac{1}{8}$) do not depend
on the precise definition of these regions. The fusion of environments
corresponds to a non-linear term in the differential equation, which
can thus be affected by a change of cut-off procedure. Quite remarkably, the
universality class at the new transition is determined by the properties
of the front solutions of this differential equation
(i.e their velocity),
that {\it do not depend} on the precise form of this non-linear term. Hence
universality in the usual RG sense finally appears from non trivial
results of front propagation in non-linear equations.

This differential RG equation for $P_{l} (z_{+},z_{-})$, together with
the screening equations for $J$ and $\sigma $, can be obtained by 
a systematic and perturbatively controlled RG procedure. One
formulation consists in using the replica trick to average the free
energy over disorder, leading to a Coulomb gas with $m$-component
charges. In this formulation, the infinite set of fugacities,
associated with the vector charges $Y[{\bf n}]$, encode the
distribution function $P_{l} (z_{+},z_{-})$ of the random fugacities
$z_{\pm}$. Hence, since this last distribution cannot be {\it a priori}
described by a finite number of moments at low temperature, we have to
consider the scaling behaviour of the fugacities of all the replica
charges with components $0,\pm 1$. The $m\to 0$ limit of this infinite
set of RG equations is then taken by using an appropriate
parametrization in terms of $P_l(z_{+},z_{-})$ which yields
the non linear RG equation for $P_l(z_{+},z_{-})$. The second, equivalent
formulation, does not rely on replica. It is constructed 
using a new expansion of physical quantities in the
``number of independent regions'', introduced in this paper.

\subsection{Disorder induced topological transitions and connection with DP}

 After some transformations, we reduce the non-linear RG equation
that governs the scale dependence of the distribution 
$P_{l} (z_{+},z_{-})$ to the 
celebrated Kolmogorov-Petrovskii-Piscounov (KPP) equation
(also named the Fisher equation). The known results on this 
equation allow us to derive the form of the tails of the distribution
and thus to obtain the phase diagram. 
In particular, the velocity of the front solutions of the KPP equation
directly determines whether the small 
parameter $P_{l} (y\sim 1)$ increases or decreases.
Hence using known results on the selection of this velocity, we
obtain the phase diagram of the figure
\ref{fig:phasediag-intro} expressed in renormalized variables.
The critical behaviour at the transition $\sigma =\frac{1}{8}$,
follows from the finite size corrections to the velocity corrections 
of the KPP front and defines a new universality class.
In particular, the correlation length $\xi$
diverges at the $\sigma =\sigma _{c}=\frac{1}{8}, T=0$ transition as  
\[
\xi \sim \exp \left(\frac{cte}{|\sigma -\sigma  _{c}|} \right)
\]
in contrast with the KT behaviour found in \cite{rubinstein83}.
Note that recent numerical simulations 
\cite{maucourt97,kosterlitz97} of this model seem to agree with
the phase diagram of fig. \ref{fig:phasediag-intro} and it 
would be interesting to also determine numerically the 
precise critical behaviour at low temperature.

 Besides the caracterization of the new critical behaviour 
our work also enables to study the freezing of vortices,
which occurs below a temperature $T_{g}$ (see figure
\ref{fig:phasediag-intro}). This freezing corresponds to a transition
for the single charge problem, whose study with our new RG technique
is presented in \cite{carpentier99ter}. In particular, this
renormalization approach draws a precise connection
between the physics of freezing of the XY defects and the problem of
branched processes (or of random directed polymers on Cayley trees 
(DPCT) for a discrete version) studied by Derrida and Spohn \cite{derrida88}.
This connection naturally emerges from our Coulomb gas RG equations
via the KPP equation which has also appeared in the exact solution of 
the DPCT problem \cite{derrida88}. It does not rely on any 
ad-hoc construction.

 The paper is organized as follows : in Section \ref{part:model}, the
random XY model is defined and its CG formulation is carefully
derived. In particular, the relation between the continuum limit and
the decomposition of the random potential $V_{{\bf r}}$ in to a local
part (random core energy) and a long-range potential is discussed in
section \ref{part:disdec}. Part \ref{part:replica} describes the
 renormalization method of the replicated Coulomb gas, and while a
direct method (without replica), which  consists in expanding the free
energy into the number of independent regions (random environments) is
presented in Section \ref{part:direct}. The RG equations are
analyzed in Section \ref{part:analysis} using results on the propagation of
KPP-like fronts. The consequences for the determination of
the phase diagram and the critical behaviour is detailed
in Section \ref{part:XYphase}. A formulation in terms of a
Sine-Gordon model is given in Section
\ref{part:sinegordon}. The comparison with previous approaches is
postponed to Section \ref{part:other-approaches}, and most of the
technical details can be found in the appendices.

\section{XY Model with random phases} \label{part:model}

\subsection{lattice Coulomb gas}

In this Section and in the following we study the
XY model with random phases  \cite{rubinstein83}.
We start with the model defined on the 2D square lattice
by its partition function:

\begin{equation}\label{xy}
 Z[A]=\prod_{i}\int_{-\pi}^{\pi} d \theta_{i}~ 
e^{- \beta H[\theta,A]} 
\quad \text{ with }\quad   H[\theta,A]=\sum_{\langle i,j \rangle} 
V(\theta_{i}-\theta_{j}-A_{ij}) 
\end{equation}
where the sum is over pairs of nearest neighbors (i.e over bonds) on
the lattice and $\beta=1/T$ is the inverse
temperature. The $A_{ij}$ are random gauge fields, independent from
bond to bond, each with gaussian distribution of variance
$\overline{A_{ij}^{2}}= \pi \sigma$. The periodic 
potential $V(\theta)$ is defined for the XY model by
$V(\theta)=- \frac{J}{\pi} \cos(\theta)$ 
where $J$ is the stiffness. In the limit $\sigma \to \infty$ this model
corresponds to the 2D ``gauge glass model''  \cite{hyman95,choi99}.
For finite $\sigma$ it was studied in Ref. 
 \cite{rubinstein83,tang96,scheidl97,korshunov96}.

The standard way to study this model is to decompose it into
spin waves and vortex degrees of freedom
$Z[A] = Z_{sw} Z_{CG}$. This decomposition can be performed
exactly (see Appendix \ref{part:CG}) for the corresponding Villain model 
defined by the potential
$$e^{- \beta V(\theta)}=\sum_{p=-\infty}^{+\infty} 
e^{-\frac{\beta J}{2 \pi} (\theta-2 \pi p)^2}$$
Technically this is the model which we study here.
It is reasonable however to expect that this Villain
model and the XY model should be in the same universality
class (as is the case without disorder). The RG
analysis contained in the following Sections is consistent with this
assumption 
\footnote{It can be shown to lead to unimportant additional random
terms in the bare fugacity}
. The vortex part is described 
in terms of a Coulomb gas
with integer charges $n_{{\bf r}}$ defined on the sites ${\bf r}$
of the dual infinite (square) lattice:
\begin{subequations}\label{square}
\begin{eqnarray}
&& Z_{CG} = Z_{\text{latt}} = \sum_{\{ n_{{\bf r}} \}} e^{- \beta H} \\ 
&& H =
- \frac{1}{2} \sum_{{\bf r} \neq {\bf r'}} 2 J n_{{\bf r}} 
  G_{{\bf r}-{\bf r}'} n_{{\bf r'}} 
- \sum_{{\bf r}} n_{{\bf r}} V_{{\bf r}}
\end{eqnarray}
\end{subequations}
where $G_{{\bf r} - {\bf r}'}=\int_{\bf k} G_{{\bf k}} (1- 
e^{i {\bf k}.({\bf r}-{\bf r}')})$ is the 2D Coulomb potential
with $G_{{\bf r} - {\bf r}'} \approx \ln|{\bf r} - {\bf r}'|$ at large distance.
$G_{{\bf k}}^{-1}=
\frac{1}{\pi} [2-\cos (k_{x}a_{o})-\cos (k_{y} a_{o})]$
is the lattice Laplacian and we denote 
$\int_{\bf k} \equiv 
\int_{-\pi/a_{o}}^{\pi /a_{o}} \int_{-\pi/a_{o}}^{\pi/a_{o}} 
\frac{dk_x dk_y}{(2 \pi)^2}$ where $a_{o}$ is the lattice spacing.
Note that the energy associated with a dipole of unit charge
of size $r$ is $2 J \ln r$.
As discussed in the Appendix \ref{part:CG} we can consider
a neutral CG (since $\int q=0$) with $\sum_{\bf r} n_{{\bf r}} =0$,
and neutrality has already been used to arrive at (\ref{square}).

In the vortex representation the random gauge fields ${\bf A}$ 
translate into {\it random dipoles} (along $z$) 
$q_{{\bf r}'}=\frac{1}{2 \pi} \nabla_{{\bf r}'} \times {\bf A}$ 
which couple to the vortex charges
via the Coulomb potential. As detailed
in Appendix \ref{part:CG}
this results in a gaussian 
bare disorder potential $V_{{\bf r}} = - 2 J \sum_{{\bf r}'}
G_{{\bf r} - {\bf r}'} q_{{\bf r}'}$. An important feature
of this problem is that the disorder potential seen by
the charges (the vortices) has logarithmic
{\it long range correlations}
\begin{eqnarray}
\overline{(V_{\bf r} - V_{{\bf r}'})^2} = 4 \sigma J^2 
\ln|{\bf r}-{\bf r}'| + O(1)
\end{eqnarray}
for ${\bf r} \neq {\bf r}'$, since 
$\overline{V_{{\bf k}} V_{-{\bf k}}}= 2 \sigma J^{2} G_{{\bf k}}$.
Note that at a given point $\overline{V_{\bf r}^2} \approx 2 \sigma J^2
\ln L$ where $L$ is the system size and that
up to now, these are exact transformations.

Before defining the continuum limit of this Coulomb gas, we note that
an alternative definition to this lattice Coulomb gas consists in
labelling the {\it non zero charges} in (\ref{square}) by their
positions ${\bf r}_{i}$ and their corresponding charge $n_{i}$.
Instead of the integer field $n_{{\bf r}}$ of (\ref{square}) defined
at each site of the lattice, a configuration is represented by the set
$\{n_{i},{\bf r}_{i} \}$. We obviously have
$n_{\bf r}=\sum_{i}n_{i}\delta_{{\bf r},{\bf r}_{i}}$. With this 
representation, the partition function of (\ref{square}) reads 
\begin{equation} \label{square-bis}
Z_{\text{latt}}= 1 + \sum_{p>0}
\sum'_{\{n_{1},\dots n_{p} \}}
\sum_{{\bf r}_{1}\neq \dots \neq {\bf r}_{p}}
e^{\beta J\sum_{{\bf r}_{i}\neq {\bf r}_{j}}
n_{i}G_{{\bf r}_{i}-{\bf r}_{j}} n_{j} +
\beta \sum_{i} n_{i}V_{{\bf r}_{i}}}
\end{equation}
 where $G_{{\bf r}_{i}-{\bf r}_{j}}$ and $V_{{\bf r}_{i}}$
 have been defined after eq. (\ref{square}) and the sum over the charge
configurations (primed sum) counts each distinct neutral configuration of 
non zero charges only once\footnote{This leads to the factor 
$1/\prod_{n} N(n)!$ in the
definition of the configuration sum of the CG, where $N(n)$ is the
total number of particles of charge $n$}.

\subsection{continuum limit and decomposition of the disorder}
\label{part:disdec}

In order to implement a renormalization procedure 
one first needs to introduce a continuum version
of this model. In the usual approach to 2D Coulomb gas
the continuum limit is obtained by replacing the 
lattice Coulomb interaction by the
approximation  \cite{kosterlitz74,jose77}
\begin{eqnarray}  \label{approx}
G_{{\bf r} - {\bf r}'} \approx 
G_{{\bf r} - {\bf r}'}^{{\rm app}} = 
\left(  \ln \left(\frac{| {\bf r} - {\bf r}'|}{a_{o}}\right)
+ \gamma\right)(1-\delta^{(a_{o})}_{{\bf r},{\bf r}'})
\end{eqnarray}
where $\delta^{(a_{o})}_{{\bf r},{\bf r}'}=1$ for 
$|{\bf r}-{\bf r}'|<a_{o}$ and
$0$ otherwise, and $\gamma=\ln(2 \sqrt{2} e^C)$ with
$C=0.577216$ is the Euler constant. This approximation is
excellent  \cite{itzykson89} on the lattice
for $|{\bf r} - {\bf r}'|  \gtrsim a_{o}$. In the standard
method the continuum CG model
is then defined by considering a gas of integer hard core charges 
$n_{{\bf r}_i}$ at point ${\bf r}_i$ of diameter
$a_{o}$ which interact with $G_{{\bf r} - {\bf r}'}^{{\rm app}}$.
Using neutrality $\sum_{{\bf r}} n_{{\bf r}}=0$ and (\ref{approx}),
allows to rewrite 
 \cite{kosterlitz74,jose77} the hamiltonian
(\ref{square-bis})
 as a sum of a simple logarithmic
interaction 
$- J \sum_{i \neq j} n_{{\bf r}_i}
\ln \left(\frac{|{\bf r}_i-{\bf r}_j|}{a_{o}}\right) n_{{\bf r}_j}$ 
between the hard core charges and
a {\it fugacity term} $\ln y \sum_{i}  n_{{\bf r}_i}^2$ of bare value
$y=e^{-\beta E_c}$ where $E_c = \gamma  J$ can be interpreted
as the bare core energy for the defects of the model (\ref{square}).

In presence of disorder, special care has to be 
taken to define properly the continuum limit for
the random potential, since its correlator is
logarithmic. Since on the lattice the
correlator of the disorder $G$ is the same 
as the Coulomb interaction, it is consistent to use
the same $G^{{\rm app}}$ for the disorder in
the continuum model. This immediately leads to
the fact that the disorder $V_{{\bf r}}$ must be separated in
two parts, using (\ref{approx}): a {\it long range correlated gaussian} part
$V^{>}_{{\bf r}}$ and a {\it local} part $v_{\bf r}$:
\begin{subequations} 
\label{decomposition}
\begin{eqnarray}
&& V_{{\bf r}} = V^{>}_{{\bf r}}  + v_{{\bf r}}  \\
&& \overline{(V^{>}_{\bf r} - V^{>}_{{\bf r}'})^2} 
=4 \sigma  J^2 \ln\left(\frac{|{\bf r}-{\bf r}'|}{a_{o}} \right)
 (1-\delta^{(a_{o})}_{{\bf r},{\bf r}'}) \\
&& \overline{v_{{\bf r}} v_{{\bf r}'}} =
2 \sigma  J^2 \gamma \delta^{(a_{o})}_{{\bf r},{\bf r}'}
\end{eqnarray}
\end{subequations} 
with no cross correlation.
Using this decomposition we can now write the
partition function of the continuum model:
\begin{eqnarray}\label{Zcont}
&& 
Z_{\text{cont}} =
1 + \sum_{p=2}^{p=\infty} 
\sum'_{n_1,\dots,n_p}
\int_{|{\bf r}_{i}-{\bf r}_{j}|\geq a_{o}} 
\frac{d^{2}{\bf r}_{1}}{a_{o}^2}  \dots \frac{d^{2}{\bf r}_{p}}{a_{o}^2} 
e^{ - \beta H[n,r] } \\
&& H[n,r] = - J 
\sum_{i \neq j} n_i \ln \left(\frac{|{\bf r}_i-{\bf r}_j|}{a_{o}}\right) n_j 
- \sum_{i} n_i V^{>}_{{\bf r}_i} - \sum_{i} \ln Y[n_i,{\bf r}_i]
\end{eqnarray}
where the primed configuration sum counts only once each distinct
neutral charge configuration. We have introduced the
{\it spatially dependent fugacity}, of bare value, from
(\ref{approx},\ref{decomposition}):
\begin{eqnarray} \label{locfug}
\ln Y[n, {\bf r}] = - \gamma \beta J  n^2  + \beta n v_{\bf r} 
\end{eqnarray}
Thus we find that disorder favors some regions resulting in a
local fugacity for $\pm 1$ charges $y({\bf r})=e^{-\beta E_c({\bf r})}$
with a core energy $E_c({\bf r}) = E_c \pm v_{\bf r}$ 
which now varies from point to point.
Thus one anticipates that problems will arise in the
conventional fugacity expansion if $y({\bf r})$ varies 
{\it strongly} from point to point. In addition, note that the local
fugacities are different for $+$ charges and $-$ charges
(although there is still a statistical $+|-$ symmetry).

We have thus defined a continuum model with a particular
cutoff procedure. This is apparent, e.g. in the decomposition 
of the disorder that we have used. This decomposition is
cutoff dependent and we have taken care to chose the same
cutoff procedure for the disorder term and the interaction.
We have chosen here a real space hard cutoff procedure, which is 
often used in CG studies. Other cutoff procedures can
be used. It is natural to expect, and we will partially 
verify that the large scale results will not depend
on the particular procedure chosen.

The alert reader will have already noticed that 
the disorders $V^{>}_{{\bf r}}$ and $v_{{\bf r}}$ as defined above
in (\ref{decomposition}) are not strictly speaking physical,
since the Fourier transform of their respective correlators is not positive.
This is an artefact of this particular choice of a real space 
hard cutoff. It is not a serious problem, and is easily cured by choosing
instead a cutoff in Fourier space. The resulting decomposition
of disorder is then completely legitimate. This is further explained
in Appendix \ref{part:cutoff}. For simplicity, we will however proceed
using the above cutoff choice, which has illustrative value, keeping
in mind that the more legitimate choice detailed in 
Appendix \ref{part:cutoff} can be used instead, completely equivalently
at all stages, for technical rigor.

We now turn to the renormalization of the model.

\section{Renormalization using replica} \label{part:replica}

In this Section we study, using replica, the renormalization group
properties of the disordered Coulomb gas defined by
(\ref{square}). In the present case the replica method is particularly
convenient in order to perform the combinatorics necessary to renormalize
consistently the model.

The strategy is first in (\ref{part:replicacg}) to transform the 
model (\ref{square})
into a vector Coulomb gas with $m$-replica charges. The fugacities
of these $m$-vector charges will then naturally encode the 
distribution of the spatially dependent fugacities defined above 
(see (\ref{locfug})).
The renormalization group (RG) equations for these fugacities are derived
in (\ref{part:replicarg}) for fixed $m$. By a suitable parametrization,
in (\ref{part:replicalimit}) we then extract in the $m \to 0$ limit the 
RG equations which 
describe the scale dependence of the full distribution $P(z_+,z_-)$ of the
local fugacities of the topological defects in the original disordered 
model (\ref{square}),
as well as the scale dependence of the stiffness $J$ and long wavelength disorder
$\sigma$.

\subsection{Replica Coulomb gas, continuum limit and fugacity expansion}
\label{part:replicacg}

We start again from the model (\ref{square}) on a square
lattice. As is well known, disordered averaged correlation
functions and free energy can be obtained by studying the
replicated partition function (generating function) 
$Z^{m}_{\text{latt}}$ in the
limit $m \to 0$. For integer $m$, $Z^{m}_{\text{latt}}$
can then be written exactly as a CG with m-component {\it vector} charges 
$n^a_{\bf r}$, $a=1,..m$ (each $n^a_{\bf r}$ is integer),
living on the sites of the lattice (note that $a$ denotes a 
replica index while ''$a_{o}$'' denotes the cutoff).
Averaging over the
bare disorder one obtains the partition sum of a fully coupled, 
translationally invariant, vector Coulomb gas on a square lattice: 
$\overline{Z^m_{\text{latt}}}=\sum_{\{n^a_{\bf r}\}}
e^{- \beta {H^{(m)}_{\text{latt}}}}$ with 
\begin{equation}  \label{latticereplica}
 \beta H^{(m)}_{\text{latt}} = -
\sum_{{\bf r} \neq {\bf r}'} K_{ab} 
~n^a_{\bf r} G_{{\bf r}-{\bf r}'} n^b_{{\bf r}'}
\end{equation}
where $K_{ab}=\beta J \delta_{ab} - \sigma \beta^2 J^2$ and summation over repeated
replica indices is assumed unless otherwise specified.

The next step is to approximate the lattice replica
model (\ref{latticereplica}) by a continuum
Coulomb gas with $m$-component vector charges. In the following
we consider a hard core cutoff in real space. The problem of the choice and
consequences of the cutoff procedure is rather subtle here
and will be discussed below. Using the approximate propagator
(\ref{approx}) we obtain the continuum hamiltonian 
\begin{subequations} \label{replicah}
\begin{eqnarray}
\beta H^{(m)}_{\text{cont}}[n,r] &=& 
-  \sum_{i \neq j} K_{ab}
n^a_i G_{{\bf r}_{i}-{\bf r}_{j}}^{\text{app}}  n^b_j\\
&=&- 
\sum_{i \neq j}  K_{ab}
 n^a_i  \ln\left(\frac{|{\bf r}_i-{\bf r}_j|}{a_{o}} \right) n^b_j
- \sum_{i} \ln Y[{\bf n}_i]
\end{eqnarray}
\end{subequations}
where the charge ${\bf n}_i$ is
located in ${\bf r}_{i}$. In the second equality we have used
the neutrality of the Coulomb gas, i.e. $\sum_{{\bf r}}n^{a}_{{\bf r}}=0$ for
each $a=1,..m$, to introduce the local fugacity
$Y[{\bf n}]$ which is a function of the whole set of components
of the {\it vector} charge ${\bf n}=(n^1,..n^m)$.
Its bare value from (\ref{approx})
is a simple quadratic function  \cite{scheidl97}:
\begin{eqnarray}
Y[{\bf n}]_{\text{bare}} = e^{- n_a \gamma K^{ab} n_b}
\end{eqnarray}
This quadratic form results from the {\it Gaussian nature of the bare local
disorder}
and corresponds to (\ref{locfug}) in the unreplicated version.
If this form was preserved by the RG, as was implicitly {\it assumed}
in  \cite{scheidl97}, one would be able to study the model using only 
two coupling constants. However this is not the case. As shown below, 
the vector charge fugacity $Y[{\bf n}]$ has a non trivial flow under RG 
and does not remain purely quadratic. The local disorder does not 
remain gaussian and we will have to follow its 
full probability distribution.

We now study the scale dependent properties of the $m$-component vector Coulomb
gas using the {\it expansion in the vector charge fugacity} 
$Y[{\bf n}]$. Although it is the natural way to study the renormalization
of a vector Coulomb gas, it may seem at this stage somewhat formal.
This is not so however since, as will become clear below, it turns out
to correspond exactly to the expansion in the number of rare favorable regions
of local disorder, which is the physically relevant (and novel) expansion for this
model. For the replicated partition function in
the continuum model this vector fugacity expansion reads:

\begin{eqnarray}  \label{expansion}
&& \overline{Z^m} = 1 + \sum_{p=2}^{p=\infty} 
\sum'_{n^a_1,\dots,n^a_p}
\int_{|{\bf r}_{i}-{\bf r}_{j}|\geq a_{o}} 
\frac{d^{2} {\bf r}_{1}}{a_{o}^2}\dots \frac{d^{2} {\bf r}_{p}}{a_{o}^2}
 e^{- \beta H^{(m)}_{\text{cont}}[n,r]}
\end{eqnarray}
which contains fugacities via (\ref{replicah})
and the primed sum over charge configurations counts each distinct 
neutral configuration only once. In presence of disorder,
infrared divergences appear everywhere in the low temperature XY phase
 \cite{korshunov93}. To treat these divergences we now turn to the RG method.

\subsection{RG equations} \label{part:replicarg}

We perform the RG analysis of the present $m$-component vector
Coulomb gas on the partition function $\overline{Z^m}$.
It is a simple extension of the analysis for the
scalar Coulomb gas  \cite{kosterlitz74,nienhuis87}. Details are presented
in the Appendix \ref{part:rgreplica} and we only sketch the
method here. For any fixed $m$ it is possible to leave the form
of the expansion (\ref{expansion}) unchanged under the
increase of the hard core cut-off $a_{o} \rightarrow a_{o} e^{dl}$
provided one defines scale dependent coupling constants $K_l^{ab}$
and fugacities $Y_{l} [{\bf n}] $. This corresponds to the (one loop)
 renormalizability of the $m$-component vector model, which we checked 
here to order $Y[{\bf n}]^2$.
The RG flow equations which determine these couplings are found as
(see Appendix \ref{part:rgreplica}):

\begin{subequations}\label{rgrep}
\begin{eqnarray}   \label{rgrep1}
&& \partial_l (K_l^{-1})_{ab} = c_1 \sum_{{\bf n} \neq 0} n^a n^b 
Y[{\bf n}] Y[-{\bf n}] \\ 
\label{rgrep2}
&& \partial_l Y[{\bf n}] = ( 2 -  n^a K_{ab} n^b ) Y[{\bf n}]
+ c_2  \sum_{
\genfrac{}{}{0pt}{}
 {{\bf n}'+ {\bf n}'' = {\bf n}} 
{{\bf n}',{\bf n}'' \neq 0 }
}
Y[{\bf n}'] Y[{\bf n}'']  
\end{eqnarray}
\end{subequations}
with $c_1= 2 \pi^2$, $c_2 = \pi$ for our hard cut-off procedure,
and the second equation (\ref{rgrep2}) is defined 
only for ${\bf n} \neq {\bf 0}$. The first equation (\ref{rgrep1}) comes 
from the 
annihilation of dipoles of opposite replica vector charges separated by 
$a_{o} \leq |{\bf r}_i-{\bf r}_j| \leq a_{o} e^{dl}$. It
gives the renormalization of the interaction (screening by small dipoles)
and of the disorder. Simple rescaling gives the first 
term of the second equation (\ref{rgrep2}), i.e.
the naive scaling dimension of $Y[{\bf n}]$. 

\begin{figure}
\centerline{\fig{14cm}{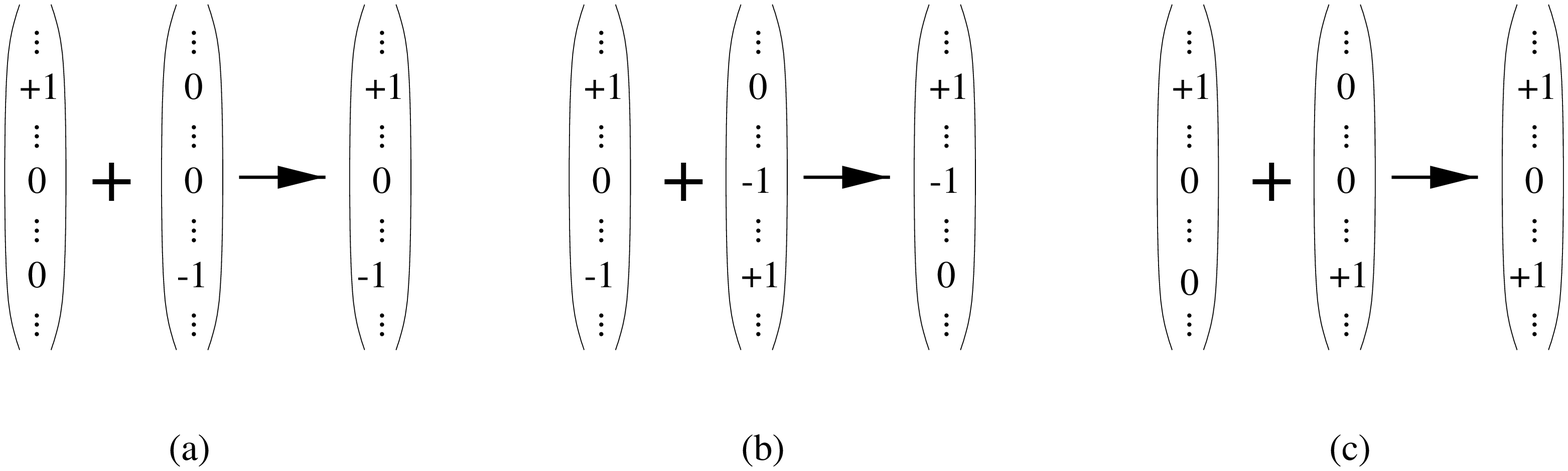}}
\caption{
Fusion of replicated charges.
In cases where the only charges which are relevant are the
single component $\pm 1$ charges the fusion (a) always leads to charges with
a higher number of components $\pm 1$ and are thus irrelevant: in that
case the fusion upon coarse graining does not play any role.
On the contrary, in the present case multicomponent charges are also
relevant and must be taken into account: fusion of charges (b) does not necessarily
increase the number of $\pm 1$ components and must be taken into 
account in the renormalization of the fugacities
of the relevant charges. (c) (and (a)) illustrate examples of fusion of vector charges 
in the replicated Coulomb gas which {\it do not} correspond a true fusion of scalar charges
in a single environment (nor to any annihilation as in (b)).
Instead they correspond to the ``fusion of environments'' 
and encode for the renormalization of the probability distribution of local fugacities
upon coarse graining (as they keep track of some non trivial correlations,
see discussion in the text).
\label{fusionCG2}}
\end{figure}

The second 
contribution in (\ref{rgrep2}) comes from the possibility of {\it fusion of
two replica vector charges} upon coarse graining (see fig.
\ref{fusionCG2}). While such term
can be neglected in the (pure) scalar Coulomb gas (as it yields less
relevant operators) it is usually crucial when studying
most vector CG models, as e.g. in the analysis of two dimensional
melting transition  \cite{nelson79}. Indeed following too closely the
analysis for the pure XY model has led previous studies
 \cite{rubinstein83,nattermann95,tang96,scheidl97,korshunov96}
to miss the possibility of fusion and thus such a contribution.
As we see in the following it has important and non trivial consequences
for the physics of the low T phase and the transition. 
For the present disordered Coulomb gas, contrarily to
the conventional analysis  \cite{nelson79}, one cannot hope
to capture the most relevant operators
by restricting to single component charges (e.g. $n^a=\pm \delta_{a a_1}$).
This was recently emphasized by Scheidl  \cite{scheidl97}.
However, since this leads to considering multicomponent 
vector charges, it is thus crucial to treat properly this 
fusion term, which was not done previously (e.g. in  \cite{scheidl97}). 
Moreover, discarding this term in (\ref{rgrep2}) leads to a set 
of RG equations which is {\it not consistent} to their lowest order 
$ O(Y[{\bf n}]^2)$.
This term may {\it a priori} modify the scaling dimensions
in a non trivial way in the $m \to 0$ limit, and it is thus
crucial to study carefully its effect.

Before doing so in the next Section,
let us give for completeness the renormalization
of the free energy density per replica,
defined as $f_m = a_{o}^2 L^{-2} \frac{1}{m} \ln \overline{Z^m}$.
It reads:

\begin{eqnarray}   \label{rgrep3}
\partial_l f_m = 2 f - \pi^2 \frac{T}{m} \sum_{{\bf n}' \neq 0} 
Y[{\bf n}'] Y[-{\bf n}']
\end{eqnarray}
from which the flow of the free energy density can be obtained as
$f=\lim_{m \to 0} f_m$.

\subsection{limit $m \to 0$ and fusion-diffusion formalism}

\label{part:replicalimit}

We now have to find an analytical continuation to
$m \to 0$ of the whole set of RG equations (\ref{rgrep}).
This is a priori a formidable task because (\ref{rgrep}) are in fact,
for arbitrary $m$, an infinite set of coupled equations.
Remarkably, in the process of performing this analytical
continuation, an appealing physical interpretation
in terms of probability distributions of local fugacities
(local disorder : see Section \ref{part:disdec}) will emerge
naturally and be our guide in the following.

\subsubsection{parametrization of the fugacities and analytical continuation}

As a first step, we will consider only charges with components
$n^a = 0, \pm 1$ in each replica. How to incorporate higher charges
(e.g. $n^a = \pm 2 \dots$) is discussed in Appendix \ref{highercharges}, 
where it is shown that they are less relevant in the
region of the phase diagram studied here.
We first remark that the possible forms of the $Y[{\bf n}]$
are severely constrained. Replica permutation
symmetry, which we will assume here and is preserved by the RG,
together with $n^a = 0, \pm 1$ implies that $Y[{\bf n}]$ depends only 
on the number
$n_{+}$ and $n_{-}$ of $+1/-1$ components of ${\bf n}$. A
natural possible parametrisation of $Y[{\bf n}] \equiv Y[n_{+},n_{-}]$
consists in introducing a function of two arguments $\Phi(z_{+},z_{-})$
such that:
\begin{eqnarray} \label{eq:ydef}
&& Y[{\bf n}] = \langle z_{+}^{n_{+}} z_{-}^{n_{-}} \rangle_\Phi =
 \langle
\prod_{a}\left[\delta_{n^{a},0}+z_{+}\delta_{n^{a},+1}+z_{-}\delta_{n^{a},-1}
\right] \rangle_\Phi 
\end{eqnarray}
where we denote 
$\langle A \rangle_{\Phi (z)} = \int_{z_{+},z_{-}} A~ \Phi (z_{+},z_{-})$.
Our strategy is to establish an RG equation for
$\Phi (z_{+},z_{-})$ (in the limit $m \to 0$)
whose solutions $\Phi_l (z_{+},z_{-})$ will parametrize solutions 
$Y_l[{\bf n}]$ of 
(\ref{rgrep1}, \ref{rgrep2}).

Let us now examine how (\ref{rgrep2}) can be transformed
in an integro-differential equation for  $\Phi (z_{+},z_{-})$.
The technical details are given
in Appendix \ref{part:replimit} . The first terms in the r.h.s. 
of (\ref{rgrep2})
translate into a differential operator $(2 + \mathcal{O}) \Phi$ 
where:
\begin{eqnarray} \label{oo}
\mathcal{O}=\beta J (2 + z_+ \partial_{z_+} + z_- \partial_{z_-}) 
+ \sigma (\beta J)^2 (z_+ \partial_{z_+} - z_- \partial_{z_-})^2
\end{eqnarray}
Using new ``core energy'' variables $u$,$v$ such that $z_{\pm}=e^{\beta (u \pm v)}$,
and the corresponding function $\tilde{\Phi}(u,v)$
such that $\tilde{\Phi}(u,v) du dv= \Phi (z_{+},z_{-}) dz_{+} dz_{-}$,
it can be interpreted as a {\it diffusion process}
since the first term of (\ref{rgrep2}) now
translates into $(2 + \tilde{\mathcal{O}}) \tilde{\Phi}$ with 
$
\tilde{\mathcal{O}}=J \partial_u  + \sigma J^2 \partial_v^2
$.

To deal with the second term we first extend the
RG equation (\ref{rgrep2}) so as to allow for zero charge ${\bf n}=0$,
since it is easier to continue analytically unrestricted sums.
 After some combinatorics (see appendix \ref{part:replimit}), we find that
using the representation
(\ref{eq:ydef}), (\ref{rgrep2}) can be rewritten completely equivalently
in terms of $\Phi$ as:
\begin{eqnarray} \label{loose}
 \partial_l \Phi(z_+,z_-) &=& (2 + \mathcal{O}) \Phi(z_+,z_-) 
- 2 c_2 \mathcal{N}\Phi(z_+,z_-) + c_2 \mathcal{N}^{2} \delta(z_+) \delta(z_-) \\
&& 
+ c_2 \int_{z'_+,z'_-,z''_+,z''_-} \Phi(z'_+,z'_-) \Phi(z''_+,z''_-) 
\nonumber \\
&&~~~\times 
\delta\left( z_+ - \frac{z'_+ + z''_+}{1 + z'_-  z''_+ + z'_+  z''_-} \right)
\delta\left( z_- - \frac{z'_- + z''_-}{1 + z'_-  z''_+ + z'_+  z''_-}\right)
\nonumber
\end{eqnarray}
where $\mathcal{N}=\int_{z_{+},z_{-}}\Phi (z_{+},z_{-})$. This equation describes
the scale dependence ($l$) of the function $\Phi_l (z_{+},z_{-})$ which
parametrizes the whole set of scale dependent fugacities $Y_l[{\bf n}]$ in
the limit $m \to 0$.

\subsubsection{RG equations and fugacity distribution}

Up to now the function $\Phi(z_+,z_-)$ has been introduced
as a generating function to parametrize the fugacities $Y[{\bf n}]$. It is
a priori an arbitrary function and in particular $\mathcal{N} = \int \Phi$ is still
undetermined. In this paragraph we will exchange
$\Phi(z_{+},z_{-})$ for a {\it physical} function $P(z_{+},z_{-})$  of norm unity, 
which will be interpreted in the following as the probability distribution
for the local fugacities $z_+,z_-$ of $\pm 1$ charges.
We start from the above equation (\ref{loose})
for $\Phi (z_{+},z_{-})$ which can be simply interpreted as 
describing the sum of two processes. Defining from the
random fugacities $z_{\pm} = e^{\beta(u \pm v)}$ the 
random core energy variables $E_c^{\pm}=- (u \pm v)$, the first
process in (\ref{loose}) corresponds to a brownian diffusion for
$v$ (i.e the local disorder potential as in (\ref{locfug}))
together with a convection for $u$.
The second process involves a fusion, with a rate $c_2$ upon increase of the
cutoff, of two sets of random variables
$(z'_+,z'_-)$, $(z''_+,z''_-)$ into a single one $(z_+,z_-)$ 
according to the transformation law:
\begin{equation}
\left( \{ z'_{\pm} \} ; \{ z''_{\pm} \} \right) \longrightarrow 
z_{\pm} = \frac{z'_{\pm} + z''_{\pm}}{1 + z'_-  z''_+ + z'_+  z''_-}
\end{equation}
as in a $A + A \to A$ reaction. The term $-2 c_2 \mathcal{N} \Phi$ in
(\ref{loose}) corresponds to a loss of two charges, while
the last term corresponds to a creation of the fused one.
The term $\delta(z_-) \delta(z_+)$ keeps track of
the ``dead charges'' which disappear by setting them to $0$
(since they decouple from the system). It is in a sense only a counting device,
since by construction $\mathcal{N} = \int \Phi$ is unchanged upon fusion.
We thus introduce $\Phi_{>}(z_+,z_-)$ restricted to $z_+>0, z_{-}>0$, such that
$\Phi = \Phi_{>} + (\mathcal{N} - \mathcal{N}_>) \delta(z_-) \delta(z_+)$
where $\mathcal{N}_> = \int_{z_+ >0, z_- >0}  \Phi$ is the total
weight of non zero charges. Integrating (\ref{loose}) over
$z_+ >0, z_- >0$ we obtain that:
\begin{equation}
\partial_l \mathcal{N}_> = 2 \mathcal{N}_> - c_2 \mathcal{N}_>^2
\end{equation}
Thus  in the presence of fusion it converges quickly towards
$\mathcal{N}_>^* = 2/c_2$. 

Since $\mathcal{N}_> = \int \Phi_{>}$ converges to a constant, this suggests
to introduce a normalized function 
$P(z_+,z_-) =\Phi_{>}/\mathcal{N}_>^* $. As shown below 
it is natural to interpret $P(z_+,z_-)$ as a {\it probability distribution}.
From (\ref{loose}) we find that it obeys the following RG equation:

\begin{multline}\label{rgeqp}
 \partial_l P(z_+,z_-) =  \mathcal{O} P - 2 P(z_+,z_-) \\
+2 \left<  \delta( z_+ - \frac{z'_+ + z''_+}{1 + z'_-  z''_+ + z'_+  z''_-})
\delta( z_- - \frac{z'_- + z''_-}{1 + z'_-  z''_+ + z'_+  z''_-})  \right>_{P' P''} 
\end{multline}
where $\langle .. \rangle_{P' P''}$ denotes 
$\int_{z'_+,z'_-,z''_+,z''_-} P(z'_+,z'_-) P(z''_+,z''_-)$
and the probability conserving diffusion operator $\mathcal{O}$ has been defined
in (\ref{oo}).

The limit $m \to 0$ of the other RG equations (\ref{rgrep1}) which
give the renormalization of the  
stiffness $J$ and disorder strength $\sigma$ is performed in Appendix
\ref{part:replimit} using  
$\Phi$. Reexpressed in terms of $P$ they read:

\begin{subequations} \label{screening}
\begin{eqnarray} 
&& T \frac{dJ^{-1}}{dl} = \frac{4 c_{1}}{c_{2}^2}
\left< \frac{ z'_+ z''_- + z'_- z''_+ + 4 z'_+ z''_- z'_- z''_+ }
{(1 + z'_+ z''_- + z'_- z''_+)^2} \right>_{PP} \\
&& \frac{d\sigma}{dl} = \frac{4 c_{1}}{c_{2}^2}
\left< \frac{ (z'_+ z''_-  - z'_- z''_+)^2 }{
(1 + z'_+ z''_- + z'_- z''_+)^2}  \right>_{P P}
\end{eqnarray}
\end{subequations}            
where we have chosen (arbitrarily) to keep $T$ fixed and 
renormalize $J$ (only the combination $K=\beta J$ flows).

The above formulae (\ref{rgeqp},\ref{screening}) forms our
complete set of RG equations. As will become clear in the
following Sections, $P(z_+,z_-)$ represents the distribution
of the fugacities $z_+,z_-$  of local environments and 
the last term in (\ref{rgeqp}) corresponds to fusion of
environments upon coarse graining. Remarkably, once expressed
in terms of $P$ the coefficients of the above RG equations exhibit
some {\it universality}. 
The factor of 2 in the last term in (\ref{rgeqp}) arises from the fraction
of environments $\partial_l V/V =2$ which are fused when increasing the cutoff.
Note also that the coefficient $c_1/c_2^2$ which naturally appears in
(\ref{screening}) is not affected by a uniform rescaling 
of the fugacities. Note that some features of these equations are cutoff dependent,
as will be discussed in a following Section.

Finally, the flow of the free energy density is found to be:
\begin{eqnarray}\label{rgfree}
\partial_l f &=& 2 f - T \pi^2 
\left< \ln (1 + z'_+ z''_- + z'_- z''_+) \right>_{\Phi \Phi} \nonumber
\\
 &=& 2 f - 4 T \left< \ln (1 + z'_+ z''_- + z'_- z''_+) \right>_{P P}
\end{eqnarray}

These RG equations will be studied in Section \ref{part:analysis}.
First we will present another renormalisation procedure,
without replicas. Although it is technically more difficult to implement,
it allows for a more direct physical interpretation, which
in turns sheds some light on the more systematic replica method
presented above. In addition it may be more appealing to the
replicaphobic Section of the community.

\subsection{validity of the method and universality}

The above method which relies on an expansion
in the vector fugacities $Y[{\bf n}]$ can be 
justified provided there are {\it few vector charges} in the system 
i.e in the dilute limit. Even though the $Y[{\bf n}]$ may appear
as formal fugacities, the above RG equations can be justified
in an expansion of the exact renormalised potential $G^{R} (r)$
seen by two test charges distant of $r$ in $a_{o}/r$, as was
emphasized by Nienhuis \cite{nienhuis87}. It is even
claimed to be exact \cite{nienhuis87} in that limit\footnote{In that case 
a cubic term should be added to them, $- h
Y[{\bf n}] \sum_{{\bf n}' \neq 0} Y[{\bf n}'] Y[-{\bf n}']$ with
$h=\pi (\frac{8}{3} \pi + \sqrt{3})/2$ in the second
equation. We omitted this term as we found that it vanishes in the
limit $m \to 0$}. As will become apparent in the following Section the 
physical meaning of this diluted limit of {\it vector charges} 
exactly corresponds to the limit of a small density of regions favorable 
to the creation of frozen defects which is the physically relevant
limit in the regimes studied in this paper.

To understand how cutoff dependence comes in the method
used here, it is instructive to study the limit of zero disorder.
One can indeed check that one recover the usual results in the
limit of the pure case $\sigma=0$. This however, requires some
careful consideration of the cutoff procedure for the
vector CG representation. As discussed in the Appendix F,
even in the pure case the distribution $\Phi(z_+,z_-)$
which parametrizes the vector fugacities solution of
the CG RG equations can be non trivial. It does satisfy $P(z_+,z_-) 
\sim \delta(z_+ - z_-)$ but the fugacity $z_+ = z_- = y$ still has a 
non trivial ''distribution'' $\Phi(y)$ for a generic choice
of cutoff. There is no paradox there and it is compatible 
with the standard Korsterlitz-Thouless RG equation of the pure case,
as is explained in Appendix F, universality being recovered 
at small fugacity.

\section{Direct method of renormalisation}\label{part:direct}

In this Section we introduce a method to study the model (\ref{square})
and more generally Coulomb gas with disorder without using
replicas. We start in 
Section \ref{part:physics} from the general motivation and 
rederive the RG equation for $P(z_+,z_-)$ in (\ref{rgeqp})
in a more physical way. We also identify the small
parameter which allows to study perturbatively the
present problem. In Section \ref{part:exp} 
we introduce a quantitative and systematic method
to expand in this small parameter. The direct renormalization
approach using this expansion is performed
in Section \ref{part:directRG}. The connection with
the replica method of Section \ref{part:replica}
is presented in Section \ref{part:connect}.

\subsection{RG method for broad disorder: physical derivation}
\label{part:physics}

Let us first explain the spirit of the direct method 
and illustrate how one is led to the RG equation (\ref{rgeqp}),
derived more quantitatively in the next Sections.
We have seen in Section \ref{part:disdec} that 
the local disorder $v_{\bf r}$ defines the site dependent
fugacities. We concentrate on $\pm 1$ charges for 
which these fugacity variables read (see (\ref{locfug}))
\begin{eqnarray}  \label{locz}
z^{\pm}_{\bf r} = y_{\bf r} \exp(\pm \beta v_{\bf r})
\end{eqnarray}
and are quenched random variables with only short range spatial correlations. 
One now studies the system under a change of cutoff
$a_{o} \to a_{o} e^{dl}$ (coarse graining) which includes an
integration over the corresponding degrees of freedom.
We find that the coarse grained model remains of the
same form as the original one, with a renormalized stiffness
$J_l$, a renormalized gaussian long range disorder strength
$\sigma_l J^2_l$ and a local disorder distribution
$P_l(z_{+},z_{-})$. Note that, although the bare local disorder
$v_{\bf r}$ is gaussian, it becomes {\it non gaussian} under
coarse graining. This is a novel feature of the present
approach, at variance with previous
attempts at renormalizing the model  \cite{korshunov96,tang96,scheidl97}.
It complicates the analysis but is necessary to capture correctly the physics of the
model which is driven by the rare events.

The RG equations (\ref{rgeqp}, \ref{screening})
for the fugacity distribution $P_l(z_{+},z_{-})$ of
local environments (higher charges are less important and considered
later), for the stifness $J_l$ and for the correlated disorder
strength $\sigma_l$ can be understood from the following considerations.
The correction to the fugacity distribution $P_l(z_{+},z_{-})$
is the sum of two contributions:

(i) {\it rescaling}: the first observation
is that upon changing the cutoff, as can be seen from
its correlator (\ref{decomposition}) and is detailed below,
the long range disorder $V^{>}$ produces an additional local disorder
contribution which can be written as a renormalisation of the
local charge fugacity:
\begin{eqnarray}
z^{\pm}_{\bf r} \to z^{\pm}_{\bf r}  
e^{ \beta ( - J dl \pm dv_{\bf r} ) }  \label{rgrescaling}
\end{eqnarray} 
where $dv_{\bf r}$ is a gaussian random variable, uncorrelated from
site to site and with
$\overline{dv_{\bf r} dv_{{\bf r}'}} = 2 J^2 \sigma dl 
\delta^{(a_0)}_{{\bf r},{\bf r}'}$.
This contribution leads to an effective diffusion and drift in the
random core energy variables $E_c^{\pm} = - (u \pm v) = - T \ln z^{\pm}$ as 
$\partial_l E_c^{\pm} = J dl \mp dv$ and thus produces the 
first terms in (\ref{rgeqp}).

\begin{figure}
\centerline{\fig{8cm}{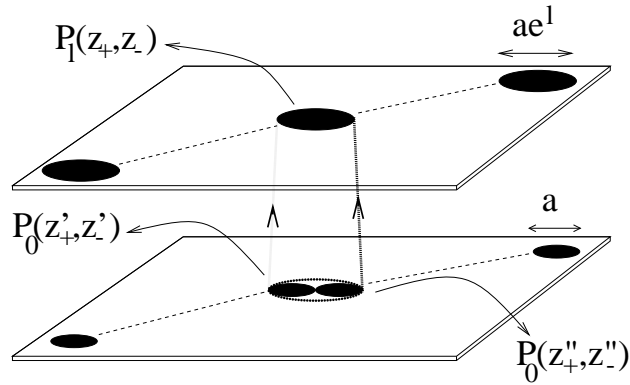}}
\caption{ \label{fig:fusion} 
schematic representation of the fusion of two local
environments: when the cutoff is increased the two regions centered around ${\bf r}_1$ 
(with local fugacities $z'_+, z'_{-}$) and 
${\bf r}_2$(with local fugacities $z''_+, z''_{-}$)
with $a_0 < |{\bf r}_1 - {\bf r}_2| < a_0 e^l$
are fused into a single region around ${\bf r} = \frac{1}{2} ({\bf r}_1 + {\bf r}_2)$
with fugacity $z_+, z_{-}$ }
\end{figure}

(ii) {\it fusion of environments}.
The second contribution comes from the fusion of environments.
Upon a change of cutoff, any two regions located around
${\bf r}_1$ and ${\bf r}_2$ with $a_0 < |{\bf r}_1 - {\bf r}_2| < a_0 e^{dl}$
have to be considered as a single region in the system with the
rescaled cutoff. As a consequence the two corresponding pairs of fugacities
$z^{\pm}_{{\bf r}_1}$ and $z^{\pm}_{{\bf r}_2}$ must be combined 
and replaced by a single pair of effective fugacity variables $z_{+},z_{-}$ 
associated with the new region at ${\bf r} =\frac{1}{2}({\bf r}_1 + {\bf r}_2)$,
as illustrated in Fig. \ref{fig:fusion}.
$z_{+}$ can be determined by estimating the relative Boltzman weight
$W_{+}/W_{0}$ to have a configuration with charge 1
(which lies either in ${\bf r}_1$ or ${\bf r}_2$)
versus a neutral one (either no charges or a dipole
in ${\bf r}_1$,${\bf r}_2$), and similarly for $z_{-}$.
This gives the fusion rule:
\begin{eqnarray}   \label{fusion1}
z^{\pm}_{\bf r} = \frac{ z^{\pm}_{{\bf r}_1}
+ z^{\pm}_{{\bf r}_2} }{1 + z^{+}_{{\bf r}_1} z^{-}_{{\bf r}_2}
+ z^{-}_{{\bf r}_1} z^{+}_{{\bf r}_2} }
\end{eqnarray}
The corresponding correction to the distribution $P_l(z_{+},z_{-})$
produces the last two terms of (\ref{rgeqp}).

Finally, the RG equation for $J_l$ and $\sigma_l$ can be obtained
from the screening by small dipoles of the effective interaction
and disorder between two infinitesimal test charges as described
in Section \ref{part:directRG2}.

Several comments are in order concerning this RG procedure.
First we note that in defining local fugacity variables
(\ref{locz}) we have added an explicit spatial dependence to
the part $y_{\bf r}$ of the fugacity which does not distinguish 
between a $+1$ and a $-1$ charge. This dependence is not
explicitly present in the bare model formula (\ref{locfug})
(although it is present if an additional small disorder in the
local stiffness $J_{ij}$ is included) 
but, as we can see from (\ref{fusion1}) it appears as soon as
fusion takes place (the fusion rule is not compatible with
a uniform $y_{\bf r}=y$). Second, there are some assumptions
underlying the RG procedure: technically we treat the local regions
as independent from point to point, we restrict $V^{>}_{{\bf r}}$ 
to be strictly gaussian, together with the usual assumptions
(e.g. short distance expansions) of the CG renormalization. These 
assumptions are consistent and
amount to discard less relevant operators. These irrelevant
operators can be identified within the
method using replica of Section \ref{part:replica}
where the above assumptions appear as standard in the RG 
of the $m$-component vector CG. For instance, the 
separation of the disorder into the two components
$V^>_{\bf r}$ and $v_{\bf r}$ corresponds in the replica method
to the natural splitting in (\ref{replicah})
between the vector fugacity local operator
$Y[{\bf n}]$ (originating from $v_{\bf r}$) and the off-diagonal 
replica Coulomb interaction
$K_{a \neq b}$ (from $V^>_{\bf r}$). This will be further apparent on the
equivalent Sine Gordon representation of the problem presented in
Section \ref{part:sinegordon}. Accordingly,
the definition of the independent local regions (and thus of the local disorder
environments and of the detailed form of the distribution
$P(z^+,z^-)$) is clearly cutoff dependent. So is the  
detailed form of the fusion rule (\ref{fusion1}). However,
universality of the physical results will be recovered in a remarkable way
in Section \ref{part:analysis}, independently
of the details of the cutoff procedure. 
As we will see, this is because, in the low temperature
limit, the above definitions and fusion rules
capture correctly (to order $P(z \sim 1)^2$)
the rare events which dominates the physics. They
correctly evaluate the universal part of $P(z^+,z^-)$ (its tails
in the low temperature region where they dominate the physics) while
they also correctly describe the weak disorder regime
at higher temperatures. 

Finally, we note that usual
charge fusion between certain types of replica charges, represented on
figure \ref{fusionCG2} corresponds,
in the method without replica, to fusion of environments. 

\subsection{Expansion of physical quantities in the number of local regions
(''number of points'')}
\label{part:exp}

To renormalize consistently the present model we need
a method which can handle in a systematic way 
broad distributions of local fugacities. We have found 
such a method, which we now introduce. It is based on 
a {\it systematic expansion of physical quantities 
in the number of independent points}. It generalises
the conventional fugacity expansion in $y$ of the pure
case, but is more powerful. In effect, it amounts to a
partial resummation of the conventional expansion. It is
versatile since, as we will see, it yields back the conventional
expansion in the pure case or at high temperature, but is also
able to handle the broad distributions which arise at low
temperature. The idea there is that only few rare regions
(favorable to the charges) in each environment will dominate the observables
and thus it becomes possible again to expand in the
density of such rare regions.

The idea underlying this expansion in the number of ``independent points''
is that the site-dependent fugacities associated with distinct ``points''
can be considered as statistically independent.
On a lattice, these ``independent points'' naturally correspond to the sites
of the lattice, while in the continuum model, their definition is necessarily cut-off
dependent. Upon coarse graining, we will be able to use this systematic
expansion to renormalize consistently the fugacity distribution associated with each
point.

For definiteness, we will show how to construct the expansion in number of points
on the free energy $F[V]=-\beta^{-1} \ln Z[V]$. It can be performed
in a given environment, keeping the full functional dependence
in the set of potentials $\{ V_{\bf r} \}$. The construction can then
be easily generalized to any physical quantity, such as arbitrary
powers of the free energy $\overline{F[V]^p}$ (which yield an expansion of
all moments averaging term by term over disorder) 
or all correlation functions of the field $n_{\bf r}$
which can be obtained from products of free
energies $\overline{F^{a_{1}}[V_{1}]\dots F^{a_{k}}[V_{k}]}$
by differentiation with respect to the potentials $V_{r}$.
In a second stage (next Section) we will use this expansion to
justify the RG equation for the disorder distribution.

We recall that the Coulomb gas model we consider is defined by 
its partition function
\begin{equation} \label{square-recall}
Z[V]  = 1 + \sum_{p>0}
\sum'_{\{n_{1},\dots n_{p} \}}
\sum_{{\bf r}_{1}\neq \dots \neq {\bf r}_{p}}
e^{\beta J \sum_{{\bf r}_{i}\neq {\bf r}_{j}}
n_{i}G_{{\bf r}_{i}-{\bf r}_{j}} n_{j} +
\beta \sum_{i} n_{i}V_{{\bf r}_{i}}}
\end{equation}
Here and below, as in (\ref{Zcont}), all formulae can be extended 
to the continuum model by replacing discrete sums over distincts sites 
${\bf r}_{1}\neq {\bf r}_{2}\dots \neq {\bf r}_{p}$ by 
integrals with, {\it e.g} hard core conditions
$|{\bf r}_{1}- {\bf r}_{2}|\geq a_{o}$ and introducing the uniform
fugacity $y$ as was done in Section \ref{part:model}.
Note that in the above
expression (\ref{free}) we do not make use of 
the decomposition (\ref{decomposition}) and $V_{\bf r}$ denotes
the original disorder.

The expansion in the number of points of the
free energy has the form:
\begin{eqnarray}
   \label{free}
F[V]&=&f^{(0)} + \sum_{{\bf r}_{1}\neq {\bf r}_{2}}
f^{(2)}_{{\bf r}_{1},{\bf r}_{2}}[V]  
+ \sum_{{\bf r}_{1}\neq {\bf r}_{2}\neq {\bf r}_{3}} 
f^{(3)}_{{\bf r}_{1},{\bf r}_{2},{\bf r}_{3}}[V]
+\dots
\end{eqnarray}
where, in the disordered case, each term $f^{(k)}_{{\bf r}_{1},\dots {\bf r}_{k}}$ depends on the values taken by the
disorder potential $V_{{\bf r}}$ exactly and only at points  
${\bf r}_{1},\dots {\bf r}_{k}$. 
The explicit construction of these terms 
is given both for the disordered and pure case ($V=0$)
in Appendix \ref{part:appexp}. From a practical point of view, the
explicit expression of $f^{(k)}_{{\bf r}_{1},\dots {\bf r}_{k}}$ is in all cases:
\begin{eqnarray}\label{fk}
f^{(k)}_{{\bf r}_{1}\dots {\bf r}_{k}}&=&
\sum_{l=0}^{k} (-1)^{k-l}\sum_{i_{1},\dots i_{l}\in [1,\dots k]}
F_{{\bf r}_{i_{1}},\dots {\bf r}_{i_{l}}}[V]
\end{eqnarray}
where $F_{{\bf r}_{i_{1}},\dots {\bf r}_{i_{l}}}[V]$ is the free energy of the
Coulomb gas defined only on the set of $l$ points 
${\bf r}_{i_{1}},\dots {\bf r}_{i_{l}}$ (instead of the full lattice)
and the summation is over all distinct subsets of the set ${\bf r}_{1}\dots {\bf r}_{k}$.
The definition (\ref{free}) is unambiguous, though subtle. Looking at
the explicit expression (\ref{fk}) as a sum over smaller subset of
points one could imagine adding other terms to the $f^{(k)}_{{\bf r}_{1},\dots {\bf r}_{k}}$
depending on less than $k$ points. This is not possible in a global way,
as the whole series must add up to the free energy, and the formula
(\ref{fk}) enforces it order by order. We refer to the Appendix
\ref{part:appexp} for further details about the precise
definition and construction. Note that the term $f^{(1)}$ vanishes here
because due to neutrality one cannot define a CG on a single site 
and that the expansion in the number
of points of powers $F^q[V]$ involves a rearrangment of the expansion
of $F[V]$. 

It is important to stress that each term $f^{(k)}$ of the above expansion
corresponds to an infinite sum over terms of {\it arbitrary high order in the conventional
fugacity $y$} (as defined in (\ref{locfug})). This can also be seen when setting
disorder to zero, where we find that each $f^{(2 k)}$ ($k \ge 1$)
and $f^{(2 k-1)}$ ($k \ge 2$) starts as $y^{2 k}$ plus an infinite
number of additional higher order terms in $y$ (see below). 
Indeed the expansion (\ref{free}) 
corresponds to a complete resummation of the conventional
fugacity expansion usually performed in Coulomb gas studies
 \cite{rubinstein83} except that it is usually performed on the partition
function $Z$ while here we perform it on the free energy.
The expansion (\ref{free}) using the 
free energy is the appropriate expansion when the fugacities
(or the core energies) are random and strongly site dependent
with a broad distribution (e.g. most of the sites have $z \sim 0$
except for a few which have $z \sim O(1)$).
Indeed, in this case the only small parameter is the probability
$P(z \sim 1)$ that a given point is favorable for a charge.
Thus the $k$-th term of the expansion (\ref{free}) is of
order $P(z \sim 1)^k$, since it is associated with $k$
independent points. We can thus consider (\ref{free})
as a perturbative expansion in the small parameter 
$P(z \sim 1)$, valid in the $XY$ phase, which replaces
the conventional expansion in $y$. 

To compute the $f^{(k)}_{{\bf r}_{1},\dots {\bf r}_{k}}$
we thus have to consider a Coulomb gas defined by the
partition function (\ref{square-recall}) 
(in the continuum limit), restricted to the
system of points ${\bf r}_{1},\dots {\bf r}_{k}$.
We will in addition restrict ourselves to charges
$n_{\bf r}= \pm 1,0$, as higher charges, examined later
will be less relevant. Let us closely examine the lowest order
terms $k=2$ and $k=3$.

\subsubsection{independent dipole approximation}

For $k=2$ we need only two points and the partition 
function (\ref{square-recall})
reads simply 
\begin{eqnarray} \label{Z-2}
Z_{{\bf r}_{1},{\bf r}_{2}}&\equiv & 1+W_{{\bf r}_{1},{\bf r}_{2}}=
1+\left(\frac{|{\bf r}_{1}-{\bf r}_{2}|}{a_{o}} \right)^{-2 \beta J}
\left ( y^{2}e^{V_{{\bf r}_{1}}-V_{{\bf r}_{2}}}+
 y^{2}e^{V_{{\bf r}_{2}}-V_{{\bf r}_{1}}}\right)
\end{eqnarray}
 The three terms of the partition
function (\ref{Z-2}) corresponds respectively to no charges or a dipole in   
$\{{\bf r}_1,{\bf r}_2 \}$, and the two possible positions for the
dipole. This results in the Boltzmann weigth for a dipole 
$W_{{\bf r}_{1}{\bf r}_{2}}$. The first terms of (\ref{free}) thus reads 
\begin{equation}\label{f-2}
f^{(2)}_{{\bf r}_{1},{\bf r}_{2}}=-\beta^{-1}
\ln (1+W_{{\bf r}_{1}{\bf r}_{2}})  
\end{equation} 
Restriction of the expansion (\ref{free}) to its first nonvanishing
term, {\it i.e} $F=\sum_{{\bf r}_{1},{\bf r}_{2}} f^{(2)}_{{\bf r}_{1},{\bf r}_{2}}$
corresponds to the so-called {\it independent dipole approximation}.
This is the approximation on which the 
analysis of  \cite{korshunov96} was based. This approximation,
which neglects all interactions between dipoles, may seem at first sight to
be a good enough approximation in the XY phase and at the transition.
Although this is reasonable in the pure XY model, this turns out to be
incorrect here: by discarding the next term 
$f^{(3)}_{{\bf r}_{1},{\bf r}_{2},{\bf r}_{3}}$ one throws away 
crucial statistical correlations. Indeed,
when renormalizing the distribution of local
fugacities, we have to take into account correlations between dipoles induced  
by the disorder, which arise as follows. Suppose that the
site ${\bf r}_{1}$ is favorable to the creation
of a $+$ defect, {\it i.e} if $z^{+}_{{\bf r}_{1}}\sim 1$
and ${\bf r}_{2}$ and ${\bf r}_{3}$ are both favorable to
creation of a $-$ defect (while other neigbouring sites
are unfavorable). Within the independent dipole approximation
the dominant configuration would be to 
put both a dipole in ${\bf r}_{1},{\bf r}_{2}$ and one on
${\bf r}_{1},{\bf r}_{3}$ to take advantage of these three favorable sites.
These are the configurations (4-6) in Fig. \ref{fig:expansion}.
However this configuration is forbidden because of the hard core
constraint (and we have restricted to $\pm 1$ charges, higher
ones being less favorable energetically).
Thus we need to take into account the effective correlations
between dipoles which arise because of rare favorable sites.
This is done by considering the second term of the expansion
$f^{(3)}_{{\bf r}_{1},{\bf r}_{2},{\bf r}_{3}}$. Furthermore,
as we will see below, consistent one loop renormalization requires to examine 
all terms in the expansion (\ref{free}) and how they change under
coarse graining, and thus to go well beyond the independent dipole
approximation.

\subsubsection{third order term and fusion}

\begin{figure}
\centerline{
\fig{6cm}{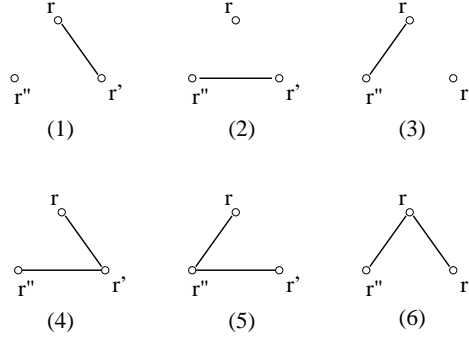}}
\caption{ 
\label{fig:expansion} 
Possible configurations for three sites: (1-3) are the only configurations
physically allowed, (4-6) are present in the independent dipole approximation
but forbidden in real model (see text). The bars denote pairs
of $\pm 1$ charges }
\end{figure}

Let us derive the explicit formula for $f^{(3)}_{{\bf r}_{1},{\bf r}_{2},{\bf r}_{3}}$.
The partition function with three sites reads:
\begin{equation}\label{Z-3}
F[V_{{\bf r}_{1}},V_{{\bf r}_{2}},V_{{\bf r}_{3}}]=
-\beta^{-1}\ln \left( 
1+W_{{\bf r}_{1}{\bf r}_{2}}
+W_{{\bf r}_{1}{\bf r}_{3}}+W_{{\bf r}_{2}{\bf r}_{3}}\right)
\end{equation}
 However, since {\it all terms of the expansion} of 
$f^{(3)}_{{\bf r}_{1}{\bf r}_{2}{\bf r}_{3}}$ in terms of $W$ depend
 exactly and only on ${\bf r}_{1}{\bf r}_{2}{\bf r}_{3}$, we have to substract 
to this free energy the terms depending on less than three sites, which can  
be identified as 
$f^{(2)}_{{\bf r}_{1}{\bf r}_{2}}+f^{(2)}_{{\bf r}_{1}{\bf r}_{3}}+
f^{(2)}_{{\bf r}_{2}{\bf r}_{3}}$.
The final expression for the second term of the expansion of the free energy
is thus 
\begin{eqnarray}\label{free3}
f^{(3)}_{{\bf r}_{1}{\bf r}_{2}{\bf r}_{3}}&=& 
-\beta^{-1}\ln \left( 
1+W_{{\bf r}_{1}{\bf r}_{2}}
+W_{{\bf r}_{1}{\bf r}_{3}}+W_{{\bf r}_{2}{\bf r}_{3}}\right)\\
\nonumber &&+\beta^{-1}
\left[
\ln (1+W_{{\bf r}_{1}{\bf r}_{2}})+
\ln (1+W_{{\bf r}_{1}{\bf r}_{3}})+
\ln (1+W_{{\bf r}_{2}{\bf r}_{3}}) \right]\\
\nonumber 
&=& -\beta^{-1}\ln \left( 
\frac{1+W_{{\bf r}_{1}{\bf r}_{2}}
+W_{{\bf r}_{1}{\bf r}_{3}}+W_{{\bf r}_{2}{\bf r}_{3}}}
{(1+W_{{\bf r}_{1}{\bf r}_{2}})(1+W_{{\bf r}_{1}{\bf r}_{3}})
(1+W_{{\bf r}_{2}{\bf r}_{3}})}
\right)
\end{eqnarray}
Let us first notice that now the three sites component of the expansion
(\ref{free} ) restricted to the first two terms is exactly 
$-\beta^{-1}\ln \left( 
1+W_{{\bf r}_{1}{\bf r}_{2}}
+W_{{\bf r}_{1}{\bf r}_{3}}+W_{{\bf r}_{2}{\bf r}_{3}}\right)$. Thus adding this
second term has cured the problem coming from the configurations (4,5,6) of
figure (\ref{fig:expansion}) which, as discussed above are not allowed.

It is interesting to note that
the term $f^{(3)}_{{\bf r}_{1}{\bf r}_{2}{\bf r}_{3}}$ is present 
(and important) in the expansion of
$\ln Z$, even though there are no actual configuration with
three charges in a given environment (from neutrality).
Indeed each term $f^{(k)}$ in the expansion (\ref{free}) contains
contributions from every even number (less or equal to $k$) of charges.
As can be seen from above (fig \ref{fig:expansion})  
it takes into account effective {\it correlations} between the
distributions of fugacities of three differents sites 
${\bf r}_{1}{\bf r}_{2}{\bf r}_{3}$ induced by the hard
core constraints. By contrast, 
there would be no such term involving three sites (because
of neutrality) in the similar expansion carried 
out on $Z$ (conventional fugacity expansion).
It was thus missed in previous studies, while in fact we show in the next Section
that it gives rise to a crucial contribution to 
the renormalization of the distribution of
disordered fugacities : the {\it fusion of environments}. 

The above defined expansion in number of independent points
(a type of cluster-virial
expansion) which we have illustrated on the first few terms,
can be performed systematically to all orders. By coarse graining 
all terms of the expansion self-consistently, we will now obtain the
renormalisation of the disorder distribution. 

\subsection{Direct renormalisation on the free energy}

\label{part:directRG}

We now propose a renormalisation scheme based on the above 
expansion (\ref{free}). From (\ref{part:physics}) one captures the
relevant physics by defining random local 
charge fugacities $z^{\pm}_{\bf r}$ at each point 
and splitting the disorder distribution as 
$P[V]=\prod_{\bf r} P(z^{+}_{\bf r}, z^{-}_{\bf r}) P[V^>_{\bf r}]$
as in (\ref{decomposition}). We know from the results of Section
\ref{part:replica} that one can renormalize the model by defining only
three scale dependent quantities: the full local disorder distribution
$P_l(z^{+}, z^{-})$, the stiffness $J_l$ and the second moment $\sigma_l J_l^2$ 
of the long range disorder $V^>$. We now separately obtain 
the renormalization of each of these quantities, in a systematic
perturbative expansion in the small parameter $P(z \sim 1)$ deduced
from the expansion in the number of points.  There are three 
types of contributions as follows.

\subsubsection{Rescaling}
\label{part:directRG0}

As can be seen from (\ref{Zcont}, \ref{decomposition})
there is an explicit dependence in the cutoff $a_0$ in the expression
of the interaction energy between charges and of the correlator of the long range 
part of the disorder $V^>$.  This dependence will appear in each term
$f^{(k)}$ of the expansion (\ref{free}). Upon changing the cutoff
this will results in contributions of order $O(dl)$ which can be
absorbed by appropriate redefinitions as follows.

The interaction term changes as:
\begin{eqnarray}
\beta J \sum_{ {\bf r} \neq {\bf r}' } n_{\bf r} \ln \left( \frac{|{\bf r} - {\bf r}'|}{a_0} \right) n_{{\bf r}'}
=
\beta J \sum_{ {\bf r} \neq {\bf r}' } n_{\bf r} \ln \left( \frac{|{\bf r} - {\bf r}'|}{a_0 e^{dl} } \right) n_{{\bf r}'}
- \beta J  dl \sum_{{\bf r}} n_{\bf r}^2
\end{eqnarray}
where we have used the neutrality condition $\sum_{{\bf r}} n_{\bf r} =0$. The first term is
the interaction term with the rescaled cutoff while the additional term produces
an additive contribution to the $\pm 1$ charge fugacity 
$z^{\pm}_r \to z^{\pm}_r  e^{- \beta J dl}$.

Similarly from (\ref{decomposition}) the correlator of $V^{>}$ 
can be rewritten as
\begin{eqnarray}  \label{change}
\overline{(V^{>}({\bf r}) - V^{>}({\bf r}'))^2} 
& = & 4 \sigma  J^2 ( \ln \frac{|{\bf r}-{\bf r}'|}{\tilde{a}_0} )
 (1-\delta^{(\tilde{a}_0)}({\bf r} - {\bf r}')) \\
\nonumber
&  & + 4 \sigma J^2 dl ~ (1-\delta^{(\tilde{a}_0)}({\bf r} - {\bf r}'))
\end{eqnarray}
explicitly as the sum of a new long range disorder 
correlator with cutoff $\tilde{a}_0=a_0 e^{dl}$ and 
a short range disorder correlator (we have discarded terms of
order $O(dl^2)$). Thus the original problem with
cutoff $a_0$ can be rewritten as one with cutoff $\tilde{a}_0$
with (i) a new gaussian long range disorder with identical
form of the correlator (\ref{decomposition}) with $a_0$ replaced
by $\tilde{a}_0$ (ii) a new local (short range)
disorder $v({\bf r}) \to v({\bf r}) + dv({\bf r})$
with $\overline{dv({\bf r}) dv({\bf r}')} = 2 \sigma J^2 dl
\delta^{(a_0)}({\bf r} - {\bf r}')$
since it is clear from (\ref{change}) that
when $a_0 \to a_0 e^{dl}$ the LR disorder 
produces an additive {\it gaussian} 
contribution $dv$ to the SR disorder.
Adding the two contributions we find that the
change of cutoff produces the total rescaling contribution
(\ref{rgrescaling}).

\subsubsection{Fusion of environments}
\label{part:directRG1}

Having introduced the expansion (\ref{free}) one can now
coarse grain its continuum version by increasing the cutoff
 $a_{0} \to \tilde{a}_{o}=a_{0} e^{dl}$ and integrating over
the corresponding degrees of freedom.
Upon this increase two points ${\bf r}_{i},{\bf r}_{j}$ from 
the $k$ points integral ($f^{(k)}$) will be fused if they satisfy 
$a_{0}< |{\bf r}_i - {\bf r}_j|<a_{0} e^{dl}$. This will produce a 
contribution which is an  integral 
at scale $\tilde{a}_{0}$ depending only on $k-1$ {\it independent points} and 
thus, by definition of the $f^{(k)}$, produces a correction of order $O(dl)$ to  
the $k-1$ terms $f^{(k-1)}$. All these corrections can be reabsorbed into
a correction to the fugacity distribution (together with an additive change 
to $f^{(0)}$ the free energy contribution of all degrees of freedom which have been
eliminated in the change of cutoff).

We now illustrate how this works on the case $k=3$, and indicate
in the Appendix \ref{app:kpoint} how it works for arbitrary $k$.
To lowest order in $dl$,
this correction is independent of $k$ and can be easily performed 
by considering the three points integral, using 
(\ref{free3}) with {\it e.g} 
$a_{0}< |{\bf r}_{1} - {\bf r}_{2}|<\tilde{a}_0 = a_{0} e^{dl}$ : 
\begin{multline}\label{correct1}
\int_{a_{0}<|{\bf r}_{1}-{\bf r}_{2}|<a_{0} e^{dl}} f^{(3)}_{\bf{r}_1, \bf{r}_2, \bf{r}_3}=
-\beta^{-1} \int_{a_{0}<|{\bf r}_{1}-{\bf r}_{2}|<a_{0} e^{dl}} 
\biggl[
 \ln \left(1+
\frac{W_{{\bf r}_{1}{\bf r}_{3}}+W_{{\bf r}_{2}{\bf r}_{3}}}
{1 + W_{{\bf r}_{1}{\bf r}_{2}}} \right)
\\
-\ln (1+W_{{\bf r}_{1}{\bf r}_{3}})-
\ln (1+W_{{\bf r}_{2}{\bf r}_{3}}) 
\biggr]
\end{multline}
where we assume $|{\bf r}_{1}-{\bf r}_{3}|\geq \tilde{a}_{0},
|{\bf r}_{2}-{\bf r}_{3}|\geq \tilde{a}_{0}$. 
Upon coarse graining, the two points ${\bf r}_{1},{\bf r}_{2}$ are fused
to a single point $\tilde{{\bf r}}= ({\bf r}_{1}+{\bf r}_{2})/2$
and one obtains a correction to 
$f^{3}_{\tilde{{\bf r}},{\bf r}_{3}}$.

Using the decomposition (\ref{decomposition}) of the disorder 
potential into a correlated component and a local part : 
$V_{{\bf r}}=V^{>}_{{\bf r}}+v_{{\bf r}}$, we can rewrite the
Boltzmann weigth for a dipole $W_{{\bf r}_{1}{\bf r}_{2}}$
,defined in (\ref{Z-2}), as  : 
\begin{eqnarray}
W_{{\bf r}_{1}{\bf r}_{2}}=
\left(\frac{|{\bf r}_{1}-{\bf r}_{2}|}{a_{o}} \right)^{-2 \beta J}
\left( 
z^{+}_{{\bf r}_{1}}z^{-}_{{\bf r}_{2}}
e^{\beta (V^{>}_{{\bf r}_{1}}-V^{>}_{{\bf r}_{2}})}
+
z^{+}_{{\bf r}_{2}}z^{-}_{{\bf r}_{1}}
e^{\beta (V^{>}_{{\bf r}_{2}}-V^{>}_{{\bf r}_{1}})} \right)
\end{eqnarray}
where we have used the definition $z^{\pm}_{\bf r} = y e^{ \pm \beta v_{\bf r}}$
for the local charge fugacities.

Several simplifications now occur in evaluating (\ref{correct1}).
We first note that in (\ref{correct1}) the integral is of order $dl$.
Thus to find the leading correction of $f^{(2)}_{\tilde{{\bf r}}{\bf r}_{3}}$ to order 
$O(dl)$, we only have to consider the integrand expanded in order $0$ in $dl$.
To this order, from the correlator (\ref{decomposition}) of the disorder
potential $V^{>}_{{\bf r}}$ and
using $|{\bf r}_{1}-{\bf r}_{2}|/a_{o}=1+\mathcal{O} (dl)$, we find that 
\begin{equation}\label{correct2}
V^{>}_{{\bf r}_{1}}=V^{>}_{{\bf r}_{2}} +\mathcal{O} (dl) = 
V^{>}_{\tilde{{\bf r}}} +\mathcal{O} (dl)
\end{equation}
Thus the weight of the fused pair can be simplified as
$W_{{\bf r}_{1}{\bf r}_{2}}= 
(z^{+}_{{\bf r}_{1}}z^{-}_{{\bf r}_{2}}+
z^{+}_{{\bf r}_{2}}z^{-}_{{\bf r}_{1}})$. Similarly,
$W_{{\bf r}_{1}{\bf r}_{3}}$ and $W_{{\bf r}_{2}{\bf r}_{3}}$ simplify
to order $0$ in $dl$ using (\ref{correct2}) : 
\begin{subequations}
\begin{eqnarray}\label{correct1-3}
W_{{\bf r}_{1}{\bf r}_{3}}&=& 
\left(\frac{|\tilde{{\bf r}}-{\bf r}_{3}|}{a_{o}} \right)^{-2 \beta J}
\left( 
z^{+}_{{\bf r}_{1}}z^{-}_{{\bf r}_{3}}
e^{\beta (V^{>}_{\tilde{{\bf r}}}-V^{>}_{{\bf r}_{3}})}
+
z^{+}_{{\bf r}_{3}}z^{-}_{{\bf r}_{1}}
e^{\beta (V^{>}_{{\bf r}_{3}}-V^{>}_{\tilde{{\bf r}}})} \right)\\
\label{correct1-2}
W_{{\bf r}_{2}{\bf r}_{3}}&=& 
\left(\frac{|\tilde{{\bf r}}-{\bf r}_{3}|}{a_{o}} \right)^{-2 \beta J}
\left( 
z^{+}_{{\bf r}_{2}}z^{-}_{{\bf r}_{3}}
e^{\beta (V^{>}_{\tilde{{\bf r}}}-V^{>}_{{\bf r}_{3}})}
+
z^{+}_{{\bf r}_{3}}z^{-}_{{\bf r}_{2}}
e^{\beta (V^{>}_{{\bf r}_{3}}-V^{>}_{\tilde{{\bf r}}})} \right)
\end{eqnarray}
\end{subequations}
Using these simplifications, we can now rewrite the first term of
(\ref{correct1}): 
we obtain for the corresponding dipole weigth : 
\begin{eqnarray}
&&
W'_{\tilde{{\bf r}},{\bf r}_{3}}=
\frac{W_{{\bf r}_{1}{\bf r}_{3}}+W_{{\bf r}_{2}{\bf r}_{3}}}
{1 + W_{{\bf r}_{1}{\bf r}_{2}}}=
\left(\frac{|\tilde{{\bf r}}-{\bf r}_{3}|}{a_{o}} \right)^{-2 \beta J}\\
\nonumber 
&&~~~~ \times 
\left( 
\frac{(z^{+}_{{\bf r}_{1}}+z^{+}_{{\bf r}_{2}})z^{-}_{{\bf r}_{3}}}
{1+z^{+}_{{\bf r}_{1}}z^{-}_{{\bf r}_{2}}+
   z^{-}_{{\bf r}_{1}}z^{+}_{{\bf r}_{2}}}
e^{\beta (V^{>}_{\tilde{{\bf r}}}-V^{>}_{{\bf r}_{3}})}
+
\frac{(z^{-}_{{\bf r}_{1}}+z^{-}_{{\bf r}_{2}})z^{+}_{{\bf r}_{3}}}
{1+z^{+}_{{\bf r}_{1}}z^{-}_{{\bf r}_{2}}+
   z^{-}_{{\bf r}_{1}}z^{+}_{{\bf r}_{2}}}
e^{\beta (V^{>}_{{\bf r}_{3}}-V^{>}_{\tilde{{\bf r}}})} \right)
\label{correct1-1}
\end{eqnarray}

It is now simple to verify that the 
change in the disorder averaged
$< \sum_{{\bf r}_{1}\neq {\bf r}_{2}}
f^{(2)}_{{\bf r}_{1},{\bf r}_{2}}[V] >_V$ 
originating from the $f^{(3)}$ term is 
correctly accounted for by following rules
for the random fugacity variables.
The last two terms of (\ref{correct1}) produce, with the simplification
(\ref{correct1-3},\ref{correct1-2}), the rules 
\begin{equation}\label{rule1}
(z^{\pm}_{{\bf r}_{1}},z^{\pm}_{{\bf r}_{2}})
\longrightarrow  z^{\pm}_{\tilde{{\bf r}}}=z^{\pm}_{{\bf r}_{1}}
~~;~~
(z^{\pm}_{{\bf r}_{1}},z^{\pm}_{{\bf r}_{2}})
\longrightarrow z^{\pm}_{\tilde{{\bf r}}}=z^{\pm}_{{\bf r}_{2}}
\end{equation}
From (\ref{correct1-1}), we find the rule corresponding to the first term of
(\ref{correct1}) :  
\begin{equation}\label{rule2}
(z^{\pm}_{{\bf r}_{1}},z^{\pm}_{{\bf r}_{2}})
\longrightarrow 
 z^{+}_{\tilde{{\bf r}}}=
\frac{(z^{+}_{{\bf r}_{1}}+z^{+}_{{\bf r}_{2}})}
{1+z^{+}_{{\bf r}_{1}}z^{-}_{{\bf r}_{2}}+
   z^{-}_{{\bf r}_{1}}z^{+}_{{\bf r}_{2}}},
 z^{-}_{\tilde{{\bf r}}}=
\frac{(z^{-}_{{\bf r}_{1}}+z^{-}_{{\bf r}_{2}})}
{1+z^{+}_{{\bf r}_{1}}z^{-}_{{\bf r}_{2}}+
   z^{-}_{{\bf r}_{1}}z^{+}_{{\bf r}_{2}}}
\end{equation}
This is true for any point ${\bf r}_{3}$ {\it different} from
${\bf r}_{1}$ and ${\bf r}_{2}$. One can thus
rewrite these fusion corrections as a correction 
to the (unnormalised) 'distribution'  of 
local fugacities $\Phi (z_{+},z_{-})$ : 
\begin{eqnarray}\label{fusion-direct}
\partial_{l}\Phi (z_{+},z_{-})=&&
c_{2}\int_{z^{\pm}_{{\bf r}_{1}},z^{\pm}_{{\bf r}_{2}}}
\Phi (z^{+}_{{\bf r}_{1}},z^{-}_{{\bf r}_{1}})
\Phi (z^{+}_{{\bf r}_{2}},z^{-}_{{\bf r}_{2}})\\
\nonumber \times&& 
\left[
\delta\left(z^{+}- \frac{(z^{+}_{{\bf r}_{1}}+z^{+}_{{\bf r}_{2}})}
{1+z^{+}_{{\bf r}_{1}}z^{-}_{{\bf r}_{2}}+
   z^{-}_{{\bf r}_{1}}z^{+}_{{\bf r}_{2}}}\right)
\delta\left(z^{-}- \frac{(z^{-}_{{\bf r}_{1}}+z^{-}_{{\bf r}_{2}})}
{1+z^{+}_{{\bf r}_{1}}z^{-}_{{\bf r}_{2}}+
   z^{-}_{{\bf r}_{1}}z^{+}_{{\bf r}_{2}}} \right)\right.\\
\nonumber &&
-\delta\left(z^{+}- z^{+}_{{\bf r}_{1}}\right)
 \delta\left(z^{-}- z^{-}_{{\bf r}_{1}}\right)
-\delta\left(z^{+}- z^{+}_{{\bf r}_{2}}\right)
 \delta\left(z^{-}- z^{-}_{{\bf r}_{2}}\right)
 \Biggl]
\end{eqnarray}
The coefficient $c_{2}$ comes from the integration over the relative position
${\bf r}_{1}-{\bf r}_{2}$. 
Thus from (\ref{fusion-direct}) and (\ref{rgrescaling}), we recover exactly the RG equation
(\ref{loose}) for the function $\Phi$. 

One can similarly check that the above rule correctly account for
the corrections to $k-1$ term from the $k$ term upon change of
cutoff. To be exhaustive, we must also consider a constant term in the expansion
(\ref{free}) of the free energy: $f_0(l)$ which satisfies
$\partial_{l}f_0 \sim \langle 
\ln(1+z'_{+} z''_{-} + z'_{-} z''_{+}))\rangle_{\Phi \Phi}$.
This term corresponds to the free energy sum of all
degrees of freedom which have been eliminated
up to scale $l$. Indeed, $P(z_+,z_-)$ (or $\Phi((z_+,z_-)$)
contains an average over all 
disorder configurations at smaller scales (all
environments which have been eliminated). A full and systematic proof
can in principle be made by considering all averages of powers
$F^q[V]$ expanded in number of points
and all fusions to order $k$ giving corrections
to order $k-1$. We will not attempt it here, as 
the systematic study to all orders is much easy to 
perform within the (equivalent) replica formalism.

\subsubsection{Screening}
\label{part:directRG2}

To derive the scaling behaviour of both the stiffness $J$ and the strength of
the correlated disorder $V^{>}_{{\bf r}}$, we consider the screening of the
interaction and the correlation of the disorder between two infinitesimal test
charges $e_{1}$ and $e_{2}$ in the sample. To implement the study of this
screening within our expansion in number of independentpoints, we first
define $F[V,e_{1},e_{2}]$ as the free energy of the disordered Coulomb gas
defined in (\ref{Zcont}) with the two additional test charges 
$e_{1},e_{2}$ fixed in ${\bf r}_{1},{\bf r}_{2}$. These test charges 
interact with the other integer charges of the Coulomb gas,
which screen the interaction between them, and the correlator of the disorder 
$V^{>}_{{\bf r}}$. From $F[V,e_{1},e_{2}]$, one can define the screened  
interaction and disorder by 
\begin{subequations}\label{def-screened}
\begin{eqnarray}
\beta G^{R}({\bf r}_{1},{\bf r}_{2}) &=&  \frac{1}{2} \left.
 \frac{ \partial^2 F[V,e_{1},e_{2}]}
{\partial e_{1} \partial e_{2}} \right|_{e_{1}=e_{2}=0}\\
-\beta V^{R}({\bf r}_{1}) &=& \left. 
\frac{ \partial F[V,e_{1},e_{2}] }{\partial e_{1}}
 \right|_{e_{1}=e_{2}=0}
\end{eqnarray}
\end{subequations}
At large distance we expect from renormalisability 
$\beta G^{R}\sim 
\beta J^{R} \ln \left(|{\bf r}_{1}-{\bf r}_{2}|/a_{o} \right)$, 
which imply the definitions for the renormalized coupling constant and 
disorder strength 
\begin{subequations}
\begin{eqnarray}
\label{scr:ctes}
J_{R}&=& \lim_{k\rightarrow 0}
 G_{{\bf k}}^{-1} G^{R}_{{\bf k}}\\
2 \sigma_{R} J_{R}^2 &=& \lim_{k\rightarrow 0} 
  G_{{\bf k}}^{-1} \overline{V^{R}_{{\bf k}}V^{R}_{-{\bf k}}}
\end{eqnarray}
\end{subequations}
where the Fourier transform of the 2D laplacian $G_{{\bf k}}^{-1}$ has been
defined after equation (\ref{square}).
These definitions can be transformed exactly using (\ref{Zcont}) into 
the relation 
\begin{subequations}
\begin{eqnarray}\label{JR}
J_{R}&=& J + 2 \pi \beta  J^2 \int d^2 {\bf r} (\hat{q}.{\bf r})^2 
  \overline{ \langle n_{\bf 0} n_{\bf r} \rangle_c} \\
\label{sR}
\sigma_{R} J_{R}^2 &=& \sigma J^2 - 2 \pi J^2 
   \int d^2 {\bf r}  (\hat{q}.{\bf r})^2 
  \overline{ \langle n_{\bf 0}\rangle \langle n_{\bf r} \rangle} \\
  &+& 4 \pi \sigma \beta J^3 
   \int d^2 {\bf r} (\hat{q}.{\bf r})^2 
  \overline{ \langle n_{\bf 0} n_{\bf r} \rangle_c}  \nonumber
\end{eqnarray}
\end{subequations}
 Up to now, these are only standard definitions of the renormalised 
stiffness $J^{R}$ and disorder strength $\sigma^{R}$.
From this one deduces the RG equations for $J$ and $\sigma$ since by coarse graining 
(\ref{JR},\ref{sR}) we obtain \cite{tang96,scheidl97,korshunov96}  
\[
T \partial_{l} J^{-1}= -2 a_0^4 
\pi^{2} \overline{\langle n_{0} n_{R=a}\rangle_{c} }
~~;~~ 
 \partial_{l}\sigma = - 2 a_0^4 \pi^{2} 
\overline{\langle n_{0} \rangle \langle n_{R=a}\rangle}
\]
where the $a_0^4$ factor arises because we are dealing with correlations
of the charge {\it density}.

The novel and most tricky part is how to evaluate the right hand side in
a systematic way. Our method is to expand these correlation functions
in the number of independent points. This can be done in a systematic way
using (\ref{free}) and the definitions
\begin{subequations}
\begin{eqnarray}
 \label{def-correl1}
\langle n_{\bf 0} n_{\bf R} \rangle_c &=& 
\frac{ \partial^2 F[V] }
{\partial V_{{\bf 0}} \partial V_{{\bf R}}} \\
 \label{def-correl2}
\langle n_{\bf 0}\rangle \langle n_{\bf R} \rangle  &=&
\frac{ \partial F[V] }{\partial  V_{{\bf 0}}}
\frac{ \partial F[V] }{ \partial V_{{\bf R}}}
\end{eqnarray}
\end{subequations}
Note that while the expansion of the first term (connected correlation)
in the number of points is readily obtained from the expansion
(\ref{free}), the expansion of the second term (disconnected correlation)
is more subtle since we need to square the derivatives of (\ref{free}) and 
rearrange it again in an expansion in the number of points.
Fortunately, to the order we are working, we need only the
two point term in the expansion. Thus we can 
restrict ourselves to terms with two points on both expansions. The
expansion of the first term (\ref{def-correl1}) follows straightforwardly from
(\ref{free}) : 
\begin{eqnarray*}
&& \frac{ \partial^2 F[V] }
{\partial V_{{\bf 0}} \partial V_{{\bf R}}}  =
\sum_{{\bf r}_{1}\neq {\bf r}_{2}}
\frac{ \partial^2 f^{(2)}_{{\bf r}_{1},{\bf r}_{2}}}
{\partial V_{{\bf 0}} \partial V_{{\bf R}}}
+\sum_{{\bf r}_{1}\neq {\bf r}_{2}\neq  {\bf r}_{3}}
\frac{ \partial^2 f^{(2)}_{{\bf r}_{1},{\bf r}_{2},{\bf r}_{3}} }
{\partial V_{{\bf 0}} \partial V_{{\bf R}}}+ \dots 
\end{eqnarray*}
and thus from (\ref{f-2}) one finds:
\begin{eqnarray*}
 \frac{ \partial^2 F[V] }
{\partial V_{{\bf 0}} \partial V_{{\bf R}}}|_{2 points} 
&&=
\frac{ \partial^2 f^{(2)}_{{\bf 0},{\bf R}}}
{\partial V_{{\bf 0}} \partial V_{{\bf R}}} \\
&&=
\frac{-\beta^{-1}}{1+W_{{\bf 0},{\bf R}}}
\frac{ \partial^2 W_{{\bf 0},{\bf R}}}
{\partial V_{{\bf 0}} \partial V_{{\bf R}}} + 
\frac{\beta^{-1}}{(1+W_{{\bf 0},{\bf R}})^{2}}
\frac{ \partial W_{{\bf 0},{\bf R}}}
{\partial V_{{\bf 0}}}
\frac{ \partial W_{{\bf 0},{\bf R}}}
{\partial V_{{\bf R}}}
\end{eqnarray*}
since we can keep only terms depending of the two points 
(${\bf 0}$ and ${\bf R}$) in the last equality (the 
other contributions vanish). Similarly one finds
that: 
\begin{eqnarray}
 \frac{ \partial F[V] }{\partial V_{{\bf 0}} }
\frac{ \partial F[V] }{ \partial V_{{\bf R}}}|_{2 points}&& =
\left(\sum_{{\bf r}_{1}\neq {\bf r}_{2}} 
\frac{ \partial f^{(2)}_{{\bf r}_{1},{\bf r}_{2}} }
     {\partial V_{{\bf 0}} }+\dots \right)
\left.
\left(\sum_{{\bf r}_{3}\neq {\bf r}_{4}} 
\frac{ \partial f^{(2)}_{{\bf r}_{3},{\bf r}_{4}} }
     {\partial V_{{\bf R}} }+\dots \right)\right|_{2 points}  \\
\nonumber
&& =
\frac{  \partial f^{(2)}_{{\bf 0},{\bf R}} }
     {\partial V_{{\bf 0}} }
\frac{  \partial f^{(2)}_{{\bf 0},{\bf R}} }
     {\partial V_{{\bf R}} } =
\frac{\beta^2}{(1+W_{{\bf 0},{\bf R}})^{2}}
\frac{  \partial W_{{\bf 0},{\bf R}} }
     {\partial V_{{\bf 0}} }
\frac{  \partial W_{{\bf 0},{\bf R}} }
     {\partial V_{{\bf R}} }  \label{expcorr}
\end{eqnarray}
where the dipole weights $W_{{\bf 0},{\bf R}} $ have been defined
in (\ref{Z-2})

These expressions further simplify since it is sufficient
to work to lowest order in $dl$, i.e to order $0$ in $dl$.
We can use that 
$V^{>}_{{\bf R}}=V^{>}_{{\bf 0}} +\mathcal{O} (dl)$ since we need only
to evaluate the correlation at $|{\bf R}|=a_{o}$. This yields 
that 
$W_{{\bf 0},{\bf R}}=z^{+}_{{\bf 0}}z^{-}_{{\bf R}}+
z^{-}_{{\bf 0}}z^{+}_{{\bf R}}$. One can also 
replace the derivatives with respect to the
full disorder $\partial_{V_{{\bf r}}}$ by derivatives with respect to the 
local disorder $\partial_{v_{{\bf r}}}$ (see (\ref{decomposition})) and thus
we find: 
\begin{subequations}
\begin{eqnarray}\label{exp:correl1}
&& 
- a_0^4 \langle n_{\bf 0} n_{R=a} \rangle_c = 
\frac{ z^{+}_{{\bf 0}}z^{-}_{{\bf R}}+
       z^{+}_{{\bf R}}z^{-}_{{\bf 0}}+
     4 z^{+}_{{\bf 0}}z^{-}_{{\bf R}} z^{+}_{{\bf R}}z^{-}_{{\bf 0}} }
{(1 + z^{+}_{{\bf 0}} z^{-}_{{\bf R}}+
      z^{+}_{{\bf R}} z^{-}_{{\bf 0}} )^2} \\
\label{exp:correl2}
 & & 
- a_0^4 \langle n_{0}\rangle \langle n_{R=a}\rangle =
 \frac{ (
z^{+}_{{\bf 0}} z^{-}_{{\bf R}}-
z^{+}_{{\bf R}} z^{-}_{{\bf 0}})^2 }{
(1 + z^{+}_{{\bf 0}} z^{-}_{{\bf R}}+
     z^{+}_{{\bf R}} z^{-}_{{\bf 0}} )^2}
\end{eqnarray}
\end{subequations}
By using this expansion into (\ref{def-correl1},\ref{def-correl2}) we find
the screening RG equations (\ref{screening}). 

Thus the expansion in the number of points has allowed us to derive
consistent RG equations {\it without using replicas}. 
Note the difference with the previous unsuccesful
attempt in  \cite{korshunov96} to derive RG equations for the stiffness
and disorder without replicas. The authors of  \cite{korshunov96}
were working in the independent dipole approximation. Thus
they computed the correlation functions as in the first line
of (\ref{expcorr}) except that they kept the full expression
$(\sum_{{\bf r}_{1}\neq {\bf r}_{2}} 
\frac{ \partial f^{(2)}_{{\bf r}_{1},{\bf r}_{2}} }
     {\partial V_{{\bf 0}} } )
(\sum_{{\bf r}_{3}\neq {\bf r}_{4}} 
\frac{ \partial f^{(2)}_{{\bf r}_{3},{\bf r}_{4}} }
     {\partial V_{{\bf R}} })$. This led
to intractable expressions involving up to {\it four} independent
points. The beauty of the present method is that we know that
we can keep, in a consistent way to this order,
only the terms in this expression
which involve two points.

\subsection{Conclusion and connection with the replica method}
\label{part:connect}

We have thus developed two RG procedures, one using replicas
based on an expansion in the {\it vector fugacity}
$Y[{\bf n}]$ which may have seemed somewhat formal, the other
one, direct without replicas, as an expansion in the
number of favorable regions $P(z \sim 1)$ (at low 
temperature, while at higher temperature it smoothly
crosses over into the usual expansion in the uniform
fugacity $y=<z>$). In fact these two approaches are {\it equivalent}.
Indeed, we have checked using some
combinatorics detailed in Appendix \ref{part:connection-replica}, 
that the fugacity expansion in $Y[{\bf n}]$
(\ref{expansion}, \ref{replicah}) is {\it identical} term by term, in the limit
$m \to 0$ to the expansion in number of points of the free energy
(\ref{free}). Thus, working with one or the other is equivalent
\footnote{the RG procedure used in the previous Section
to derive the screening equations without replicas was found
to be consistent with the more direct RG of the replicated CG
explained in Appendix 
\ref{part:rgreplica}.}.
It also shows that the limit of diluted {\it vector charges}
physically corresponds to the limit of rare favorable regions.

\section{Analysis of the RG equations} 

\label{part:analysis}

\label{part:KPP}

In this Section we analyse the RG equations derived in the
previous Sections and we obtain the phase diagram and 
the critical behaviour of the XY model with random phases.

\subsection{general analysis of the full RG equations}

\label{part:numerics}

Our task now is to study the closed set of
RG equations (\ref{rgeqp},\ref{screening}) for the scale dependent
distribution of fugacities $P_l(z_+,z_-)$, the long range disorder strength
$\sigma_l$ and the stiffness $J_l$. Since we study below the zero temperature
limit, we have chosen to keep $T$ fixed and renormalize $J$ 
(only the combination $K=\beta J$ flows).

We started by studying the set (\ref{rgeqp},\ref{screening}) 
numerically. Although we will not reproduce here the details
we describe the main results. They confirmed
the overall physical picture and led us to introduce an approximation,
described below, which allows for an analytical solution.

We found that at low $T$ and for $\sigma$ smaller than some critical
value $\sigma_c$ an XY phase exists (as in Fig. \ref{fig:phasediag-intro}),
where both the stiffness $J$ and $\sigma$ converge to finite non zero
values at large $l$:
\begin{eqnarray}
J_l \to J_{R} >0   \qquad \sigma_l \to \sigma_{R} 
\label{xydef}
\end{eqnarray}

\begin{figure}[htb]
\centerline{\fig{8cm}{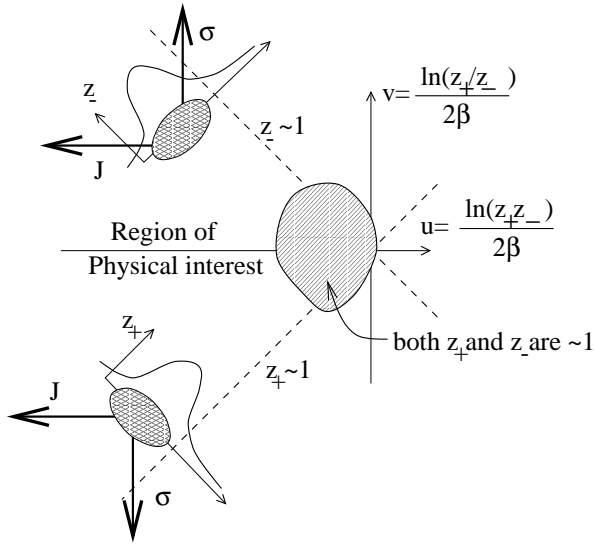}}
\caption{\label{fig:distrib-2D} 
Schematic representation of the scale
dependent behavior of the distribution $P (z_{+},z_{-})$ in the plane
of core energy variables $(u,v)$ and the regions
discussed in the text). The stifness $J_l$ produces a convection
towards the left (negative $u$ axis) while at the same time the
long range disorder $\sigma$ produces a diffusive spreading 
of the probability weight along the (local disorder) 
$v$ axis (in both directions).
For $\sigma > \sigma_c$ diffusion wins and
the distribution spreads. For smaller $\sigma < \sigma_c$ 
the weight of the distribution remains confined
within the left quadrant $z_{\pm} < 1$. In general it develops a two
dimensional front solution.}
\end{figure}

In general, at low $T$ we found that, starting at scale $l=0$ from
(\ref{locfug}) with a small initial averaged fugacity (and small 
local disorder) the distribution 
$P_l(z_+,z_-)$ becomes broad and develops power law tails in 
the variables $z_+$, $z_-$ which quickly extend up to fugacities 
$z_\pm \sim O(1)$. However, in the XY phase (\ref{xydef}) we
found that the typical $z_\pm$ goes to zero and that 
the concentration of rare favorable regions, after an
initial increase, ends up decreasing towards zero at large $l$.
It is useful to define the {\it single fugacity distribution}:
\begin{eqnarray}
P_l(z) = \int dz_+ P_l(z_+,z) = \int dz_- P_l(z,z_-)
\label{singledef}
\end{eqnarray}
which does not satisfy a closed RG equation. Then we find 
in the XY phase that $P_l(z \sim 1)$ decreases at large $l$
(or equivalently $\int_{z>z^*} P_l(z)$ where $z^*$ is some arbitrary 
threshold). Thus the probability that either $z_+$ or $z_-$ is 
of order $1$ decreases. On the other hand, for $\sigma > \sigma_c$
we find that $P_l(z \sim 1)$ ends up increasing at large scale and 
the whole distribution $P_l(z)$ ends up drifting towards increasing
$z$ values. This corresponds to the disordered phase where $J_l$ vanishes at
large scale ($J_R=0$).
The most interesting flow occurs in the critical regime
near the transition. There, if one interprets the fraction of favorable
regions $P_l(z \sim 1)$ as the perturbative parameter, the
structure of the flow near $\sigma = \sigma_c$ is reminiscent of the RG flow 
a la Kosterliz-Thouless with a separatrix, in the plane
$(P_l(1),\sigma)$ between the XY and the
disordered phase. Accordingly, exactly at the transition,
i.e on the separatrix (which, to be accurate, is a critical manifold in the
space of distributions and couplings $P_l(z_+,z_-), J, \sigma$)
the distribution $P(z_+,z_-)$ becomes very broad and develops a
fixed shape, with $P_l(z \sim 1)$ slowly decreasing to zero
(which makes the critical regime perturbative in $P_l(1)$).
A schematic representation of the distribution is given in
Fig. \ref{fig:distrib-2D}.

Let us close this paragraph by noting that at higher $T$, by contrast,
we found that the distribution remains peaked and that the overall
picture becomes more consistent with considering a uniform average
fugacity (as in  \cite{rubinstein83}).
The XY transition then occurs when this uniform average fugacity
ceases to flow to zero at large scale.

To quantify these (mostly numerical) observations, we now turn to the
single fugacity approximation.

\subsection{Single fugacity RG equation} 

\label{part:rg1d}

We now argue that, given the structure of the RG flow observed numerically,
we can, with no loss of accuracy in all the regimes of interest (i.e within and near
the boundaries of the XY phase) approximate the full RG equations
(\ref{rgeqp}) for $P_l(z_{+},z_{-})$ by a simpler equation for the single 
fugacity distribution $P_l(z)$, and similarly for (\ref{screening}).

Let us first focus on the low $T$ regime, where the distribution
$P_l(z_+,z_-)$ is broad and the physics is dominated by rare favorable regions.
For such a broad distribution the parameter which allows to
organize perturbation theory is:
\begin{eqnarray}
P_l(1) \equiv P_l(z \sim 1) \sim 
P_l(z_{+} \sim 1, z_{-}\sim 0) = P_l(z_{+} \sim 0, z_{-} \sim 1)
\end{eqnarray}
where we distinguish symbolically the rare configurations $z_\pm \sim 1$ 
from the typical ones where $z_\pm \sim 0$.
The first observation is that $z_+$ and $z_-$ are even more rarely simultaneously
$\sim 1$. We note symbolically:
\begin{eqnarray}
P_l(1,1) \equiv P_l(z_{+}\sim 1,z_{-}\sim 1) \sim P_l(1)^2
\label{p11}
\end{eqnarray}
Indeed such configurations, absent from the start, are generated
from the fusion term in (\ref{rgeqp}). To generate from the
fusion rule $z_{\pm}
= (z'_{\pm}+z''_{\pm})/(1+z'_{+}z''_{-}+z'_{-}z''_{+})$ a configuration
where both $z_{+} \sim z_{-} \sim O(1)$ one clearly 
needs at least that either $z'_{+} \sim z''_{-} \sim 1$ or
$z'_{-} \sim z''_{+} \sim 1$, which is of probability 
of order $P_l(1)^{2}$ (symbolically one can write 
$\partial_{l} P_l(1,1)\sim P_l(1)^{2} + ..$). Thus, and this can be further checked,
the estimate (\ref{p11}) is valid (the diffusion term does not 
change it). One can now inspect
the corrections to the stiffness $J$ and to the long range disorder $\sigma$
in (\ref{screening}) which come mainly from configurations where 
either $z'_{+} \sim z''_{-} \sim 1$
or $z'_{-} \sim z''_{+} \sim 1$. This leading contribution is
thus of order $\sim P_l(z \sim 1)^2$.
Other fugacity configurations, e.g. with both $z'_{+} \sim z'_{-} \sim 1$
contribute in (\ref{screening}) but, from
the above estimate (\ref{p11}), their probability is
$\sim P_l(1,1) P_l(1) \sim P_l(1)^{3}$ and are thus subleading
in an expansion in $P_l(1)$. To summarize, 
one can see schematically the RG equations
(\ref{screening}) as a correction of leading order 
$P_l(1)^2$ to $J_l$ and $\sigma_l$ and the RG equation
(\ref{rgeqp}) as a correction to $P_l(1)$ of order $P_l(1)$ (diffusion term)
and $P_l(1)^2$ (fusion term).

Guided by these observations, we now approximate the full equation 
(\ref{rgeqp}) by a RG equation for a function of a single variable 
which should correctly describe the behaviour of $P (z_{+},z_{-})$
around $z_{\pm} \sim 1$. This approximation amounts to neglect the denominators
in the fusion rule of (\ref{rgeqp}). Indeed, in the spirit of our previous
estimates, this denominator contributes only 
when $z'_{+} \sim z''_{-} \sim 1$ (or $z'_{-} \sim z''_{+} \sim 1$) 
and then yields a contribution
of order $P_l(1)^2$ but which corrects only $P_l(1,1)$ since this fusion
produces both a $z_{+} \sim z_{-} \sim 1$. This corresponds to 
a correction of order $P_l(1)^{3}$ to $P_l(1)$
(which we neglect, since the leading order of the correction to $P_l(1)$
coming from the fusion term is $P_l(1)^{2}$). This can be seen
e.g. since $P_l(1,1)$ itself in turns contributes to $P_l(1)$ 
to order $P_l(1)^{2}$ (through integration over one of the two
variables), or by more detailed arguments, 
not reproduced here. Neglecting the denominators gives the equation for
the probability distribution   
\begin{eqnarray}  \label{rgeqp-approx}
 \partial_l P(z_+,z_-) &=&  \mathcal{O} P - 2 P(z_+,z_-)  \\
\nonumber
&& +2 \left<  
\delta( z_+ - (z'_+ + z''_+))
\delta( z_- - (z'_- + z''_-))  \right>_{P' P''} 
\end{eqnarray}
This equation can now be explicitly integrated over one variable,
say $z_{-}$ and yields a closed equation\footnote{
Note similarly that (\ref{rgeqp-approx}) admits solutions
of the form $P_l(z_+,z_-)= P_l(z_+) P_l(z_-)$.}
for the single fugacity distribution $P_l(z_{+})$ defined in 
(\ref{singledef}):
\begin{eqnarray}\label{RG-1D}
\partial_l P(z) &=&  
\left[
\beta J (1 + z \partial_{z})
+ \sigma (\beta J)^2 (1 +z \partial_{z})^2
 \right]P(z) - 2 P(z)  \\
\nonumber
&& +2 \int_{z',z''>0}dz'dz''  
\delta( z - (z' + z'')) P (z') P (z'')
\end{eqnarray}

We can now obtain the corresponding equations for
the corrections to $J$ and $\sigma$. In (\ref{screening})
one keeps only the configurations corresponding 
to either $z'_+ \sim z''_- \sim 1$ or $z'_- \sim z''_+ \sim 1$
but discard the much less probable configurations where all four fugacities
are $\sim 1$. The expressions then nicely factor out in terms of
$P_l(z)$ and one obtains:
\begin{subequations} \label{screening2}
\begin{eqnarray} 
&& T \partial_l J^{-1} = \frac{8 c_{1}}{c_{2}^2}
\int_{z'_+,z''_-} P(z'_+) P(z''_-) \frac{ z'_+ z''_- }
{(1 + z'_+ z''_- )^2}  \\
&& \partial_l \sigma  = \frac{8 c_{1}}{c_{2}^2}
\int_{z'_+,z''_-} P(z'_+) P(z''_-)
\frac{ (z'_+ z''_- )^2 }{(1 + z'_+ z''_- )^2}
\end{eqnarray}
\end{subequations}
where the additional factor of $2$ counts the equivalent 
integral with $+$ and $-$ exchanged.

We now summarize the obtained closed set of RG equations 
in terms of the variable $u$ defined as:
\begin{eqnarray}
z = e^{\beta u}
\end{eqnarray}
of distribution $\tilde{P}(u) du = P(z)dz$. Physically the random variable $u$
can be interpreted as the random local core energy 
$u_{\bf r} = - E_c({\bf r})$. One has:
\begin{subequations}
\label{Rg-eq-1D-u}
\begin{eqnarray}   \label{toy}
 \partial_l \tilde{P}(u) &=&  
\left( J \partial_{u} + \sigma J^2 \partial^{2}_{u} \right) 
\tilde{P}  - 2 \tilde{P} 
 \\ \nonumber 
&& + 2 \int_{u',u''}\tilde{P}(u') \tilde{P}(u'')
\delta\left(u - \frac{1}{\beta}\ln (e^{\beta u'} + e^{\beta u''}) \right) \\
\label{toy2} 
\partial_{l} (J^{-1}) &=&  \frac{8 c_{1}}{c_{2}^2}
\left< 
\frac{\beta e^{-\beta (u'+ u'')}}{ (1 + e^{-\beta (u'+ u'')} )^2} 
\right>_{\tilde{P} (u') \tilde{P} (u'')} 
=
\frac{8 c_{1}}{c_{2}^2}
\left< 
\frac{\beta e^{-\beta u}}{ (1 + e^{-\beta u} )^2} 
\right>_{\tilde{P}*_u\tilde{P}} \\
\label{toy3} 
\partial_{l} (\sigma) &=&  \frac{8 c_{1}}{c_{2}^2} 
\left< 
\frac{1}{(1 + e^{-\beta (u'+ u'')})^2 } 
\right>_{\tilde{P} (u') \tilde{P} (u'')}
= \frac{8 c_{1}}{c_{2}^2}
\left< 
\frac{1}{(1 + e^{-\beta u})^2 } \right>_{\tilde{P}*_u\tilde{P}}
\end{eqnarray}
\end{subequations}
and thus the corrections to $J$ and $\sigma$ involve only the
convolution $\tilde{P}*_u \tilde{P}$.

We have thus justified, in the 
low temperature regime, that these approximate RG equations should
describe the tails of the fugacity distribution exactly to the order
$P_l(1)^2$ in the expansion in $P_l(1)$ to which we are working.
Indeed, since we have studied the RG in the previous Sections 
only to order $Y[{\bf n}]^2$ (in the replica formulation, corresponding 
to $P_l(1)^2$ in the rare events formulation) 
it probably does not make sense at this stage to try to be more accurate.
\footnote{Although we believe the arguments based on an expansion in
$P_l(1)$ are correct, only an exhaustive analysis of the two
dimensional front solutions of (\ref{rgeqp}), clearly a complex task
which goes beyond this paper, could confirm the exactness 
of the approximation and the validity of the arguments based on
organizing the perturbation theory using a single small parameter
$P_l(1)$.}.

Finally, let us note that 
although we have focused until now on the low $T$ regime
it is rather obvious that
the approximation of discarding the denominators will be even more
valid at higher temperature since in the conventional fugacity expansion
(in a uniform averaged fugacity) these terms corresponds to 
$O(y^4)$ terms while we work to $O(y^2)$. So these new RG equations
interpolate all limits correctly.

\subsection{RG equation at $T=0$}\label{zeroT}

One can now easily see on the form (\ref{Rg-eq-1D-u}) that the RG equations
admit a well defined $ T \to 0$ limit. 
In the equation (\ref{toy}), the fusion rule
$u = \frac{1}{\beta}\ln (e^{\beta u'} + e^{\beta u''})$
now becomes a max rule $u=\max (u',u'')$  and (\ref{toy}) transforms as 
\begin{equation}\label{rg1d-1}
\partial_l \tilde{P}_l(u) =
\left( J_l \partial_{u} + \sigma_l J_l^2 \partial^{2}_{u} \right) 
\tilde{P}_l  - 2 \tilde{P}_l 
+ 4 \tilde{P}_l(u) \int_{u \geq u''} \tilde{P}_l(u'')
\end{equation}
Similarly, the function of $u',u''$ in 
(\ref{toy2},\ref{toy3}) become respectively a delta 
and a theta function when $ T \to 0$, so that in that
limit one has:
\begin{subequations}
\label{rg1d-2}
\begin{eqnarray} 
\label{rg1d-2.1} 
\partial_{l} (J^{-1}) &=&  \frac{8 c_{1}}{c_{2}^2}
\int du \tilde{P}_{l} (u)\tilde{P}_{l} (-u)\\
\label{rg1d-2.2} 
\partial_{l} (\sigma) &=&  \frac{8 c_{1}}{c_{2}^2} 
\int_{u'+u''\geq 0} du'du'' \tilde{P}_{l} (u')\tilde{P}_{l} (u'')
\end{eqnarray}
\end{subequations}
Note that the last integral exactly evaluates the probability to find
two local regions with a total negative core energy, energetically favorable 
for a dipole, in agreement with the physical picture valid at low $T$.

We thus obtain a close set of equations (\ref{rg1d-1},\ref{rg1d-2}) which
describes the scaling behaviour of the system at zero temperature. The
equation for the distribution (\ref{rg1d-1}) can be conveniently simplified
using the parametrization  
\begin{equation}\label{G-T0}
G_{l}(x)=\int_{x-E_{l}}^{+\infty}\tilde{P}_{l} (u)du
\end{equation}
with $E_{l}=\int_{0}^{l} J_{l'}dl'$. The function $G_{l}(x)$ then satisfies 
\[
\frac{1}{2}\partial_{l}G =
\frac{\sigma J^{2}}{2}\partial^{2}_{x}G
+G (1-G) \]
In the case where $\sigma$ and $J$ are $l$-independent,
this equation is known as the Kolmogorov-Fisher equation and we will recall
some results on this equation in the next Section. Before turning to its
study, let us notice that the screening equations   
(\ref{rg1d-2}) can be rewritten using this new parametrization
simply substituting $\tilde{P}_{l} (u) = - \partial_{x} G (u+E_{l})$.

\subsection{Parametrization at $T >0$}\label{part:G-Tfinie}

Although (\ref{Rg-eq-1D-u}) at finite temperature $T>0$ 
could be in principle studied directly, it is much more convenient
to introduce the generalization of the $T=0$ parametrization
(\ref{G-T0}) as 
\begin{eqnarray}\label{G-Tfinie}
G_l(x)&=&1-\int_{z>0} dz ~ P_l(z) ~ 
\exp(-z e^{-\beta(x-E_{l})}) \\
&=& 1 -
\int_{-\infty}^{\infty} du ~ \tilde{P}_{l}(u)~
\exp(-e^{\beta (u-x+E_{l})})
\nonumber 
\end{eqnarray}
with $E_{l}=\int_{0}^{l} J (l')dl'$. In the limit $\beta\to \infty$, this
function reduces to the previous one (\ref{G-T0}).
Using this new function and variable
the integral equation (\ref{toy}) can again be
transformed exactly into a simpler differential
equation which, interestingly, remains exactly the same
as in the $T=0$ case:
\begin{equation}\label{kolmogorov}
\frac{1}{2}\partial_{l}G =
\frac{\sigma J^{2}}{2}\partial^{2}_{x}G
+G (1-G)
\end{equation}
Only the initial condition explicitly depends on temperature
(see below).

Thus solving the KPP equation (\ref{kolmogorov}) allows us in principle to 
obtain the scale dependent fugacity distribution $P_{l} (z)$
at any $T$. However the relation between this function $G_{l}(x)$ 
and the distribution $P_{l}(z)$ becomes much more involved at $T>0$.
As a result, and contrarily to the $T=0$ case, the screening
equations do not admit a simple expression in term of $G_{l}(x)$.

\begin{figure}
\centerline{\fig{6cm}{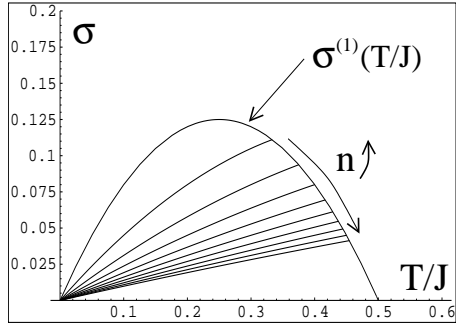}}
\caption{\label{fig:y-moments} Critical disorder $\sigma^{(n)} (T/J)$ above which
the successive moments $\langle z^{n}\rangle$ diverge as the scale
$l  \to +\infty$.}
\end{figure}

For any fixed $l$ one can reconstruct the integer moments $\langle z^{n}\rangle_{P_l}$ 
of the distribution $P_{l}(z)$ by simply expanding the generating function 
in powers of $e^{-\beta x}$ since:
\[
G_{l} (x)= \sum_{n=1}^{+\infty} \frac{(-1)^{n}}{n!}\langle z^{n}
\rangle_{P_{l}} e^{-n\beta (x-E_{l})} 
\]
Taking back this expansion in the KPP equation (\ref{kolmogorov}) and
identifying the coefficient of the exponentials $e^{-n\beta x}$, we find the
exact RG equations for the moments $\langle z^{n}\rangle$ of the distribution 
$P_{l}(z)$ : 
\begin{eqnarray}\label{RG-moments}
\partial_{l}\langle z^{n}\rangle &=& 
\left(2-n\beta J +\sigma (n\beta J)^{2}\right)\langle z^{n}\rangle\\
\nonumber &&+
\int dz'dz'' \left[(z'+z'')^{n}-(z')^{n}-(z'')^{n} \right]P_{l} (z')P_{l} (z'')
\end{eqnarray}
Starting from a reasonable initial distribution with finite
moments, the moments remain finite for finite $l$,
but, as we now discuss, increase quickly as $l \to +\infty$.
Let us examine the scaling dimension of the moments,
neglecting for now the bilinear (fusion) term.
We find that for fixed $T$ and $J$, each successive moment $\langle z^{n}\rangle$ 
diverges with the scale $l$ when the long range disorder $\sigma$ 
becomes larger than a critical disorder $\sigma >  
\sigma^{(n)}= (T/nJ)-2 (T/nJ)^{2}$ (see Fig.
\ref{fig:y-moments}) (these values will be slightly renormalized 
by the screening equations but the conclusion will be similar).
Putting back the fusion term simply implies that the moments
with $n \ge 2$ also diverge when the first one does, as indicated
in the Fig. \ref{fig:y-moments}.

The divergence of the first moment $y_l = \langle z \rangle_{P_l}$
which can be identified with the uniform fugacity $y$ of Rubinstein et al.
yields the (incorrect) reentrant phase diagram of  \cite{rubinstein83}
where the XY phase is destroyed above the line $\sigma^{(1)} (T/J)$.
Indeed it is already clear that the uniform fugacity approximation cannot work 
since we now see that higher moments diverge for even smaller disorder
strengths. Thus even within the XY phase this result for the moments
show that the distribution $P_{l}(z)$ rapidly becomes broader and 
broader. Atypical sites where $z$ is large appear and 
dominate the behaviour of the higher moments. Thus it becomes
meaningless to study the scale dependence of the integer moments,
but rather we must now consider the whole probability distribution $P_{l}(z)$
and in particular understand its tail.
This can be achieved with the help of the known solutions 
of the KPP equation which we briefly review in the next Section, before 
coming back to describing the scaling
behaviour of $P_{l} (z)$.  

It is interesting to note at this stage that the above 
KPP equation also arises in the problem of the directed
polymer with quenched disorder on the Cayley Tree (DPCT)
\cite{derrida88}. There the variable $l$ corresponds to the
number of generations and $\tilde{P}_l(u)$ to the distribution
of free energy $-u=-T \ln z$. Thus we have demonstrated in an explicit 
and non trivial way, i.e at the level of the RG equations,
that there are close connections
between the two problems, which both exhibit a similar freezing transition.
There are also notable differences between the two problems.
For instance, while the diffusion coefficient $D$ is constant in
the DPCT studied in \cite{derrida88}, in the disordered CG, $D_l$ depends 
on the scale and, in a self consistent manner, 
on the solution of the KPP equation itself,
via the equations (\ref{screening2})
which describes the physics of screening,
absent in the DPCT. As a result, additional phase 
transitions exist here as will be detailed in Section
\ref{part:XYphase}. 

\subsection{The Kolmogorov-Petrovskii-Piscounov
equation: some known useful properties}

\label{part:rappel-KPP} 

We recall in this Section
some known facts on the Kolmogorov-Petrovskii-Piscounov (KPP)
equation (also known as the Kolmogorov-Fisher equation) 
which we will need in the following Sections. The 
equation reads, in a general form:
\begin{equation}\label{KPP-general}
\frac{1}{2}\partial_{l} G =D \partial_{x}^{2} G   +f (G)
\end{equation}
where the diffusion coefficient is constant, the 
function $f (G)$ satisfies $f (0)=f (1)=0$, 
$f$ positive between $0$ and $1$ and 
$f' (0)=1, f' (G)\leq 1$ between $0$ and $1$. The usual case corresponds to  
$f (G)=G-G^{2}$. 
This equation  has been applied to 
a wide range of problems, from chemistry to hydrodynamic
instabilities or to the propagation of the Meissner phase 
into the normal phase in a superconductor \cite{saarloos98,dibartolo96}. 
It is the prototype of equations describing the diffusive invasion 
of an unstable state by a stable one. This can be seen by writing it as
a Landau equation $\partial_{l} G =-\partial F/\partial G$ 
whose free energy $F$ takes by construction its local maximum in $G=0$ (the
unstable state) and its minimum in $G=1$ (stable state). One usually chooses an
initial condition $G_{l=0}(x)$ monotonously decreasing from
$G_{l=0}(x \to -\infty)=1$ to $G_{l=0}(x \to +\infty)=0$.
For a large class of initial conditions,
the solutions of the KPP equations are
known to converge uniformly towards traveling waves solutions
of the form:
\begin{eqnarray}
G_l(x) \to h(x - m_l)
\label{front}
\end{eqnarray}
with $h(x \to +\infty)=0$ and $h(x \to -\infty)=1$.
However the question of the determination, given an initial condition,
of the asymptotic traveling wave $h$ and its velocity
$c = \lim_{l \to +\infty} \partial_l m_l$ 
has been largely debated for KPP equations or for
similar more complex non linear equations, and is still of current interest.
It is known as the front {\it velocity selection problem}.

\begin{figure}
\centerline{\fig{6cm}{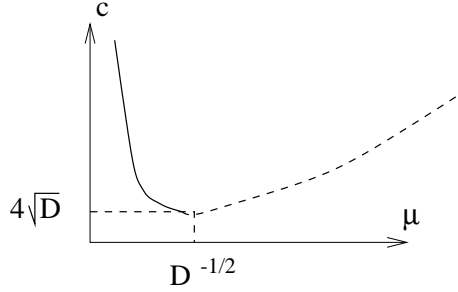}}
\caption{\label{fig:vitesse} Velocity of the front solution of KPP equation, 
as a function of its exponential decay ahead of the front.}
\end{figure}

It can be illustrated as follows. A family of possible 
front solutions exists, parametrized by the velocity $c$
and noted $h_c(x-m_l)$, as can be seen by substituting 
(\ref{front}) in (\ref{KPP-general}). Constraints exist
for the velocity. Indeed, one can linearize the KPP equation
in the region ahead of the front for large
positive $x - m_l$ where $h$ is very small.
As discussed below it is in fact this region which determines the velocity
and is universal. In this region one has:
\begin{eqnarray}
D h'' + \frac{c}{2} h' + h = 0
\end{eqnarray}
and thus $h$ is a superposition of 
two exponentials $e^{-\mu (x-cl)}$ with (see figure \ref{fig:vitesse})  
\begin{eqnarray}
c (\mu) =2 (\mu^{-1}+D\mu)
\label{velocity}
\end{eqnarray}
with $c \ge c^* = 4 \sqrt{D}$. The large $x$ behaviour is dominated
by the smaller $\mu$ and thus correspond to the left branch of the curve 
$c(\mu)$ in figure \ref{fig:vitesse} when $c > c^*$. In the marginal
case where $\mu=\mu_{c}=1/\sqrt{D}$ 
and $c=c^*$, the two eigenvalues are degenerate and the front
thus has the asymptotic behaviour for $x \to +\infty$:
\begin{eqnarray}
h(x) \sim (a x + b) e^{- \mu_c x}
\end{eqnarray}
Let us now give the known results for the selection 
of the asymptotic front among the family of possible $h_c$.

{(i) \it Velocity selection.}

Although for more complex equations one relies on 
stability analysis and a ''marginal stability criterion''
\cite{saarloos89}, in the case of the KPP equation a 
rigorous result is available.
A theorem due to Bramson \cite{bramson83} shows that the 
asymptotic traveling wave is determined by the behaviour at
$x \to + \infty$ of the initial condition $G_{l=0}(x)$
in the following manner. 
If $G_{l=0}(x)$ decays fast enough, as $G_{l=0}(x) \sim e^{-\mu x}$ for $x \to +\infty$
with $\mu \geq \mu_{c}=1/\sqrt{D}$ (or faster)
theorem B of  \cite{bramson83} states that 
$G_l(x)$ uniformly converges towards the traveling wave solution 
$h_{c^*} (x - m_l)$ of velocity 
$c=\lim_{l \to +\infty} \partial_l m_l=c^*=4\sqrt{D}$.
If $G_{l=0}(x)$ decays slower, as $G_{l=0}(x) \sim e^{-\mu x}$
with $\mu < \mu_{c}=1/\sqrt{D}$ then
$G_l(x)$ uniformly converges towards $h_{c(\mu)} (x - m_l)$
of velocity $c(\mu)$ continuously depending on 
$\mu$ and given by (\ref{velocity}).  
The asymptotic velocity is thus given by $c=4\sqrt{D}$ for  
steep enough initial condition $\mu\geq \mu_{c}=1/\sqrt{D}$ and 
$c=2 (D\mu+\mu^{-1})$ otherwise.

Moreover, the leading corrections to the velocity are also
given by the theorem of Bramson and 
are {\it independent of the function} $f(G)$ in (\ref{KPP-general}).
The corresponding position of the front $m_l$ is given by 
\cite{bramson83,brunet97,ebert98}
\begin{subequations}\label{velocity-corrections}
\begin{align}
m_l = &
\sqrt{D}\left(4l -\frac{3}{2}\ln l +\mathcal{O} (1) \right)
& \text{for} \quad  \mu > D^{-\frac{1}{2}}
\label{3demi} \\
= & 
\sqrt{D}\left(4l -\frac{1}{2}\ln l +\mathcal{O} (1) \right)
& \text{for} \quad  \mu = D^{-\frac{1}{2}}
\label{1demi}\\
= &
2 (D\mu + \mu^{-1})l
& \text{for} \quad  \mu < D^{-\frac{1}{2}}
\end{align}
\end{subequations}
Note that cases (\ref{3demi}) and (\ref{1demi})
differ only by the velocity corrections but not 
by the asymptotic front shape which is $h_{c^*}$ 
in both case.

Let us emphasize again the remarkable universality which 
arises in this problem. Clearly the detailed shape of the asymptotic front 
depends on the detailed form of the non linear term $f(G)$ 
in the KPP equation. However, the selection itself, the
selected velocity, its corrections, 
and the tail of the selected front function $h$
are all independent of $f(G)$. It is also natural physically
that it is the region in which the front penetrates
(region ahead of the front $x \to +\infty$) 
which determines the selection. This universality has been
explored further \cite{ebert98,ebert98b} and it has even been
shown that the next leading corrections to
$\partial_l m_l$ are also universal. Finally the marginal
case which interpolates between (\ref{3demi}) and
(\ref{1demi}) has been also explored\footnote{It was found in \cite{brunet97} 
that when $G_{l=0}(x) \sim x^{a} e^{-\mu_c x}$ 
with $-2 < a < 0$ one has: $
m_l = \sqrt{D} \left( 4 l - \frac{1-a}{2} \ln l + O(1) \right)$}.
The asymptotic front shape is the same for all these cases, but 
not the detailed finite $l$ shape as we now discuss.

{(ii) \it Shape of the front for finite $l$} 

The problem has also been studied for finite but large $l$
\cite{brunet97,ebert98,ebert98b}. We start with the case where $G_{l=0}(x)$
decays fast for $x \to +\infty$ as  $G_{l=0}(x) \sim e^{-\mu x}$ 
with $\mu > \mu_{c}=1/\sqrt{D}$, most relevant 
for the following Sections. Then one must distinguish two regions in the traveling
wave solution as illustrated in figure \ref{fig:front-shape}.
The central region is the ``bulk'' or ``interior front'' 
for $|x - m_l|$ fixed and finite. There, the shape of 
the front corresponds to its asymptotic form
$G_l(x) \approx h_{c^{*}} (x-m_l)$ and the center moves
as $\partial_{l}m(l)=c^{*}-\frac{3\sqrt{D}}{2l}+{\it O} (l^{-\frac{3}{2}})$
with $c^* = 4\sqrt{D}$. 
Far ahead of this ``interior front'', $x - m_l \gg 1$,
$G_l(x)$ still decays faster that in the asymptotic solution $h_{c^{*}} (x-m_l)$.
Thus there exist an intermediate region (which we call the ``intermediate front region'') 
which matches between the interior region and the region at infinity.
In this intermediate region which corresponds to $x - m_l \sim \sqrt{D l}$
one must take into account diffusion, and one can solve the linearized KPP equation 
without assuming a front solution, but assuming a scaling form.
It is found \cite{ebert98,brunet97} that in this scaling ``far front'' region $x - m_l \sim \sqrt{D l}$,
the solution $G_l(x)$ 
behaves as\footnote{Similar results where obtained for the marginal cases
$G_{l=0} \sim x^{a} e^{-\mu_c x}$ with $-2 < a < 0$
with a scaling function noted 
$G_{\frac{1-a}{2}}((x-m (l))/(2 D l))$ in \cite{brunet97}
(for $a=-2$ one gets the above result, but the form 
for general $a$ is more complicated).}:
\begin{subequations}
\begin{eqnarray}\label{front-intermediate}
&& G_l(x) \approx g_l(x-m_l) \\
&& g_l(x) = A \left(\frac{x}{\sqrt{D}}+cte + \mathcal{O} (l^{-\frac{1}{2}}) \right)
e^{-\mu_{c} x} e^{-\frac{x^{2}}{8Dl}}
\end{eqnarray}
\end{subequations}
Note that later on we will need to distinguish a region even further
ahead of the front for $x - m_l \sim l$. 


Let us close this Section by returning to the question of the
dependence of the RG equation of the random XY model 
in the cutoff procedure.
We can show that the cutoff procedure will determine the 
function $f(G)$. Let us consider for instance again the limit
of zero disorder, discussed in Appendix F. Since in that
case $D=0$ one should have:
\begin{eqnarray}
\frac{1}{2} \partial_l G_l(x) = f(G)
\label{zerodis}
\end{eqnarray}
with in that case $1 - G_l(x) = <\exp( y e^{-\beta (x-E_l)})>_{P_l(y)}$
and $P$ characterizes the ``disorder'' associated with the
choice of cutoff procedure. An interesting choice corresponds
to $P_l(y) = \delta(y-y_l)$ where $y_l$ satisfies the Kosterlitz
Thouless RG equation $\partial_l y_l = (2-K) y_l$.
It is easy to see that in that case $G_l(x)$ satisfies 
(\ref{zerodis}) with $f(G) = - (1-G) \ln (1- G)$. 
This is further discussed in \cite{carpentier99ter}

\section{XY phase and critical behaviour at low temperature}

\label{part:XYphase}

\subsection{Phase diagram from the scale dependent fugacity distribution}

\label{part:distrib-l} 

We now use the solutions of the KPP equation discussed in the previous
Section to determine the phase diagram of the XY model with random phase
shift. Let us first look for the 
XY phase where we expect that 
$J_l$ and $D_l=\frac{1}{2} \sigma_l J_l^2$ reach limits,
respectively $J_R$ and $D_R = \frac{1}{2} \sigma_R J_R^2$ at large
$l$. Thus in the XY phase at large $l$ the KPP equation with constant
$D=D_R$ can be used. Precise behaviour near the phase transition away from
the XY phase, as well as intermediate scale dependence inside the
XY phase {\it a priori} requires taking into account the $l$ dependence of
$D_l$ which will be done in the following Sections. Note that $l$ dependence of $J_l$
itself results only in a shift that can always be trivially taken into account.

\begin{figure}
\centerline{\fig{11cm}{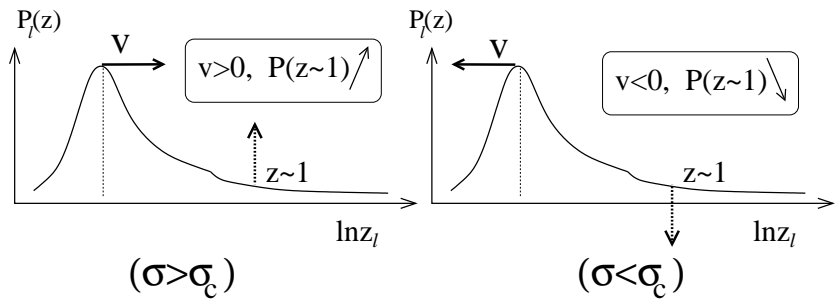}}
\caption{\label{fig:relative-velocity} 
Schematic representation of the connection between the sign of the total velocity of the
front $v=\partial_{l}X_{l} = c - J$ and the increase 
or decrease of $P_{l} (1)$.}
\end{figure}

First we note that the phase diagram will be entirely determined by the
velocity selected in the KPP equation. Indeed, we know from the previous
section that $G_l(x)$ converges to a traveling front solution
$G_l(x) \to h_c(x-m_l)$ of velocity $c$. The parametrization
(\ref{G-Tfinie}) then implies that the distribution $P_l(z)$ of vortex fugacity
for the random XY model, itself converges to a traveling front solution,
more conveniently
expressed in the random core energy variable $u=\ln z$ as: 
\begin{eqnarray}
 \tilde{P}_{l} (u)  \to_{l \to +\infty}~~ \tilde{p}(u-X_{l})  ~~~~,~~~~~ X_{l}=m_l-E_{l}
\end{eqnarray}
where $m_l$ is given in (\ref{velocity-corrections}) 
and $E_l = \int_0^l J_{l'} dl'$. 
The center of the front of $\tilde{P}_{l}(u)$, located in
$u_{typ} \approx X_l + O(1)$ corresponds to the 
maximum of the distribution $\tilde{P}(u)$ 
(as can be easily seen on (\ref{G-T0})) and to the typical
values of the random variable $u$. The front shape of
$\tilde{P}_{l}(u)$ is simply related to the 
KPP front solution $h_c$ through $\int_u \tilde{p}(u) \exp(e^{\beta(u-x)})=
1-h_c(x)$. The asymptotic velocity of the front of $\tilde{P}_{l}(u)$ is thus:
\begin{eqnarray}
- \partial_l E^c_{typ} = \partial_l u_{typ} = \partial_l X_l = \partial_l m_l - J_l
\to_{l \to +\infty}~~ c - J_R 
\end{eqnarray}
The total velocity in the $u$ variable is thus the KPP velocity 
minus the stiffness. The former comes from the spread of the 
distribution due to disorder, while the latter from the effect of
interactions. In previous Sections we have explained that the XY phase
corresponds to a decrease of the density of favorable regions 
$P_l(1) = P_l(z \sim 1)$ (i.e $u\sim 0$)
and to the absence of topological defects at large scale.
In particular, for $J$ and $D$ to reach finite 
asymptotic values $J_R$ and $D_R$, it is necessary that 
$P_{l}(z\sim 1)$ decreases fast enough. $P_l(1)$ can be 
estimated crudely\footnote{this estimate turns out 
to be accurate only near the transition as discussed below, 
but this does not change the conclusions} as $\tilde{p}(-X_l)$.
Thus, as shown in
fig. \ref{fig:relative-velocity}, when 
the total velocity $\partial_{l}X_{l}$ is negative (XY phase),
the front moves to the left (large $z$ or large $u$) and thus the
probability of events $z\sim 1$ decreases while if
$\partial_{l}X_{l}$ is positive, $P_{l}(1)$ increases asymptotically
(disordered phase). The XY phase thus corresponds to the region where the 
total velocity is negative and thus to:
\begin{eqnarray}
c - J_R > 0
\end{eqnarray}
in which case the whole probability distribution of the core
energy $E^c_{typ}=-u$, its typical and average values 
drift to $+\infty$ at larger scale. The transition
line between the disordered and the XY phase can be located, in the
plane $J_R$, $\sigma_R$, by finding the
line where this relative velocity $\partial_{l} X_{l}$ vanishes.

\begin{figure}[thb] 
\centerline{\fig{8cm}{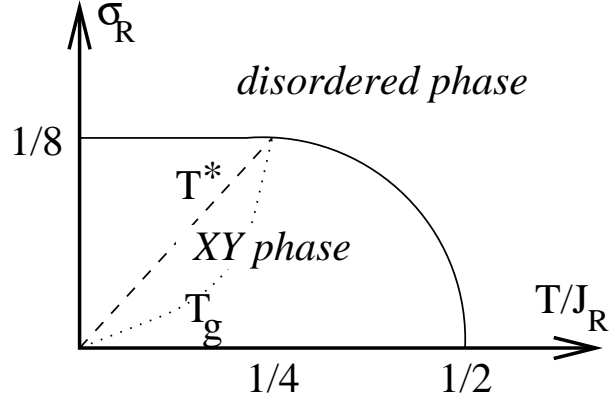}}
\caption{\label{fig:phase-diag} Phase diagram as a function 
of the renormalized disorder strength $\sigma_R$ and the temperature $T/J_R$.
$T_{g}$ corresponds to the freezing temperature and the transition line
for $T < T_g$ is strictly horizontal when drawn in the renormalized
parameters (and slightly curved when drawn in the bare parameters).} 
\end{figure}

The phase diagram can now be obtained by determining the velocity $c$
as a function of $T$ and $\sigma$. The crucial observation is
that by construction (as is the case in \cite{derrida88})
the initial condition $G_{l=0}(x)$ decays for large $x$ as:
\begin{eqnarray}
G_{l=0}(x) \sim_{x \to +\infty}  <z>_{P_0} e^{- \beta x}
\end{eqnarray}
thus, since $<z>_{P_0}$ is finite, it decays exponentially
and one can apply Bramson's results (\ref{velocity-corrections})
detailed in the previous
Sections with the identification:
\begin{eqnarray}
\mu = \beta
\end{eqnarray}
Thus in effect it is the temperature which selects the velocity $c$.
We find that, as depicted in fig \ref{fig:phase-diag}:

(i) for $\beta < \beta_g = \mu_c = 1/\sqrt{D_R}$, i.e at high enough
temperature $T > T_g= J_R \sqrt{\sigma_R/2}$, the velocity continuously
depends on temperature:
\begin{eqnarray}
c = c(\beta) = 2 (D_R \beta + \beta^{-1}) = T 
\left(2 + \frac{\sigma_R J_R^2}{T^2}
\right)
\end{eqnarray}
and thus, in that regime, the XY phase exists for:
\begin{eqnarray}
\partial_{l} X_{l} = T \left( 2 - \frac{J_R}{T} + \frac{\sigma_R J_R^2}{T^2}
\right)  < 0 
\label{xhight}
\end{eqnarray}
which is exactly the condition which would be
obtained from the averaged fugacity 
(at least when expressed in the renormalized parameters)
and leads to the transition line of Rubinstein et al.
as we have discussed earlier (see
eq.(\ref{RG-moments})).

(ii) the velocity of the front {\it freezes} at  $\beta=\beta_g$ and
for $\beta \ge \beta_g$, i.e at low temperature $T \le T_g$,
where it becomes temperature independent:
\begin{eqnarray}
c = c^* = 4 \sqrt{D_R}= J_R \sqrt{8 \sigma_R}
\end{eqnarray}
and the total velocity of the front $\tilde{P}_l(u)$ is now
\begin{eqnarray}
\partial_{l}X_{l} = J_R (\sqrt{8 \sigma_R} - 1 ) 
\label{xlowt}
\end{eqnarray}
Thus we obtain that below this freezing temperature at:
\begin{eqnarray}
T_g = J_R \sqrt{\frac{\sigma_R}{2}}
\end{eqnarray}
the transition between the XY phase and the disordered phases occurs at:
$\sigma_{R}=\sigma_{c}=\frac{1}{8}$. 
For $\sigma_R > \frac{1}{8}$ the system is unstable to 
the proliferation of topological defects induced by disorder.

\subsection{study of the XY phase}

To describe the XY phase it is important to understand the 
various regions of the scale dependent fugacity distribution $P_l(z)$.
Indeed the fugacity distribution also gives the distribution
of the charge correlation functions in the Coulomb gas 
from the relations (\ref{exp:correl1}).

\subsubsection{shape of the fugacity distribution in the XY phase}

\label{part:distrib-shape-l}

\begin{figure}[thb] 
\centerline{\fig{10cm}{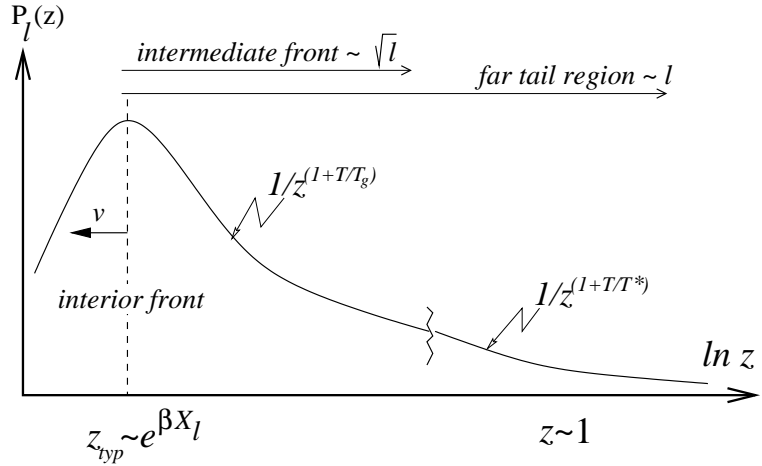}}
\caption{  \label{fig:front-shape} Structure of the 
scale dependent fugacity distribution $P_{l} (z)$ 
in the XY phase at low $T$ and the several regions 
discussed in the text.}
\end{figure}

In the XY phase one can describe the large scale behaviour of the
fugacity distribution by neglecting the slow scale dependence of 
$\sigma$ and $J$ (adiabatic approximation). The behaviour of 
$\tilde{P}_l(u)$ can be obtained by studying $\tilde{G}_l(x)=G_l(x+E_l)$ 
since one has $1-\tilde{G}_l(x) = \int_u \tilde{P}_l(u) \exp(e^{\beta (u-x)})$
with $E_l = \int_0^l dl' J_{l'}$. At $T=0$ this simplifies into 
$\tilde{P}_l(u)=-\tilde{G}_l'(u)$. It is also useful to note that
the Laplace transform of $P_l(z)$ is $\hat{P}_l(s) = 1 - \tilde{G}_l(-T \ln s)$.

At low $T$ in the XY phase $P_l(z)$ becomes very broad
and one must distinguish for large but finite $l$ 
several different regions
represented in fig. \ref{fig:front-shape}. Using the
results of the preceding Sections on the KPP equation 
we now describe these regions in details:

{\it interior front}: the first region corresponds to the bulk of the
probability distribution centered around the typical value
of the fugacity $z_{typ} \sim e^{\beta X_l}$. From the results 
for the front velocity (\ref{xlowt},{\ref{xhight}) 
one thus obtains that in the XY phase
at high temperature $T>T_g=J \sqrt{\sigma/2}$ the {\it typical} renormalized fugacity 
decays with the scale as:
\begin{eqnarray}
z^{typ}_l \sim e^{(2 - \frac{J}{T} + \frac{\sigma J^2}{T^2}) l}
\end{eqnarray}
which in that case is also the scaling of the average fugacity $\langle
z \rangle_{P_l}$.
On the other hand, below the freezing temperature $T<T_g$ it decays as:
\begin{eqnarray}
z^{typ}_l \sim e^{- \beta J (1 - \sqrt{8 \sigma}) l}
\label{typ2}
\end{eqnarray}
The bulk of the distribution is thus well described
by the asymptotic front of the KPP equation as
$\tilde{G}_l(x) \approx h(x-X_l)$ and 
$\tilde{P}_l(u) \approx \tilde{p}(u-X_l)$.
Its precise shape is of course non universal
as it depends on the details of the definition of the
fugacity at the level of the cutoff (e.g. the function $f(G)$ 
in (\ref{KPP-general})). However, since the physically interesting
region $z > z_{typ}$ corresponds to the region ahead of the
KPP front (even though the {\it total} velocity $v=c-J$
is negative, see fig. \ref{fig:front-shape}) universal results about the
KPP equation can be used. In particular the {\it near tail} (i.e the 
scaling region $u-X_l$ fixed but large) is universal
and can be obtained since $h(x) \sim  x e^{- x/T_g}$.
Through Laplace inversion one gets that the fugacity distribution
in that near tail region behaves as a power law:
\begin{eqnarray}
P_l(z) dz \sim   \left( \frac{z^{typ}_l}{z} \right)^{1+\frac{T}{T_g}} 
\ln \left( \frac{z}{z^{typ}_l} \right) ~~ \frac{dz}{z^{typ}_l}
\label{typical}
\end{eqnarray}
which is valid for $0<T<T_g$ and in the regime $z/z^{typ}_l$ fixed (as $l$ grows) 
but large. The freezing which occurs at $T_g$
thus concerns {\it typical regions} which, deep in the XY phase 
have a very small fugacity $z^{typ}_l$. It corresponds to
the temperature at which the first moment of the distribution 
on the r.h.s. of (\ref{typical}) becomes infinite. Thus for
$T<T_g$ the true average fugacity $\langle z \rangle_{P_l} \gg z^{typ}_l$
and becomes dominated by rare local environments corresponding to
values of $z$ outside of the bulk of $P_l(z)$
(where (\ref{typical}) is invalid).

{\it intermediate front}: the second region is the ``intermediate front region''
(see fig. \ref{fig:front-shape}) and corresponds to 
$u - X_l \sim \sqrt{l}$. There we know from (\ref{front-intermediate})
that $\tilde{G}_l(x) \approx g_l(x - X_l)$ with 
$g_l(y) \sim \frac{y}{\sqrt{D}} \exp (-\frac{y}{\sqrt{D}}-\frac{y^{2}}{8Dl})$.
This region will be crucial to describe the critical behaviour in
the following Section.

{\it far tail region}: finally the most important region to describe the
XY phase is the ``far tail region''
far ahead of the front with $u - X_l \sim l$
(see fig. \ref{fig:front-shape}). Indeed, to obtain the
renormalization of $J$ and $\sigma$ (via the screening
equations (\ref{screening2})) 
we are interested in the rare events $z \sim 1$ of small
probability $P_l(1)$, but which dominate the {\it average} 
correlations. These events correspond to the region of fixed $u$ and 
thus to $u - X_l \sim - X_l \sim (J-c) l$.
Fortunately, this region is so far ahead of the front
that $G_l(x)$ can be obtained with excellent accuracy
by solving the linearized KPP equation:
\begin{eqnarray}
\label{linkpp}
\frac{1}{2} \partial_l G = D \partial^2_x G + G
\end{eqnarray}
by straightforward integration from from $l=0$ to $l$. This leads to:
\begin{eqnarray}  \label{kpplinear}
G_l(x) =  \frac{e^{2l}}{\sqrt{8 \pi \int_0^l D_{l'} dl'}} \int_{x'}
e^{- \frac{(x-x')^2}{8 \int_0^l D_{l'} dl'}} G_{l=0}(x')
\end{eqnarray}
We also notice that in this regime
$\tilde{P}_l(u)$ can also be obtained explicitly. Indeed,
since the relation between $\tilde{P}_l(u)$ and
$G_l(x)$ is a simple linear ($l$-dependent) convolution
it is straighforward to show (e.g. via $x$ Fourier space)
that $G_l(x)$ obeying the linearized KPP equation 
(\ref{linkpp}) is equivalent to $\tilde{P}_l(u)$ 
satisfying:
\begin{eqnarray}   \label{linearizedpu}
&& \partial_l \tilde{P}(u) =  
\left( J \partial_{u} + \sigma J^2 
\partial^{2}_{u} \right) \tilde{P}  + 2 \tilde{P} 
\end{eqnarray}
This linearized equation was not entirely obvious to guess directly from
(\ref{Rg-eq-1D-u}) and justifies, even in this simplest regime,
the detour through the rigorous results for the KPP equation. 
Using $E_l = \int_0^l dl' J_{l'}$ it yields:
\begin{eqnarray}  \label{kpplinearu}
\tilde{P}_l(u) =  \frac{e^{2 l}}{\sqrt{8 \pi \int_0^l D_{l'} dl'}}
 \int_{u'} e^{- \frac{(u + E_l - u')^2}{8 \int_0^l D_{l'} dl'}} \tilde{P}_0(u')
\end{eqnarray}

It is interesting to note that for {\it fixed} $x$ and $u$, 
$\tilde{G}_l(x)$ and $\tilde{P}_l(u)$ can be estimated
in the large $l$ limit as simple exponentials:
\begin{eqnarray}
&& \tilde{P}_l(u) \sim_{l\to \infty}
C \frac{e^{2 l (1-\frac{1}{8\sigma})}}{\sqrt{4 \pi \sigma J^2 l}} 
e^{- \frac{u}{2\sigma J}} \label{papp}
\end{eqnarray}
with $C = \int_{x'} e^{\frac{x'}{2\sigma J}} P_{l=0}(x')$
and an identical expression for $\tilde{G}_l(x=u)$ with a
different prefactor $C' = \int_{x'} e^{\frac{x'}{2\sigma J}} G_{l=0}(x')$.
These forms, which are valid in the region of fixed fixed $u \sim O(1)$ 
(and fixed $x \sim O(1)$ respectively) will be useful to
estimate the renormalization of $J$ and $\sigma$ 
below\footnote{they were obtained from (\ref{kpplinear}) 
and (\ref{kpplinearu}) by approximating 
$\exp(- (x^2+{x'}^2)/(8 Dl)) \approx 1$, and similarly for $u$
(to be precise the $D$ and $J$ which 
appear here are $1/l \int_0^l D_l$ and $1/l \int_0^l J_l$ respectively).
They are valid only as long as the integrals which defines
$C$ and $C'$ are convergent. Since $G_{l=0}(x') \sim e^{- \beta x'}$ for large
positive $x'$ the estimate for $\tilde{G}_{l}(x')$ is valid only as long as
$T < T^* = 2 \sigma J$. for $T > T^*$ one must instead 
perform saddle point estimate, see below}.

Thus at low temperature we find the following decay
of the probability of rare favorable local environments:
\begin{eqnarray}
P_{l}(z\sim 1) \sim A l^{-1/2} e^{2 l (1 - \frac{1}{8\sigma})}
\label{pl1}
\end{eqnarray}
with $A=C/\sqrt{4 \pi \sigma J^2}$. We also find that the 
distribution $P_{l}(z)$ has an algebraic tail at low temperature
in the region $z \sim 1$ (i.e. $z$ fixed as $l \to \infty$)
as:
\begin{eqnarray}
P_{l}(z) \approx P_{l}(1) \frac{1}{z^{1+T/T^{*}}} \qquad , \qquad T^* = 2 \sigma J
\end{eqnarray}
with is a different power law 
behaviour\footnote{note that for $T > T^*$ the averaged 
fugacity $\langle z \rangle_{P_l}$ is controlled by $z \ll 1$,
while for $T < T^*$ is controlled by $z \gg 1$} than the one 
which characterizes typical fugacities (\ref{typical}).
These two different power law behaviours are represented in figure
\ref{fig:front-shape}.

Finally, one can check that the three regions match
properly and thus we have a fairly complete description of $P_l(u)$.
For instance, using the expression valid in the region
(ii) for {\it fixed} $x$ at large $l$ one gets:
\begin{eqnarray*}  
 \tilde{G}_l(x) \approx g_l(x-X_l) \sim_{l\to \infty}
&&- \frac{A X_l}{\sqrt{D}} 
e^{\frac{X_l}{\sqrt{D}} - \frac{X_l^2}{8 D l} - x( \frac{1}{\sqrt{D}} 
+ \frac{X_l}{4 D l})}\\
&&= \frac{l^{X_l/(8 \sqrt{D} l)} e^{2l (1-\frac{1}{8\sigma})}}{\sqrt{4 \pi 
\sigma J^2 l}} 
e^{- \frac{x}{2\sigma J}}
\end{eqnarray*}
which always gives the result (\ref{papp}) up to the logarithmic corrections
$l$ prefactors factors, and reproduces even the 
$l$ prefactors correctly for $X_l/l \to 0$, which is 
when we expect the matching to become exact\footnote{this type of matching 
has been used to derive the $\frac{3}{2}$ coefficient in
the logarithmic correction
to the velocity \cite{ebert98}.}.

\subsubsection{Screening and correlations in the XY phase}

To study the screening equations (\ref{screening2}) in the XY phase we need
the distribution $\tilde{P}_l^d(u_d) = \tilde{P}_l *_u \tilde{P}_l$
of dipole core energy $u_d = u' + u''$.
Fortunately in this regime it can be computed simply from
(\ref{kpplinearu}) and is given simply by:
\begin{eqnarray}  \label{kpplinear-bis}
&& \tilde{P}^d_l(u) =  \frac{e^{4 l}}{\sqrt{16 \pi \int_0^l D_{l'} dl'}}
 \int_{u'} e^{- \frac{(u + 2 E_l - u')^2}{16 \int_0^l D_{l'} dl'}} 
\tilde{P}^d_0(u') \sim 
C'' \frac{e^{4 l (1-\frac{1}{8 \sigma})}}{\sqrt{8 \pi \sigma J^2 l}} 
e^{- \frac{u}{2 \sigma J}}
\end{eqnarray}
the last equality being valid for fixed $u$ and $l \to +\infty$,
and $C''=\int_{u'} \exp(\frac{u'}{2 \sigma J}) \tilde{P}^d_0(u')$
in terms of the initial
dipole core energy distribution. Substituting this last form
in (\ref{screening2}) we obtain, in the large $l$ limit:

\begin{subequations} \label{screeningc}
\begin{eqnarray}
&& \partial_{l} (J^{-1}) = 
\frac{8 c_{1}}{c_{2}^2} 
\frac{e^{4 ( 1 - \frac{1}{8 \sigma})l}}{\sqrt{8 \pi \sigma J^2 l}}
C'' \int_{-\infty}^{+\infty} du \frac{\beta e^{- \beta u} e^{-\beta^* u}}{(1 + e^{-\beta u})^2}
= 
C_s I_J(\beta) l^{\frac{1}{2}} ~ P_l(1)^2
\\
&& \partial_{l} \sigma =
\frac{8 c_{1}}{c_{2}^2}
\frac{e^{4 ( 1 - \frac{1}{8 \sigma})l}}{\sqrt{8 \pi \sigma J^2 l}}
C'' \int_{-\infty}^{+\infty} du \frac{e^{-\beta^* u}}{(1 + e^{-\beta u})^2}
=
C_s I_\sigma(\beta) l^{\frac{1}{2}} ~ P_l(1)^2 
\end{eqnarray}
\end{subequations}
with $C_s = 8 c_{1} C'' J \sqrt{\pi \sigma}/(
\sqrt{2} c_{2}^2 C^2)$ a constant roughly independent of
the temperature. The two above integrals
\begin{eqnarray}
I_J(\beta) = \frac{T}{T^*} \frac{\pi}{\sin(\frac{T}{T^*} \pi)} \quad, \quad
I_\sigma(\beta) = T \frac{(1-\frac{T}{T^*}) \pi}{\sin(\frac{T}{T^*} \pi)}
\end{eqnarray}
are convergent respectively only for $T<T^*$ and $T< 2 T^*$,
in which cases it is indeed legitimate to replace $\tilde{P}_l^d(u)$
by its above asymptotic form. Note that the $T=0$ limit is well
defined since in that case $I_J =1$ and $I_\sigma =T^* = 2 \sigma J$.

At higher temperature $T>T^*$ the full front solution controls
the renormalization of $J$. Using the simple above gaussian form
yields that the $u$ integral is dominated by the saddle point
$u = u' - 2 J l + 4 \sigma J^2 l$ which gives:
\begin{eqnarray}
\partial_{l} (J^{-1}) \sim e^{ 2 (2 l - \frac{J}{T} + 
\frac{\sigma J^2}{T^2}) } \qquad T > T^* 
\end{eqnarray}
which corresponds for $T>T^*$, to the behaviour of $<z>^2$
(and for $T>T_g$ of $z_{typ}^2$), as expected.
A similar, though more involved, analysis can be performed for $\sigma$.

Thus we have obtained the equations (\ref{screeningc})
for the renormalization of $J$ and $\sigma$ in the XY phase.
Since we have shown that $P_l(1)$ decreases exponentially
as (\ref{pl1}) we conclude that $J$ and $\sigma$
reach their finite limits $J_R>0$ and $\sigma_R < +\infty$
as power laws of the systems size.

This analysis also yields the full distribution of the correlation
function of the charges in the XY phase. Indeed, let us recall that:
\begin{subequations}
\begin{eqnarray}
&& 
- \tilde{a}_0^4 \langle n_{\bf 0} n_{R=\tilde{a}_0} \rangle_c = 
\frac{ z^{+}_{{\bf 0}}z^{-}_{{\bf R}}+
       z^{+}_{{\bf R}}z^{-}_{{\bf 0}}+
     4 z^{+}_{{\bf 0}}z^{-}_{{\bf R}} z^{+}_{{\bf R}}z^{-}_{{\bf 0}} }
{(1 + z^{-}_{{\bf 0}} z^{-}_{{\bf R}}+
      z^{+}_{{\bf R}} z^{-}_{{\bf 0}} )^2} \\
 & & 
- \tilde{a}_0^4 \langle n_{0}\rangle \langle n_{R=\tilde{a}_0}\rangle =
 \frac{ (
z^{+}_{{\bf 0}} z^{-}_{{\bf R}}-
z^{+}_{{\bf R}} z^{-}_{{\bf 0}})^2 }{
(1 + z^{-}_{{\bf 0}} z^{-}_{{\bf R}}+
     z^{+}_{{\bf R}} z^{-}_{{\bf 0}} )^2}
\end{eqnarray}
\end{subequations}
Thus by following the RG up to the scale $l^* = \ln(R/a_0) = \ln(\tilde{a}_0/a_0)$ we
can obtain from $P_{l^*}(z)$ the distribution of the charge
correlations at large $R$. 

For instance, from the above and (\ref{typ2}) one finds the following decay of the 
{\it typical} thermal and disorder correlations:

\begin{subequations}
\begin{eqnarray}
&&  - \langle n_{\bf 0} n_{R} \rangle_c  \bigr|_{typ} 
\sim \left(\frac{a_0}{R} \right)^{4 + \frac{2 J}{T} (1 - \sqrt{8 \sigma})} \\
&&  - \langle n_{0}\rangle \langle n_{R}\rangle \bigr|_{typ} \sim 
\left(\frac{a_0}{R} \right)^{4 + \frac{4 J}{T} (1 - \sqrt{8 \sigma})}
\end{eqnarray}
\end{subequations}
for $\sigma < \sigma_c$ and $T<T_g$. For $T>T_g$
one has instead $\langle n_{\bf 0} n_{R} \rangle_c |_{typ}
\sim (a_0/R)^{2J/T - 2 \sigma J^2/T^2}$.

The averaged moments can be obtained as above by substituting the
exponential form of the distribution $P^d_{l^*}(u)$ for $u \sim 0$.
Performing the corresponding integrals, one gets:
\begin{subequations}
\begin{eqnarray}
&&  \overline{ \langle n_{\bf 0} n_{R} \rangle_c^p }
\sim A_p(T) \left(\ln \frac{R}{a_0} \right)^{-\frac{1}{2}}
\left(\frac{a_0}{R} \right)^{4 p + 4 (\frac{1}{8 \sigma} - 1)} \\
&&  \overline{ \langle n_{0}\rangle^p \langle n_{R}\rangle^p }  \sim 
B_p(T) \left(\ln \frac{R}{a_0} \right)^{-\frac{1}{2}}
\left(\frac{a_0}{R} \right)^{4 p + 4 (\frac{1}{8 \sigma} - 1)}
\end{eqnarray}
\end{subequations}
with
\begin{subequations}
\begin{eqnarray}
&& A_p(T) = \frac{C''}{\sqrt{8 \pi \sigma J^2}}
T ~\frac{\Gamma\left[p + \frac{T}{T^*}\right] 
\Gamma\left[p -\frac{T}{T^*}\right]}{\Gamma[2 p]}  \\
&& B_p(T) = \frac{C''}{\sqrt{8 \pi \sigma J^2}}
T ~\frac{\Gamma\left[\frac{T}{T^*}\right] 
\Gamma\left[2 p - \frac{T}{T^*}\right]}{\Gamma[2 p]} 
\end{eqnarray}
\end{subequations}
These formulae are valid only at low enough temperature 
$T < T^*(p)$.

Finally, although we have not attempted a precise RG calculation
of the XY order correlation functions, the following behaviour 
should hold in the quasi ordered XY phase:
\begin{subequations}
\begin{eqnarray}
&& \overline{ <e^{i (\theta_{\bf R} - \theta_{\bf 0})} > } \sim R^{-\eta} \\
&& \overline{ <e^{i \theta_{\bf R} } > <e^{- i \theta_{\bf 0}} > } 
\sim R^{- \overline{\eta}} 
\end{eqnarray}
\end{subequations}
with $\eta = \frac{1}{2} ( \sigma_R + \frac{T}{J_R} )$ 
and $\overline{\eta} = \frac{1}{2} \sigma_R$. At the zero temperature
transition point $\sigma_R=\sigma_c=1/8$, $T=0$, the value of these
exponents are universal $\eta = \overline{\eta} = 1/16$.

\subsection{Critical behaviour at zero temperature}

\label{part:critical}

We now study the transition from the XY to the
disordered phases at low temperature at $\sigma_R = 1/8$.
From the previous Sections we know that it occurs 
when the total velocity of the front of the
distribution $\tilde{P}_l(u)$ vanishes, i.e the critical
region is defined by 
\begin{eqnarray}
\partial_{l} X_l = J (\sqrt{8 \sigma} - 1) \approx 0 
\end{eqnarray}
in a large $l$ regime. While in the XY phase the physics
was dominated by the {\it far tail} of the traveling front
$u - X_l \sim l$ in the critical regime 
it is the {\it interior front} $u - X_l \sim O(1)$, as well as 
the intermediate front region $u - X_l \sim \sqrt{l}$
(see fig. \ref{fig:front-shape})
which controls the transition. Thus the correct description of the transition
{\it requires} the knowledge of the KPP physics in an essential
way, and is thus entirely novel.

For simplicity, we will only present the analysis at $T=0$.
We will also work to first order in $\sigma - \sigma_c$. 
We first assume that the coefficient $D = \frac{1}{2} \sigma J^2$
varies sufficiently slowly near the transition so that 
the results from the KPP equation with a constant $D$ can
be used. This assumption will be self consistently verified
at the end.

Near the transition the center of the front is located at
$u=X_l$ with $X_l \approx (4 \sqrt{D}- J) l - \frac{3}{2} \sqrt{D} \ln l + X_0$
as indicated by (\ref{velocity-corrections}). Thus in the
critical regime one still has $X_l \to - \infty$, although 
logarithmically in $l$ exactly at the transition. Thus this
critical regime can still be studied perturbatively,
as $P_l(1)$ remains very small. The front velocity has the following
scale dependence:
\begin{eqnarray}
\partial_l X_l = 4 \sqrt{D} - J - \frac{3 \sqrt{D}}{2 l} + h.o.t
\label{velcritical}
\end{eqnarray}
where the h.o.t. contains the universal \cite{ebert98} $O(l^{-3/2})$ subdominant
corrections to velocity in the KPP equation as well as 
additional subdominant corrections originating from 
the slowly varying $D_l$. In the critical region at
$T=0$ we can use the KPP front solution 
$\tilde{P}_l(u) = \tilde{p}_l(u - X_l)$
with, from (\ref{front-intermediate}):
\begin{eqnarray}
\tilde{p}_l(x) = - g_l'(x) \sim_{x \gg 1}
A \frac{x}{D} e^{- \frac{x}{\sqrt{D}}}
e^{- \frac{x^2}{8 D l}}
\end{eqnarray}
an expression valid as long as $x/l \to 0$.
The RG equations for $J$ and
$\sigma$ in the critical region thus read:

\begin{subequations} \label{sct0}
\begin{eqnarray} \label{sct0.1}
&& \partial_{l} (J^{-1}) = \frac{8c_{1}}{c_{2}^{2}}
\int du p_l(u - X_l) p_l(- u - X_l) \\
&& \partial_{l} \sigma = \frac{8c_{1}}{c_{2}^{2}}
\int_{u+u'>- 2 X_l} p_l(u) p_l(u') \label{sct0.2}
\end{eqnarray}
\end{subequations}

We need to evaluate these expressions for $X_l$
large and negative (typically either as $\sim - \ln l$ on the critical
manifold or as $\sim (\sigma - \sigma_c) l$ very close to it). At criticality
they behave approximately as $\exp(2 X_l/\sqrt{D})$, which can be guessed
by setting $u \sim 0$ in the first expression.
A more accurate estimate of the above integrals is performed in the
Appendix \ref{part:integrals}. The end result is that, in the regime of interest
(where we can discard terms of order $(\sigma-\sigma_c)^2$) one has:
\begin{eqnarray}
&& \partial_{l} (J^{-1}) \sim \frac{C}{\sqrt{D}}
X_l^3 e^{2 X_l/\sqrt{D}} \qquad , \qquad
 \partial_{l} \sigma \sim C X_l^3 e^{2 X_l/\sqrt{D}}
\end{eqnarray}
where $C$ is a constant. Introducing the small parameter:
\begin{eqnarray}
g_l = e^{\frac{X_l}{\sqrt{D}}}
\end{eqnarray}
the density of favorable regions reads:
\begin{eqnarray}
P_l(1) \sim A \frac{X_l}{D} e^{\frac{X_l}{\sqrt{D}}} e^{- \frac{X_l^2}{8 D l}} 
\sim X_l g_l
\end{eqnarray}
since we can discard terms of order $(\sigma-\sigma_c)^2$.
From (\ref{velcritical}) one finds that $g_l$
satifies precisely the RG flow equation:
\begin{eqnarray}  \label{eq-g}
&& \partial_l g = \left( 16 (\sigma-\sigma_{c}) - \frac{3}{2 l} + h.o.t \right) g 
\end{eqnarray}
to lowest order in $\sigma-\sigma_{c}$. Note the
$\frac{3}{2}$ {\it universal} factor which arises from
the universal velocity corrections in the KPP equation
(\ref{velocity-corrections}).

A natural choice of parameters to describe the universal behaviour
around the transition is thus $g$ (which up to logarithmic corrections
is equal to the density of favorable sites $P_{l} (1)$) and $\sigma$
which satisfies:
\begin{equation}\label{eq-sigma-g}
\partial_{l}\sigma \sim C \left(\ln \frac{1}{g} \right)^3 g^{2}
\end{equation}
These two equations 
(\ref{eq-g}, \ref{eq-sigma-g}) form our complete set of RG equations
projected on the plane $\sigma,g$. They are somewhat analogous 
to the one describing a Kosterlitz-Thouless type transition,
with an important difference.
Here one readily finds that the separatrix is {\it vertical}. 
Introducing the deviation from criticality, 
$\epsilon=\sigma_c - \sigma_{\infty}$ 
(where $\sigma_{\infty} \equiv \sigma_R$) one has for
$\epsilon=0$ the flow:
\begin{eqnarray}
&& g_l \sim l^{-3/2}  \qquad, 
\qquad \sigma = \sigma_R - \gamma (\ln l)^3 l^{-2}
\end{eqnarray}
and thus satisfy $\sigma_{\infty} - \sigma_l \sim g_l^{4/3} (\ln (1/g_l))^3$.
Thus our RG equations yields
$g_{l}\sim l^{-\frac{3}{2}}$ at 
criticality\footnote{this indicate that at $T=0$
$\overline{ \langle n_{0}  n_{R} \rangle  }  \sim 
R^{-4} (\ln R)^{-3}$ at criticality} and a correlation length:
\begin{eqnarray}
\xi \sim \exp(\frac{\text{cst}}{ |\sigma-\sigma_{c}|})
\end{eqnarray}
since starting away from criticality $\epsilon>0$
one finds\footnote{the critical finite size corrections 
to $\sigma$ read $\sigma_{\infty} - \sigma_{l}
= \epsilon^2 F[\epsilon l]$ with
$F[x] = \int_x^{+\infty} dx x^{-3} e^{-32 x}$
up to logarithmic corrections.} $g_{l} \sim l^{-\frac{3}{2}} e^{16 \epsilon l}$.
This critical behaviour correspond to a new
universality class which is different from Kosterlitz-Thouless
and from the prediction of \cite{scheidl97,tang96}.
Note the crucial role of the $\frac{3}{2}$ universal factor.
Replacing it by any number less than $1$ would have led to
usual KT behaviour.

We can now check that the variations of $D$ should
be unimportant at the transition. Indeed one
finds\footnote{note the following interesting cancellation
in the variations of $D_l$. Writing $\partial_{l} \sigma = A_l[X_l]$
as a function of $X_l$, we get that $\partial_{l} (J^{-1}) = \frac{1}{2} A'[X_l]$
since $X_l$ appears only explicitly in the bound of the integral
in (\ref{sct0}). Thus one finds $\partial_l D = \frac{1}{2} J^2 A[X_l] 
(1 - J \sigma \frac{d}{d X_l} \ln A[X_l])$. Using that
using that $\frac{d}{d X_l} \ln A[X_l] = 2/\sqrt{D} + o(1)$
one gets that the leading variations cancel 
at the transition point $\sigma=1/8$.} that
$D_{\infty}- D_l \sim (\ln l)^2 l^{-2}$
at most. On the other hand 
we can estimate that if $D_{\infty}- D_l$ varies faster
than $\partial_l m_l \sim 1/l$ the KPP results should 
not be affected (see the scaling with $l$ of all terms in e.g. Eq (A7) 
in \cite{brunet97})

Let us close by noting again how universality
appears in the derivation of the critical behaviour.
Although most details of the fugacity distribution
$P(z_{\pm})$ (e.g. its bulk, the fusion rule..)
depend on the cutoff procedure the universality 
in the XY transition appears in a remarkable way.
It arises from the independence \cite{bramson83,brunet97,ebert98}
of the velocity, the velocity
corrections, and front tail on the precise form $f[G]$ of 
the non linear term in (\ref{KPP-general}).
This gives us confidence in the method developped here.

\section{Equivalent Sine Gordon model}
\label{part:sinegordon}

As is well known the Coulomb gas can also be equivalently formulated
as a Sine Gordon model  (see {\it e.g} \cite{zinn93}). In this
section we identify the random version of the Sine Gordon model
to which our analysis applies. 

The scalar Sine Gordon model, of partition function
$Z_{sg} = \int D\phi e^{- S(\phi) }$ is defined, in the absence of disorder
by the action:
\begin{eqnarray}
S(\phi) = \int d^2 {\bf r} ( \frac{1}{8 \pi K} 
(\nabla_{\bf r} \phi)^2 - g \cos(\phi)) 
\end{eqnarray}
where $\phi({\bf r}) \in [-\infty, +\infty]$. As is well known it
is equivalent to a Coulomb gas since, expanding in $g$ one has:
\begin{eqnarray} \label{sgexp}
Z_{sg} &=& \sum_{p=0}^{+\infty} \frac{1}{(2 p)!} 
g^{2p} \langle (\int d^2 {\bf r} \cos(\phi({\bf r})))^{2 p} \rangle_0
\\
\nonumber 
&=& \sum_{p=0}^{+\infty} \frac{1}{(2 p)!} 
g^{2p} C_{2 p}^p \langle \prod_{\alpha=1}^{p} \int d^2 {\bf r}_\alpha
d^2 \tilde{{\bf r}}_\alpha
e^{i ( \phi({\bf r}_\alpha) - \phi(\tilde{{\bf r}}_\alpha))} \rangle_0
\\
\nonumber 
&=& \sum_{p=0}^{+\infty} \frac{1}{p!^2} 
g^{2 p}  \prod_{\alpha=1}^{p} \int d^2 {\bf r}_\alpha d^2 
\tilde{{\bf r}}_\alpha 
 e^{ \tilde{K} \sum_{\gamma \neq \delta} n_\gamma n_\delta 
G(r_\gamma - r_\delta)}
\end{eqnarray}
where $\langle .. \rangle_0$ denotes averages with respect only to the quadratic
gradient part.
The interaction $G(r) = \Gamma(0) - \Gamma(r) \sim \ln|r|$ 
has been defined in (\ref{part:model}). The above partition sums involve
only $p$ neutral pairs of dipoles since, in the large size limit 
$\Gamma(0) \to \infty$. Finally $\gamma=1,..2p$
in the last line involves a summation over all distinct charges with
$n_\gamma=1$ for $\gamma=1,..p$ and $n_\gamma=-1$ for $\gamma=p+1,..2p$.

Let us turn to the disordered version of the model. To reproduce the 
bare version (\ref{Zcont}) of the model, one must first add a short range
correlated random {\it imaginary} field as follows:
\begin{eqnarray}
S(\phi,{\bf A}) = \int d^2 {\bf r} \left( \frac{1}{8 \pi K} 
(\nabla_{\bf r} \phi)^2 - i \frac{1}{2 \pi} {\bf A} \cdot \nabla_{\bf r} \phi
- g \cos(\phi) \right)
\label{sgaction}
\end{eqnarray}
with correlations $\overline{{\bf A}_q {\bf A}_{-q}}= \pi \sigma$.
Since it imposes now that $\langle i \nabla_{\bf r} \phi \rangle_0 = 2 K {\bf A}$,
each factor $\exp(i n_{\bf r} \phi({\bf r}))$ in the $g$ expansion in (\ref{sgexp})
yields an additional $\exp(n_{\bf r} \beta V({\bf r}))$ where
$\beta^2 \overline{V_q V_{-q}} = \frac{4 \pi \sigma K^2}{q^2}$
and thus reproduces the bare CG version (\ref{Zcont}) of our model.

Note that the above model still contains a uniform ``fugacity'' $g$.
It thus corresponds to the version studied in  \cite{rubinstein83} 
by Rubinstein et al. 
However we know that this cannot be the correct form under renormalization.
The first obvious idea is to generalize $g \to g({\bf r})$, i.e a 
disordered fugacity (the above expansion can be immediately
generalized to this case). This term will be generated, but is not the
end of the story. Indeed let us consider the symmetries of the above
action $S(\phi)$ (even in the presence of a $g({\bf r})$).
When ${\bf A}({\bf r})=0$ the action is real and invariant through 
$\phi({\bf r}) \to - \phi({\bf r})$. In the presence of the random
field this symmetry is broken (as $\phi({\bf r})$ acquires a non zero,
disorder environment dependent average) and the action is complex.
Thus nothing prevents that under coarse graining the action will
become:
\begin{multline}
 S(\phi,{\bf A},z_{+},z_{-}) = \\
\int d^2 {\bf r} \left( \frac{1}{8 \pi K} 
(\nabla_{\bf r} \phi)^2 - i \frac{1}{2 \pi} {\bf A} \cdot \nabla_{\bf r} \phi
- z_{+}({\bf r}) e^{ i \phi({\bf r}) } -
z_{-}({\bf r}) e^{ - i \phi({\bf r}) } \right)
\label{sgcoarse}
\end{multline}
and this is indeed precisely what we have found in the CG formulation,
i.e each sign of charge acquires a {\it different} local random fugacity.

There is however a symmetry constraint on the distribution of the local
random fugacities. Indeed in the above bare model (\ref{sgaction})
the full partition sum is real, as there is still a
{\it statistical symmetry}:
\begin{eqnarray}
S(\phi,{\bf A}) = S(\phi,- {\bf A})^*
\end{eqnarray}
for any configuration $\phi({\bf r})$ and environment ${\bf A}({\bf r})$.
Since the probabilities
of ${\bf A}({\bf r})$ and $- {\bf A}({\bf r})$ are the same, all physical
averages will be real. In the coarse grained model (\ref{sgcoarse}),
since we have that
\begin{eqnarray}
S(\phi,{\bf A},z_{+},z_{-}) = S(\phi,- {\bf A},z_{-},z_{+})^*
\end{eqnarray}
the probability (over environments) of the random fugacity disorder configuration
$\{ z_{+}({\bf r}) , z_{-}({\bf r}) \}$ should be equal 
to the probability of $\{ z_{-}({\bf r}) , z_{+}({\bf r}) \}$.

We can now check that this random Sine Gordon model (\ref{sgcoarse})
is indeed equivalent to the coarse grained disordered CG considered
in this paper by expanding its the partition sum in a given environment,
which yields:
\begin{multline}
Z_{sg}({\bf A},z_{+},z_{-}) =
 \sum_{p=0}^{+\infty} \frac{1}{p!^2} 
\prod_{\alpha=1}^{p} \int d^2 {\bf r}_\alpha d^2 \tilde{{\bf r}}_\alpha
z_{+}({\bf r}_\alpha) z_{-}(\tilde{{\bf r}}_\alpha)\\
\times \exp \left(
 K \sum_{\gamma \neq \delta} n_\gamma n_\delta 
G({\bf r}_\gamma - {\bf r}_\delta) 
+ \beta V^>({\bf r}_\gamma) n_\gamma \right)
\end{multline}
and thus the ${\bf A}$ random field disorder is associated to 
the long range part $V^>$ of the disorder in the CG
(see (\ref{decomposition})) 
while the random couplings constants $z_{\pm}({\bf r})$
in (\ref{sgcoarse}) are associated to the local random fugacities
in the CG.

Similarly one can establish a correspondence directly on the replicated
versions of both models. Replicating the above SG model and averaging
over disorder yields (the limit $m \to 0$ is implicit) :

\begin{multline}
\overline{Z_{sg}^m} = 
\int \prod_{a} D\phi_a e^{ - \int d^2 {\bf r} \frac{1}{8 \pi} 
\sum_{a,b} (\frac{1}{K} \delta_{ab} + \sigma)
\nabla \phi_a \cdot \nabla \phi_b }  \\
\times  
\left< e^{\int d^2 {\bf r} z_{+}({\bf r}) \sum_a e^{ i \phi_a({\bf r}) } -
z_{-}({\bf r}) \sum_a e^{ - i \phi_a({\bf r}) } } \right>_{z_+,z_-} 
\label{averagesg}
\end{multline}
Expanded to first order and treating the $z_{+}({\bf r})$ and $z_{-}({\bf r})$ as
uncorrelated in space and reexponentiating yields the replicated SG model
defined as:
\begin{subequations}
\begin{eqnarray}
&& \overline{Z_{sg}^m} = \int \prod_{a} D\phi_a e^{ - S_{rep}(\phi_a) } \\
&& S_{rep}(\phi_a) = 
\int d^2 {\bf r} \left( \frac{1}{8 \pi} 
\sum_{a,b} (\frac{1}{K} \delta_{ab} + \sigma)
\nabla \phi_a \cdot \nabla \phi_b  \right) 
- \sum_{{\bf n} \neq {\bf 0}} Y[{\bf n}] 
e^{i {\bf n} \cdot {\bf \phi }} 
\label{replicatedsg}
\end{eqnarray}
\end{subequations}
where ${\bf n} \cdot {\bf \phi} = \sum_a n^a \phi_a$ is the scalar
product in replica space and $Y[{\bf n}]$ are analogous
to the vector fugacities introduced previously in the replicated vector
CG model. Although the bare model obtained from 
(\ref{averagesg}) contains only single component replicated charges,
all multicomponent charges will be generated upon coarse graining, as in the
replicated vector CG. As we have seen all these charges should be
taken into account and thus the generic replicated SG model 
should contain all possible $Y[{\bf n}]$ with ${\bf n} \neq 0$.

This SG model can also be studied using RG, either in its
replicated form (\ref{replicatedsg}), or directly 
(\ref{sgcoarse}), very similarly to the vector CG studied
in this paper. Although variations in definitions of the fugacities
$Y[{\bf n}]$ and different cutoff procedures can induce some irrelevant
differences in the details of the renormalization (and different
RG equations), the two models
have the same physics. The Sine gordon formulation has several advantages,
such as exhibiting by construction the decomposition of the disorder in two 
physically very different components (see \ref{decomposition}).
It is also more amenable to replica variational methods than the CG,
left for future study.

\section{comparison with previous approaches}
\label{part:other-approaches}

Let us compare our method and results with the
work of Scheidl \cite{scheidl97}. In this work
the multicomponent charges (restricted to $0,\pm 1$
in each replica) were considered, but the fusion was not
taken into account. Also only dipole fugacities were
introduced. 

We can recover the RG equations and results of Scheidl 
within our approach by (i) artificially setting the fusion 
coefficient $c_2$ to $0$ in (\ref{loose}) (ii) assuming a log-normal
distribution $P=\Phi/ \int \Phi$ for the fugacities (which is consistent 
only when fusion is neglected) (iii) defining random ``dipole fugacities''
as $z_d^+ = z'_{+} z''_{-}$ and $z_d^- = z'_{-} z''_{+}$
(or equivalently $u=u'+u''$) identical to the ``dummy gaussian variables''
introduced by Scheidl. Within our diffusion formalism (see 
Section(\ref{part:replicalimit})) 
we find that the norm $\int \Phi$ diverges exponentially, 
which allows to recovers the extra factors $e^{4 l}$ appearing in
Scheidl's screening equations. Going from our diffusion formalism 
back to the replica formulation yields back the RG equations
of \cite{scheidl97}}.

In presence of a fusion term $c_2 >0$, the equations become
a priori very different. The fugacity distribution does
not remain log-normal as we do not assume a priori the form
of the distribution. Interestingly, although 
the log-normal does not reproduce correctly the true
distribution of fugacities (and misses connections such
as the one with the freezing of the DPCT via the KPP equation),
some of our results deep in the XY phase, such as the renormalization of $K$ and
$\sigma$ (see (\ref{screeningc})), agree with the one of Scheidl. It was not obvious a priori 
that the approach without fusion in the XY phase did not miss extra relevant 
physics in this model, and indeed near the transition
fusion appears to be crucial and must be taken into account.

\section{Conclusion}\label{part:conclusion}

To conclude, we have constructed in this paper a novel renormalization
group method which allows to study perturbatively, and in a consistent
way, a large class of disordered models which can be formulated as two dimensional 
Coulomb gases with quenched disorder. We have applied it specifically to the 
XY model with quenched random phases. We have obtained the phase diagram 
for this model, confirmed the existence of a low temperature XY phase,
and elucidated the critical behaviour at the transition where 
topological defects proliferate. It would be interesting to 
check our predictions in numerical 
simulations\footnote{in particular the full distribution of local fugacities
for finite sizes should be directly measurable in simulations.}

The present RG method is not based on the conventional perturbative expansion 
in a vortex fugacity $y$ spatially uniform over the system, which, as
we have shown, is only justified for pure models or for disordered models
at high enough temperature. Instead, it is constructed by first
defining the local random vortex fugacity (or core energy) and
then following its full probability distribution under coarse graining.
Below a freezing temperature this distribution becomes very broad
and cannot be followed by conventional CG methods. 
Our renormalization procedure allows to follow this broad distribution
in a controlled way, by defining the new perturbative parameter
as the concentration $P_l(z \sim 1)$ of rare regions favorable 
to vortices. This running parameter flows to zero in the XY phase 
and at the transition (marginal flow). Hence both can be studied
perturbatively. The underlying physical picture obtained here is that
the transition is controlled, as the scale increases, by the proliferation 
of vortices in less and less rare favorable regions. We find 
that it has some features reminiscent of a Kosterlitz-Thouless
transition, with important differences, such as the scaling of the
correlation length.

To derive the RG equation for the distribution of vortex fugacities,
we have introduced two equivalent methods. One is based on the replicated
vector Coulomb gas version of the model, and in an expansion in
the vector fugacities. The second one is direct, with no use of
replicas, and is based on a systematic expansion of all physical
quantities in the number of points, i.e in independent local regions
and thus, in the end, in powers of the concentration $P_l(z \sim 1)$
of rare favorable regions. As we have shown these two methods are 
fully equivalent: the first one being more systematic and the second one 
allowing for a clear physical understanding of the problem in terms
of probability distributions. These two expansion methods are 
highly non perturbative in the original uniform fugacity variable $y$.
Since they are constructed from charge fugacities they can be made
fully consistent (by contrast with previous approaches based on dipole
fugacities).

Our method sheds light on the broader issue of universality in
random systems. The spirit of the RG method is that at large scale
most information about the system is irrelevant and can be discarded.
In constructing our RG procedure we have first shown, using the
very special properties of logarithmic interactions, that it is enough 
to follow, in addition to the two parameters $K$ and $\sigma$, the distribution 
of only one or two local (i.e uncorrelated) random variables $P_l(z_+ , z_-)$.
At this stage it is clear that we still keep too much information.
Indeed we have found that the precise form the non linear RG equation obeyed by
$P_l(z_+ , z_-)$ as well as the detailed shape of this distribution are largely 
cutoff dependent. However, this complicated looking
RG equation can be generally recast, 
up to irrelevant terms, as a well known 
non linear front propagation of the KPP type. Using 
the known remarkable universality property of this type 
of non linear equations, we found that all the information needed
to describe the universal properties of the XY phase and of the transition 
is indeed independent 
of the detailed shape of the fugacity distribution, or of the
precise form of its RG equation. Only its tails,
and the finite size corrections to the front velocity seem to be needed
to determine the physical quantities and the critical behaviour.

Since we are following a full distribution of local disorder,
the present method could be termed a functional RG. We can indeed
draw some parallel 
between this functional RG procedure for the disordered CG and 
two other known examples where the universal behaviour of a disordered
finite-range system\footnote{and statistically translational invariant
by opposition to e.g. Migdal Kadanoff RG, or Cayley tree recursion
relations, which are exact only 
on very special hierachical lattices.} is extracted 
from a functional RG equation. The first
one is the asymptotically exact real space RG in $d=1$
\cite{fisher95,fisher98} well suited to ``infinite disorder''
fixed points. As emphasized in \cite{carpentier99ter} 
there are indeed similarities when treating the single vortex,
$d=1$ version of the present model. The method of 
\cite{fisher98} can be applied for the Sinai potential
which has correlations growing as a power of distance.
For the present case of weaker, logarithmic correlations,
these methods can also be applied in principle (at least around 
zero temperature where disorder is still very broad) 
but become very hard to work out analytically.
The other example of functional RG appears in a dimensional expansion 
for the problem of an interface in a random potential \cite{fisher86},
which has infinite number of marginal directions at the upper
critical dimension $d=d_c$. An important question is whether
in the problem studied in this paper
there is also an infinite number of marginal
directions. 
As discussed above, the results extracted from the
KPP equation seem to suggest that a smaller amount of information 
than the exact full distribution may be needed here.
Thus one can speculate that the critical theory
studied here could be equivalently formulated as a 
more conventional field theory, yet to be identified,
with a small number of marginal or relevant operators.
In any case the RG method developed here should provide
a physically transparent guide to study the system.
Given the wide applications of Coulomb gases in two dimensions
it is likely that other two dimensional disordered models
can be studied using methods similar to the one introduced here.

Finally, another outcome of the present work has been to unveil some
interesting connections between the renormalization of a
disordered system and the universal features of the propagation
of invading fronts in non linear systems. The existence
of intriguing relations between freezing transitions
and velocity selection in non linear fronts was noticed 
previously by Derrida and Spohn \cite{derrida88} in their study of the DPCT.
Here, we found an even deeper and puzzling connection, 
betwen the ``universality'' in these selection mecanisms 
and the universality in the critical phenomena 
captured by the renormalization group\footnote{the dimensions of operators, in the
RG sense, being directly related to the selected velocity of 
non linear fronts}. Since attempts have been made to construct renormalization
methods in order to extract the universal features 
of such non linear equations \cite{goldenfeld94}, 
this suggests that a common framework
could be developed in connexion with two dimensional 
disordered models.

{\bf Acknowledgements}

It is a pleasure to thank B. Derrida, V. Hakim and W. Van Saarloos
for useful discussions.

\newpage

\appendix

\section{Disordered lattice Coulomb Gas}\label{part:CG}

\begin{figure}
\centerline{\fig{6cm}{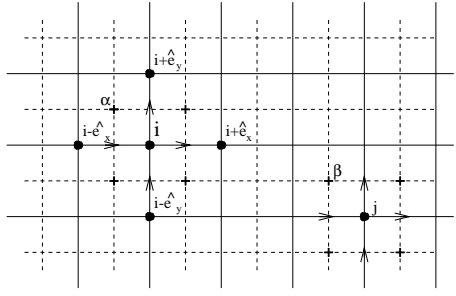}}
\caption{Representation of the two networks. The plain and dashed 
lines correspond respectively to the direct and dual networks.}
\label{reseau}
\end{figure}

 In this appendix we derive, by exact transformations, a lattice disordered 
Coulomb gas formulation of the Villain form of (\ref{xy}), 
extending to this 
disordered case the approach of Kadanoff \cite{kadanoff78}. An alternative
route to  \cite{kadanoff78} in the pure case in the method used in 
 \cite{jose77}.
 For simplicity we firt turn to the Villain version of (\ref{xy}) 
before making the duality transformations, although the inverse procedure 
could be used (see \cite{jose77}). This Villain model corresponding to
(\ref{square}) is  \cite{villain75} : 

\begin{eqnarray}
Z_{\text{Villain}}=\sum_{[l_{i,x},l_{i,y}]\in \mathbb{Z}} \prod_{i}\int_{-\pi}^{+\pi}
\frac{d \theta_{i}}{2 \pi} \exp \left[ - \frac{K}{2\pi  } \sum_{\langle i,j
\rangle}(\theta_{i}-\theta_{j}-2\pi l_{ij}-A_{ij} )^{2}\right]
\end{eqnarray}
 where $K=\beta J$ and we used 
the notation $l_{i,x/y}=l_{i,i+\hat{e}_{x/y}}$ (see
fig. \ref{reseau}). This Villain partition function indeed corresponds to a 
$\mathbb{Z}$-gauge theory, since the action is invariant under 
$
\theta_i  \rightarrow  \theta'_{i}= \theta_i+2\pi q_i ,
l_{ij}    \rightarrow  l_{ij}'= l_{ij} + q_i -q_j $ with  
$q_i \in \mathbb{Z}$.
  Choosing the {\it gauge field} $q_i$ such that for each horizontal  
link $l_{i,x}= q_{i+\hat{e}_{x}}-q_i$ ({\it i.e} $l'_{i,x}=0$)
(free boundary conditions along $x$)
we obtain a partition function of a gaussian field 
$\phi_{i}=\theta_{i}/2\pi$  
($\sum_{l_{i,x}}\int_{-\pi}^{+\pi}
\frac{d \theta_{i}}{2 \pi}=\int_{-\infty }^{\infty}d\phi_{i}$):

\begin{eqnarray*}
&& Z_{\text{Villain}}=\sum_{[l_{i,y}]} \prod_{i}\int_{-\infty}^{+\infty}
d \phi_{i} e^{- 2\pi\beta J  \sum_{i}
\left(
\left( \nabla_{i,y}\phi +  l_{i,y} -\frac{1}{2\pi} {\bf A}_{i,y} \right)^{2} 
+\left( \nabla_{i,x}\phi -\frac{1}{2\pi} {\bf A}_{i,x} \right)^{2} \right) }
\end{eqnarray*}
with the notation for the discrete gradient
$\nabla_{i,y}\phi=\phi_{i+\hat{e}_{y}}-\phi_{i}$. Integrating over this
gaussian field yields the Coulomb gas action 

\begin{eqnarray}
Z_{\text{Villain}}&=&\sum_{[l_{i,y}]} 
e^{
- \beta J \sum_{i,j}
\left[(l_{i-\hat{e}_{y},y}-l_{i,y})-
\frac{1}{2\pi}{\bf \nabla}_i.{\bf A}\right] 
 G_{ij}
\left[(l_{j-\hat{e}_{y},y}-l_{j,y})- 
\frac{1}{2\pi}{\bf \nabla}_j.{\bf A}\right] 
- \sum_i  2 \pi K~ l_{i,y}^2}\nonumber \\
&=& \label{latticeCG}
\sum_{n_{\alpha}}
e^{
- \beta J \sum_{\alpha,\beta} (n_{\alpha}-q_{\alpha}) \Gamma_{\alpha \beta} 
(n_{\beta}-q_{\beta})
}
\end{eqnarray}
 where we have discarded the $A/A$ self interaction and we 
defined the dual lattice charges $n_{\alpha}$ with
$\alpha=i+ (-\frac{1}{2},\frac{1}{2})$ (see fig. \ref{reseau}) : 
\[
n_{\alpha}=\hat{e}_{z}.({\bf \nabla}_{i}\times l)=
 l_{i,y}-l_{i-\hat{e}_{x},y} 
\]
 and the quenched random dipoles 
\[
q_{\alpha}=\frac{1}{2\pi}{\bf \nabla}_i.{\bf A}=
\frac{1}{2\pi}
( A_{i}^y-A_{i-\hat{e}_{y}}^y+A_{i}^x-A_{i-\hat{e}_{x}}^x)
\]

 Note that the neutrality of disorder charges $\sum_{\alpha}q_{\alpha}=0$
follows directly from the definition of $q$
(e.g. taking $A$ to vanish at the boundary) and implies neutrality for the
integer charges $\sum_{\alpha} n_{\alpha}=0$
(from the divergence of $\Gamma(0)$). 
The {\it Coulomb potential} $\Gamma$ is defined on the lattice by 
\begin{equation}
\triangle_i \Gamma = 
\sum_{j=i\pm \hat{e}_x,i\pm \hat{e}_y }\Gamma_{j}
-4 \Gamma_i =  
 \nabla_x \nabla_x  \Gamma_{i-\hat{e}_x} + \nabla_y \nabla_y
\Gamma_{i-\hat{e}_y} = 2 \pi  ~\delta_{i,0}
\end{equation}
 which gives the expression 
\begin{eqnarray}
\Gamma_{i-j} & = & -  2 \pi \int \frac{ d^2 {\bf q}}{ (2 \pi)^2} 
\frac{ e^{i {\bf q}.(i-j)} }{ 4 -2 \cos(q_x) -2 \cos(q_y)}\\ 
 &=&  \Gamma(0) + \ln |i-j| + \ln (2\sqrt{2}e^{\gamma}) 
+ O\left(\frac{1}{|i-j|}\right)
\end{eqnarray}
 where the $\gamma$ is the Euler constant. 
 Using the neutrality of the charges and the definition
$G_{\alpha-\beta}=\Gamma_{0}-\Gamma_{\alpha-\beta}$, and  
 the definition of the disorder
potential $V_{\alpha}=-2 J \sum_{\beta}G_{\alpha\beta}q_{\beta}$, the 
lattice disordered CG action reads 
\begin{equation}
A_{CG}=-\beta H_{CG} = \beta J 
\sum_{\alpha\beta} n_{\alpha}G_{\alpha\beta}n_{\beta}
+ \beta \sum_{\alpha}n_{\alpha} V_{\alpha}
\end{equation}

\section{Smooth cutoff procedure}  \label{part:cutoff}

The Coulomb gas, or its equivalent Sine Gordon version,
can also be renormalized using smooth cutoff
procedures. We will not give details here but refer the reader
to \cite{knops80} (see also \cite{carpentier98}).
Let us simply point out how, in that case, the full 
disorder $V_{\bf r}$ can indeed be decomposed, as
in (\ref{decomposition}), into two bona fide disorders.

In the case of a soft cutoff, the continuum approximation of the
lattice Coulomb interaction reads (instead of (\ref{approx})):

\begin{eqnarray}
G^{(a)}_{{\bf r}} = 2 \pi \int \frac{d^2 {\bf q}}{(2 \pi)^2} 
\frac{\phi(a q)}{q^2} (1 - \cos({\bf q} \cdot {\bf r}))
\end{eqnarray}
where $\phi(0)=1$ and we will choose $\phi(x)$ a positive monotonously 
decreasing function of $x$. One has the asymptotic large $r$ behaviour
$G^{(a)}_{{\bf r}} \sim \ln r + C(\phi) + O(1/r)$ (see e.g. \cite{carpentier98}).
One also has:

\begin{eqnarray}
G^{(a)}_{{\bf r}} = G^{(\tilde{a})}_{{\bf r}}  -
2 \pi dl \int \frac{\d^2 {\bf q}}{(2 \pi)^2} 
\frac{a q \phi'(a q)}{q^2} (1 - \cos({\bf q} \cdot {\bf r})) + O(dl^2)
\end{eqnarray}
where $\tilde{a}=a e^{dl}$. Thus
one can write the cutoff dependent decomposition 
$V_{{\bf r}} = V^{>,(a)}_{{\bf r}}  + v^{(a)}_{{\bf r}}$
and one gets upon increase of cutoff:

\begin{eqnarray}
&& \overline{(V^{>,(\tilde{a})}_{\bf r} - V^{>,(\tilde{a})}_{{\bf r}'})^2} 
=4 \sigma  J^2 G^{(\tilde{a})}_{{\bf r}-{\bf r}' } 
\end{eqnarray}
and $v^{(\tilde{a})}_{{\bf r}}= v^{(a)}_{{\bf r}} + dv_{{\bf r}}$ 
with:

\begin{eqnarray}
\overline{dv_{{\bf r}} dv_{{\bf r}'}} = 4 \pi \sigma  J^2  dl
\int \frac{\d^2 {\bf q}}{(2 \pi)^2} 
\frac{- a q \phi'(a q)}{q^2} e^{i {\bf q} \cdot {\bf r} } 
\end{eqnarray}

Thus both $V^{>}_{{\bf r}}$ and $dv_{{\bf r}}$ are well defined physical gaussian 
disorders since their correlators have positive Fourier transform. In addition
$dv_{{\bf r}}$ is short range correlated. For instance, taking $\phi(x)=e^{-x^2/2}$
one finds:

\begin{eqnarray}
\overline{dv_{{\bf r}} dv_{{\bf r}'}} = \frac{2 \sigma J^2}{(2 \pi)^2} dl ~
e^{- ({\bf r} - {\bf r}')^2/(2 a^2)}
\end{eqnarray}

\section{Renormalisation of the replicated Coulomb gas}\label{part:rgreplica}

In this appendix, for completeness, we explicitly renormalize 
the vector Coulomb gas defined by (\ref{expansion}). 
This amounts to extend to $m$-component vector charges the renormalisation 
of scalar CG \cite{nienhuis87}. The partition function reads:

\begin{eqnarray}\label{app-Zrep}
Z_{\text{latt}}= \sum_{N\geq 2} \sum_{\{{\bf n}_{1}\dots {\bf n}_{N} \}}' 
\prod_{i=1}^{N} \int_{h.c} \frac{d^{2}{\bf r}_{i}}{a_{o}^2} Y[{\bf n}_{i}]
e^{\mathcal{A}_{a_{o}} [{\bf n}_{1}\dots {\bf n}_{N}]}
\end{eqnarray}
where as usual the primed sum is over all distinct neutral charge 
configurations and the notation h.c. stand for all the hard core
constraints $|{\bf r}_i - {\bf r}_j| \geq a_0$ for all pairs $i,j$,
implicit in the following.
The action $\mathcal{A}_{a_{o}}$ is defined by  
\begin{equation}\label{app-A}
\mathcal{A}_{a_{o}}[{\bf n}_{1}\dots {\bf n}_{N}]=\frac{1}{2}
\sum_{i\neq j} 2K_{ab}~
n_{i}^{a}\ln \left(\frac{ |{\bf r}_{i}-{\bf r}_{j}|}{a_{o}} \right) n_{j}^{b}
\end{equation}

A common way to renormalise usual Coulomb gas consists in 
coarse graining the partition function, leaving the expansion in
number of fugacity (here $Y$) unchanged. This amounts to define scale
dependent replica stiffness $K_{ab} (l)$ and fugacities $Y_{l}[{\bf n}] $. We
will follow this scheme in this appendix. Note that another equivalent way
would be to renormalise the correlation function directly, following {\it e.g}
 \cite{carpentier97}. As in the scalar case \cite{kosterlitz74,nienhuis87},
renormalisation of the generalised vector Coulomb
gas \cite{bulgadaev81,boyanovsky89} proceeds in the same three steps :
rescaling, fusion and annihilation (screening) of small dipoles. We now turn
to the description of these three contributions : first we increase the
hard-core cut-off $a_{o}\rightarrow \tilde{a_{o}}=a_{o}e^{dl}$ 
with $dl\ll 1$.

\subsection{Rescaling}
 This increase of cut-off produces a naive rescaling : 
\begin{equation}\label{app:rescaling}
\prod_{i=1}^{N} \int \frac{d^{2}{\bf r}_{i}}{a_{o}^2} Y[{\bf n}_{i}]
e^{\mathcal{A}_{a_{o}}[{\bf n}_{1}\dots {\bf n}_{N}]} =
 \prod_{i=1}^{N}\int \frac{d^{2}{\bf r}_{i}}{\tilde{a}_{o}^2} Y[{\bf n}_{i}]
e^{\mathcal{A}_{\tilde{a}_{o}}[{\bf n}_{1}\dots {\bf n}_{N}]}
e^{dl (2-K_{ab}n_{i}^{a} n_{i}^{b})}
\end{equation}
 where we used the neutrality $\sum_{i} n_{i}^{a}=0$ to express the correction
coming from the action. We can absorb the extra factor in
(\ref{app:rescaling}) to all order in $Y$ by the change of fugacities
corresponding to the equation 
\begin{equation}
\partial_{l} Y[{\bf n}]= (2-K_{ab}n^{a}n^{b}) Y[{\bf n}]
\end{equation}

\subsection{Annihilation (screening) and Fusion of charges}
 Upon the increase of cut-off, two charges ${\bf n}_{p}$ and ${\bf n}_{q}$
have to be coarse grained if they are located in 
${\bf r}_{p}$ and ${\bf r}_{q}$
 with $a_{o}\leq |{\bf r}_{p}-{\bf r}_{q}|\leq a_{o} e^{dl}$. 
Within the small charge
density hypothesis, we consider only one such pair. For a dipole, these two
charges have to be integrated out at scale $\tilde{a_{o}}$ : 
this corresponds to the annihilation, while for a non neutral pair, 
the coarse grained charge is simply the sum of the two charges at 
scale $a_{o}$ (fusion). 
In both cases the partition function splits into $Z=Z' + Z_{p,q}$ where 
$Z_{p,q} \sim \mathcal{O}(dl)$ involves configurations with one pair of charges
${\bf n}_{p}, {\bf n}_{q}$ distant of less than
$\tilde{a}_0$ while $Z'$ doesn't. 
$Z_{p,q}$ can be written as 
\begin{multline}
Z_{p,q}=
\sum_{\{ {\bf n}_{1}\dots {\bf n}_{N} \}}' 
\prod_{i=1,..N, i \neq p,q} \int 
\frac{d^{2}{\bf r}_{i}}{\tilde{a}_{o}^2} Y[{\bf n}_{i}]  \\
\times \sum_{{\bf n}_{p},{\bf n}_{q}}
\int_{a_{o}\leq |{\bf r}_{p}-{\bf r}_{q}|\leq  a_{o}e^{dl}}  
\frac{d^{2}{\bf r}_{p}}{\tilde{a}_{o}^2}
\frac{d^{2}{\bf r}_{q}}{\tilde{a}_{o}^2} Y[{\bf n}_{p}]Y[{\bf n}_{q}]
e^{\mathcal{A}_{p,q}[{\bf n}_{1}\dots {\bf n}_{N}]}
\end{multline}
 where, with the notation 
$\alpha_{i,j}=\sum_{a,b} n_{i}^{a} 2 K^{ab} n_{j}^{b}$, the action reads 
\begin{equation}
e^{\mathcal{A}_{p,q}}=
\left( \frac{|{\bf r}_{i}-{\bf r}_{p}|}{a_{o}} \right)^{\alpha_{pq}}
\prod_{i \neq p,q}
\left( \frac{|{\bf r}_{i}-{\bf r}_{p}|}{a_{o}} \right)^{\alpha_{ip}}
\left( \frac{|{\bf r}_{i}-{\bf r}_{q}|}{a_{o}} \right)^{\alpha_{iq}}
\prod_{i < j \neq p,q}
\left( \frac{|{\bf r}_{i}-{\bf r}_{j}|}{a_{o}} \right)^{\alpha_{ij}}
\label{expans}
\end{equation}
 We must now distinguish between a neutral pair and a non neutral one.

\subsubsection{Fusion}
In this case the small pair of charges $({\bf n}_{p},{\bf n}_{q})$
at scale $a_{o}$ gives an effective charge 
${\bf n}_{p}+{\bf n}_{q}$ at scale $\tilde{a_{o}}$, located in 
${\bf R}=({\bf r}_{p}+{\bf r}_{q})/2$. Thus we must integrate over 
the relative 
position of the two charges ${\bf \rho}={\bf r}_{p}-{\bf r}_{q}$ when 
coarse graining the Coulomb Gas. 
 To obtain the corresponding correction to $Z'$ (of order 
one in $dl$), it is enough to expand (\ref{expans}) to order 0 in 
${\bf \rho}$. This expansion reads  
\begin{multline*}
\sum_{\{ {\bf n}_{1}\dots {\bf n}_{N} \}}' 
\prod_{i=1,..N, i \neq p,q} \int 
\frac{d^{2}{\bf r}_{i}}{a_{o}^2} Y[{\bf n}_{i}]
\sum_{{\bf n}_{p},{\bf n}_{q}}\int 
\frac{d^{2}{\bf R}}{a_{o}^2}~ Y[{\bf n}_{p}]Y[{\bf n}_{q}]
\left(\int_{a_{o}\leq \rho\leq  a_{o} e^{dl}}  
\frac{d^{2}{\bf \rho}}{a_{o}^2} \right) \\
\times 
\prod_{i \neq p,q}
\left( \frac{|{\bf r}_{i}-{\bf R}|}{a_{o}}\right)^{\alpha_{ip}+\alpha_{iq}}
\prod_{i < j  \neq p,q}
\left( \frac{|{\bf r}_{i}-{\bf r}_{j}|}{a_{o}}\right)^{\alpha_{ij}}
\end{multline*}
 Using $\alpha_{i,p}+\alpha_{i,q}=\alpha_{i,p+q}$, we can rewrite this
correction to $Z'$ as a single contribution to the fugacity 
$Y[{\bf n}_{p}+{\bf n}_{q}]$
of the {\it non zero} charge ${\bf n}_{p}+{\bf n}_{q}$ :
\begin{equation}\label{RG-fusion}
\partial_{l}Y[{\bf n}_{p}+ {\bf n}_{q}]=
2 \pi \sum_{{\bf n}_{p},{\bf n}_{q}} Y[{\bf n}_{p}]Y[{\bf n}_{q}]
\end{equation}
Note that in this expression ${\bf n}_{p}$ and ${\bf n}_{q}$ are
distinguishable charges : this explains the factor 2 between 
(\ref{RG-fusion}) and (\ref{rgrep2}). 

\subsubsection{Annihilation}
 In this case the small dipole of size $\rho$ is integrated out at scale
$\tilde{a_{o}}$. When coarse graining, we sum over both ${\bf \rho}$ and 
${\bf R}$, yielding a factor $a_{o}^{-4}$ from the integration measure. Thus 
diverging contributions correspond to the expansion of (\ref{expans})   
to order 2 in ${\bf \rho}$. Using $\alpha_{i,p}=-\alpha_{i,q}$, this
expansion of $Z_{p,q}$ can be expressed as 
\begin{multline*}
\sum_{\{ {\bf n}_{1}\dots {\bf n}_{N} \}}' 
\prod_{i=1,..N, i \neq p, q}  \int 
\frac{d^{2}{\bf r}_{i}}{a_{o}^2} Y[{\bf n}_{i}]
\prod_{i < j \neq p, q}
\left( \frac{|{\bf r}_{i}-{\bf r}_{j}|}{a_{o}} \right)^{\alpha_{ij}}
\sum_{{\bf n}_{p}}Y[{\bf n}_{p}]Y[-{\bf n}_{p}]\\
\times 
\int_{a_{o}\leq \rho\leq  a_{o} e^{dl}}  \frac{d^{2}{\bf \rho}}{a_{o}^2}
\int  \frac{d^{2}{\bf R} }{a_{o}^2}
\left(
1+\sum_{i, j \neq p,q}^{N}
\alpha_{i,p} \alpha_{j,p} \frac{1}{4}
{\bf \rho}.{\bf \nabla}\ln ({\bf r}_{i}-{\bf R})
{\bf \rho}.{\bf \nabla}\ln ({\bf r}_{j}-{\bf R})
 \right)
\end{multline*}
 Performing the integral and reexponentiating the last term using the
neutrality of the configuration $\{{\bf n}_{1},\dots {\bf n}_{N} \}$, we
 get the correction to $Z'$ coming from $Z_{p,-p}$:
\begin{multline*}
\mathcal{C}
\sum_{\{ {\bf n}_{1}\dots {\bf n}_{N} \}}' 
\prod_{i=1,..N, i \neq p, q} \int 
\frac{d^{2}{\bf r}_{i}}{a_{o}^2} Y[{\bf n}_{i}] \\
\times
\prod_{i < j \neq p, q}
\left( \frac{|{\bf r}_{i}-{\bf r}_{j}|}{a_{o}}\right)^{\alpha_{ij}-\pi^{2}dl 
\sum_{{\bf n}_{p}} \alpha_{i,p}\alpha_{j,p}~Y[n_{p}]Y[-n_{p}]
}
\end{multline*}
 The constant $\mathcal{C}$ corrects the intensive free energy while the 
second term can be absorbed in a correction to the coupling constant 
\[
\partial_l K_{ab}^{-1}  = 2 \pi^2  \sum_{[n]} n^c n^d Y[{\bf n}] Y[- {\bf n}]
\]

The three above contributions can be summarized into the set of 
RG equations given in the text (\ref{rgrep}) valid for 
{\it all non zero vector charge}, and for all $m$.   
The coefficient $c_{1}$ and $c_{2}$ depends on the IR regularisation.
For our hard cut-off, we find within our procedure that $c_{1}=2\pi^{2}$ and 
$c_{2}=\pi$. Note that the ratio $c_{1}/c_{2}^{2}$ is
independent of a uniform rescaling of the fugacities and
is known, in the case of single component charges 
to be universal at a transition \cite{carpentier97}.  
For a discussion of regularisation of replicated Coulomb Gas, see discussion 
in appendix \ref{app-cutoff}.

\section{$m \to 0$ limit of the replica RG equations}
\label{part:replimit}

In this Appendix we explicitly perfom the $m\to 0$ limit of the 
whole set of RG equations (\ref{rgrep1},\ref{rgrep2}) with the
restriction that the vector charges have only $0,\pm1$ components.
This limit  is taken using the parametrisation (\ref{eq:ydef}) of the 
fugacities $Y[n]$ in terms of the function $\Phi (z_{+},z_{-})$. 
In the three different terms of  (\ref{rgrep1},\ref{rgrep2}),
corresponding to rescaling, annihilation and
fusion contributions ( see appendix \ref{part:rgreplica}), we first 
modify sums to include 
fugacities for null charge, translate the expression in terms of $\Phi$ and
naturally take the $m \to 0$ limit. We will use the notation 
$\langle A \rangle_{\Phi}= 
\int_{z_{+},z_{-}}A~ \Phi (z_{+},z_{-})$.

\subsection{Rescaling}

 The term corresponding to rescaling in (\ref{rgrep2}) is 
\begin{eqnarray} \label{dim-y}
\partial_l Y[{\bf n}] = (2 - n_a K^{ab} n_b ) Y[{\bf n}] 
\end{eqnarray} 
 which holds for any vector charge ${\bf n}$. The second term can be expressed
in terms of $\Phi$ using (\ref{eq:ydef}) and 
$K^{ab}=K\delta^{ab}-\sigma K^{2}$ with $K=\beta J$.
However the expression is much simpler
if one uses, instead of (\ref{eq:ydef}), the equivalent parametrisation 
$Y[{\bf n}]=\int_{u,v} e^{\beta pu}e^{\beta qv} \tilde{\Phi} (u,v)$ 
with $p=n_{+}+n_{-}$, 
$q=n_{+}-n_{-}$ and 
$z_{\pm}=e^{\beta (u\pm v)}$.
With $\tilde{\Phi}$, this yields

\begin{eqnarray}
\sum_{a,b}n_a K^{ab} n_b ~ Y[{\bf n}]&=&
(\beta J p - \sigma \beta^2 J^2 q^2 ) \int_{u,v} \tilde{\Phi}(u,v)  
e^{\beta pu} e^{\beta qv}  \\
&=& \int_{u,v} \tilde{\Phi}(u,v) \left( J \frac{\partial}{\partial u} 
        - \sigma  J^2 \frac{\partial^2}{\partial v^2}
         \right) e^{\beta pu} e^{\beta qv}  \\ 
&=& - \int_{u,v} e^{\beta pu} e^{\beta qv} 
        \left(  J \frac{\partial}{\partial u} 
        + \sigma J^2  \frac{\partial^2 }{\partial v^2}
         \right)\tilde{\Phi}(u,v) 
\label{lim-repar}
\end{eqnarray}
As this is true for any vector charge $n$ satisfying $n^a=0,\pm1$,
{\it i.e} for any $p,q$ or
equivalently $n_{+},n_{-}$, to satisfy (\ref{dim-y})
we can search for a function $\tilde{\Phi}$ such that:
$$ \partial_l \tilde{\Phi}(u,v) = \left(2+  J \frac{\partial}{\partial u} 
        + \sigma  J^2 \frac{\partial^2}{\partial v^2}
         \right)\tilde{\Phi}(u,v)$$
In terms of $\Phi (z_{+},z_{-})$, this corresponds to the equation:
\begin{eqnarray}
&& \partial_{l} \Phi (z_{+},z_{-}) = (2+\mathcal{O})  \Phi (z_{+},z_{-})\\ 
&& \equiv
\left(2+\beta J (2+z_{+}\partial_{z_{+}}+z_{-}\partial_{z_{-}})
+\sigma \beta^{2} J^{2} (z_{+}\partial_{z_{+}}-z_{-}\partial_{z_{-}})^{2}
 \right)
\Phi (z_{+},z_{-}) \nonumber
\label{def-O}
\end{eqnarray}
where we have used that $\tilde{\Phi}(u,v)=2 z_+ z_- \Phi(z_+,z_-)$.
Note that for this process, the integral of $\Phi$, $\mathcal{N} = \int \Phi$
satisfies simply $\partial_l \mathcal{N}  = 2 \mathcal{N} $.

\subsection{Fusion}

In (\ref{rgrep2}), the fusion term  of two charges 
${\bf n}'+ {\bf n}'' = {\bf n}$ is
restricted to ${\bf n}' \neq 0$ and ${\bf n}''\neq 0$.  Furthermore 
since ${\bf n} = 0 $ corresponds to the annihilation which is
treated separately in (\ref{rgrep1}) (see below) and must not be counted twice,
the equation (\ref{rgrep2}) can be used only for ${\bf n} \neq 0 $.
It is convenient to extend the equation (\ref{rgrep2}) to include
$Y[{\bf n} = 0] =\int \Phi = \mathcal{N}$ 
for which the fusion contribution should be
absent. Thus the extended equation corresponding to fusion
reads:

\begin{eqnarray}
\label{eqphi1}
&& \partial_{l} Y[{\bf n}]=  c_{2}  \left( \sum_{ 
\genfrac{}{}{0pt}{}{{\bf n}'+ {\bf n}'' = {\bf n}}{{\bf n}',{\bf n}''\neq 0} } 
Y[{\bf n}'] Y[{\bf n}'']  
- \delta_{{\bf n},0} \sum_{{\bf n}' \neq 0} Y[{\bf n}'] Y[-{\bf n}'] \right)  \\
&& =
c_{2}  \left( \sum_{ {\bf n}'+ {\bf n}'' = {\bf n} } Y[{\bf n}'] Y[{\bf n}'']
- 2 Y[{\bf 0}] Y[{\bf n}] -  \delta_{{\bf n},0} \left( \sum_{{\bf n}'} 
Y[{\bf n}'] Y[-{\bf n}'] - 2 Y[0]^2 \right) \right) \nonumber
\end{eqnarray}
where in the second equality we allow for ${\bf n}'=0$ or ${\bf n}''=0$.
Turning to the representation in terms of $n_{+},n_{-}$ we have:

\begin{eqnarray*}
&&\sum_{ {\bf n}'+ {\bf n}'' = {\bf n} } Y[{\bf n}'] Y[{\bf n}'']\\ 
&& = \left<   \prod_a  \Bigl[ 
 ( 1+ z'_-  z''_+ + z'_+  z''_- ) \delta_{n_a,0} 
+  ( z'_+ + z''_+) \delta_{n_a,+1} 
+ ( z'_- + z''_-)  \delta_{n_a,-1}   \Bigr]  \right>_{\Phi,\Phi}
\end{eqnarray*}

Using permutation symmetry we find that (\ref{eqphi1}) can be written as:

\begin{eqnarray*}
&& \partial_{l}Y[{\bf n}] = \\
&& c_2
\left<  (1 + z'_-  z''_+ + z'_+  z''_- )^{m} 
\left(\frac{z'_+ + z''_+}{1 + z'_-  z''_+ + z'_+  z''_-}\right)^{n_{+}}
\left(\frac{z'_- + z''_-}{1 + z'_-  z''_+ + z'_+  z''_-}\right)^{n_{-}} 
 \right>_{\Phi \Phi}  \\
&& - 2 c_2 \mathcal{N} \left<  z_+^{n_+} z_-^{n_-} \right>_{\Phi}
- c_2 \delta_{n_+,0} \delta_{n_-,0} 
\left( \left<  (1 + z'_-  z''_+ + z'_+  z''_- )^{m} -1 
\right>_{\Phi \Phi} - \mathcal{N}^2 \right)
\end{eqnarray*}

The following choice for $\partial_{l} \Phi (z_{+},z_{-})$ allows to satisfy the
above equation for all ${\bf n}$ (up to now $m$ is still arbitrary):

\begin{eqnarray} \label{loosem}
&& \partial_l \Phi(z_+,z_-) = \\  \nonumber
&&c_2 \left< (1 + z'_-  z''_+ + z'_+  z''_- )^{m} 
\delta\left( z_+ - \frac{z'_+ + z''_+}{1 + z'_-  z''_+ + z'_+  z''_-} \right)
\delta\left( z_- - \frac{z'_- + z''_-}{1 + z'_-  z''_+ + z'_+  z''_-} \right)
\right>_{\Phi \Phi}  \\ \nonumber
&&
- 2 c_2 \mathcal{N} \Phi(z_+,z_-) 
- c_2 \delta(z_+) \delta(z_-) 
\left( \left<  (1 + z'_-  z''_+ + z'_+  z''_- )^{m} -1 \right>_{\Phi \Phi} - \mathcal{N}^2 \right)
\end{eqnarray}

We can now take the limit $m \to 0$ {\it explicitly} on this equation,
which yields:

\begin{eqnarray} 
 \partial_l \Phi(z_+,z_-) &=&  
c_2 \left<
\delta\left( z_+ - \frac{z'_+ + z''_+}{1 + z'_-  z''_+ + z'_+  z''_-} \right)
\delta\left( z_- - \frac{z'_- + z''_-}{1 + z'_-  z''_+ + z'_+  z''_-} \right)
\right>_{\Phi,\Phi}  \nonumber \\
&&
- 2 c_2 \mathcal{N} \Phi(z_+,z_-) 
+ c_2 \delta(z_+) \delta(z_-) \mathcal{N}^2 \label{loose2}
\end{eqnarray}

\subsection{Annihilation}

 The screening equation (\ref{rgrep1}) for the replica coupling 
constant reads 
\begin{equation}
\partial_{l} (K^{-1})^{ab} = 
c_{1} \sum_{{\bf n}\neq 0} n^a n^b  Y[{\bf n}] Y[-{\bf n}]
\end{equation}
 which corresponds to the RG equations for $\sigma$ and $J$ 
\begin{eqnarray}
&& \partial_{l} (T J^{-1} + \sigma) = 
 \frac{c_{1}}{m}\sum_{a=1}^{m}
 \sum_{{\bf n}\neq 0} n^a n^a  Y[{\bf n}] Y[- {\bf n}] 
\label{screen-rep1}\\
&& \partial_{l} \sigma = 
 \frac{c_{1}}{m (m-1)}\sum_{a\neq b=1}^{m}
\sum_{{\bf n}\neq 0} n^{a} n^{b\neq a}  Y[ {\bf n}] Y[- {\bf n}] 
\label{screen-rep2}
\end{eqnarray} 
 The sums over the charge ${\bf n}$ in 
(\ref{screen-rep1},\ref{screen-rep2}) can be easily expressed in terms of
$\Phi$ using the variable $n_{+},n_{-}$. The sum in (\ref{screen-rep2}) thus
reads 
\begin{eqnarray*}
&& \sum_{a\neq b=1}^{m} \sum_{{\bf n}\neq 0} n^{a} n^{b\neq a}  
Y[{\bf n}] Y[-{\bf n}]\\
&=&
\sum_{0\leq n_{+}+n_{-}\leq m} C_{m}^{n_{+},n_{-}}
\left[(n_{+}-n_{-})^{2}- (n_{+}+n_{-}) \right] 
\left< (z_{+}'z_{-}'')^{n_{+}}(z_{-}'z_{+}'')^{n_{-}} ) 
\right>_{\Phi (z')\Phi (z'')} \\
&=&
\left< \left[
(z'_{+}\partial_{z'_{+}}-z'_{-}\partial_{z'_{-}})^{2}-
(z'_{+}\partial_{z'_{+}}+z'_{-}\partial_{z'_{-}}) \right] \right.\\
&&~~~~~~~~~~ \times \left.
\sum_{0\leq n_{+}+n_{-}\leq m} C_{m}^{n_{+},n_{-}}
 (z_{+}'z_{-}'')^{n_{+}}(z_{-}'z_{+}'')^{n_{-}} ) 
\right>_{\Phi (z')\Phi (z'')}\\
&=&
\left< \left[
(z'_{+}\partial_{z'_{+}}-z'_{-}\partial_{z'_{-}})^{2}-
(z'_{+}\partial_{z'_{+}}+z'_{-}\partial_{z'_{-}}) \right]
 (1+z_{+}'z_{-}''+z_{-}'z_{+}'')^{m}
\right>_{\Phi (z')\Phi (z'')} \\
&=& m (m-1)
\left< (z_{+}'z_{-}''-z_{-}'z_{+}'')^{2} 
(1+z_{+}'z_{-}''+z_{-}'z_{+}'')^{m-2}
\right>_{\Phi (z')\Phi (z'')}
\end{eqnarray*}
where $ C_{m}^{n_{+},n_{-}} = m!/ n_+! n_-! (m-n_+-n_-)! $. Taking the $m \to 0$ 
limit of the last equation yields directly with (\ref{screen-rep2})
\begin{equation}
\partial_l \sigma = c_{1}
\left< \left(
\frac{z_{+}'z_{-}''-z_{-}'z_{+}''}{1+z_{+}'z_{-}''+z_{-}'z_{+}''} 
\right)^2 \right>_{\Phi (z')\Phi (z'')}
\end{equation}
 By the same method we get for the scaling equation of $K^{-1}$ :
\begin{equation}
\partial_l (T J^{-1}) = c_{1}
\left< 
\frac{z_{+}'z_{-}''+z_{-}'z_{+}''+4z_{+}'z_{-}''z_{-}'z_{+}''}
{(1+z_{+}'z_{-}''+z_{-}'z_{+}'')^{2}} 
\right>_{\Phi (z')\Phi (z'')}
\end{equation}

\subsection{Final set of RG equations}

 Putting all these contributions together, we obtain 
the scaling equations for the coupling constant $J$, the correlated disorder
strength  $\sigma$ and the distribution of local disorder 
$\Phi (z_{+},z_{-})$ :
\begin{subequations}
\begin{eqnarray}
\partial \sigma &&= c_{1}
\left< \left(
\frac{z_{+}'z_{-}''-z_{-}'z_{+}''}{1+z_{+}'z_{-}''+z_{-}'z_{+}''} 
\right)^2 \right>_{\Phi (z')\Phi (z'')}\\
\partial K^{-1} &&= c_{1}
\left< 
\frac{z_{+}'z_{-}''+z_{-}'z_{+}''+4z_{+}'z_{-}''z_{-}'z_{+}''}
{(1+z_{+}'z_{-}''+z_{-}'z_{+}'')^{2}} 
\right>_{\Phi (z')\Phi (z'')}\\
 \partial_l \Phi(z_+,z_-) &&=  (2 + \mathcal{O}) \Phi(z_+,z_-) \\\nonumber
+ c_2&&  \left<
\delta\left( z_+ - \frac{z'_+ + z''_+}{1 + z'_-  z''_+ + z'_+  z''_-} \right)
\delta\left( z_- - \frac{z'_- + z''_-}{1 + z'_-  z''_+ + z'_+  z''_-} \right)
\right>_{\Phi,\Phi}  \\\nonumber
&&
- 2 c_2 \mathcal{N} \Phi(z_+,z_-) 
+ c_2 \delta(z_+) \delta(z_-) \mathcal{N}^2 
\end{eqnarray}
\end{subequations}
 where the diffusion operator has been defined in (\ref{def-O}).

\section{Connection between the direct and the replica method}
\label{part:connection-replica}

In this appendix we explore the connections between the expansion in number of
sites of the free energy, and 
the expansion of the replicated partition function 
in power of composite charge fugacities
$Y[{\bf n}]$. As we will see, both expansions coincide exactly, and we can thus
consider the expansion of the replicated partition function as a generating
functional of the site-expansions.  

\subsection{The 2 point free energy $f^{(2)}$}
\label{sec:nattermann}

Fist we consider the first term $f^{(2)}$ given by (\ref{Z-2}) 
of the expansion in number of independent sites (\ref{free}).
This term corresponds exactly to the approximation of
independent dipoles considered in \cite{korshunov96} : dipoles do not
interact with each other, and the free energy is thus the sum over all
positions of the pairs of the free energy of a pair :  
$F=\sum_{{\bf r},{\bf r}' } f^{(2)}_{{\bf r},{\bf r}' }$, and the free energy
of a dipole is simply 
\begin{align*}
 - \beta f^{(2)}_{{\bf r},{\bf r}' }=\ln (1+W_{{\bf r},{\bf r}'})~;~~ 
& W_{{\bf r},{\bf r}'}= y^{2}
\left(\frac{|{\bf r}-{\bf r}'|}{a_{o}}\right)^{-2 \beta J} 
	(w_{{\bf r},{\bf r}'}+w_{{\bf r}',{\bf r}}) \\
& w_{{\bf r}',{\bf r}}=e^{\beta (V_{{\bf r}}-V_{{\bf r}'})}
\end{align*}
 Using the decomposition of the disorder (section \ref{part:disdec}) 
$V_{{\bf r}}=V^{>}_{{\bf r}}+v_{{\bf r}}$, and averaging over the correlated
disorder $V^{>}_{{\bf r}}$ using replica, we obtain
\begin{eqnarray}
&& \overline{\ln (1+ W_{r,r'})}^{V^{>}}
\simeq_{m\to 0} \overline{(1+W_{r,r'})^{m}-1 }^{V^{>}} \\
&& = \sum_{1 \leq p+q \leq m} 
(z_{+}^{{\bf r}}z_{-}^{{\bf r}'})^{p} 
(z_{-}^{{\bf r}}z_{+}^{{\bf r}'})^{q} 
\left(\frac{|{\bf r}-{\bf r}'|}{a_{o}}
\right)^{- 2 (p+q)\beta J+ 2 (p-q)^{2}\sigma \beta^{2}J^{2}}
\end{eqnarray}
 where we used the non local correlation of the disorder 
$\overline{(V^{>}_{{\bf r}}-V^{>}_{{\bf r}'})^{2}}= 
4\sigma J^2 \ln \left(\frac{|{\bf r}-{\bf r}'|}{a_{o}}\right)$
 and the definition 
$z_{\pm}^{{\bf r}}=ye^{\pm \beta v_{{\bf r}}}$. 
 With the bare definition of the replica charge fugacities, this finally gives
after average over $v_{{\bf r}}$,  
the expected first term of the expansion of $\overline{Z^{m}}$ : 
\begin{eqnarray}
\overline{\ln (1+ W_{r,r'})}^{V^{>}}
\simeq_{m\to 0}&& \sum_{[{\bf n} \neq 0]} Y[{\bf n}] Y[-{\bf n}]
 \left(\frac{|{\bf r}-{\bf r}'|}{a_{o}}\right)^{-2 n^{a}K^{ab}n^b}
\end{eqnarray}

\subsection{The 3 points free energy $f^{(3)}$}
\label{sec:3point}

\begin{figure}[htbp]
\centerline{
\fig{2cm}{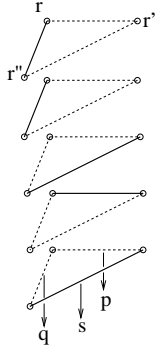}}
\caption{possible configurations of neutral triplet of vector
charges and parametrisation of this triplet by the 
numbers of pairs of components $(+1,-1)$: (p,q,s)}
\label{fig:secondorder}
\end{figure}

The methods is the same as in previous Section, but the calculations are 
slightly more tedious. We consider the second term of the free energy
 expansion : 
\begin{equation}
\ln \left(\frac{1+W_{{\bf r},{\bf r}'}+W_{{\bf r},{\bf r}''}
+W_{{\bf r}',{\bf r}''}}
{ ( 1+W_{{\bf r},{\bf r}'})( 1+W_{{\bf r},{\bf r}''})
( 1+W_{{\bf r}',{\bf r}''})}  \right)
\label{logs}
\end{equation}
 With the help of the previous Section, it is enough to consider
only 
$$\ln \left(1+W_{{\bf r},{\bf r}'}
+W_{{\bf r},{\bf r}''}+W_{{\bf r}',{\bf r}''} \right)$$
Using the
decomposition of figure (\ref{fig:secondorder}), we obtain the sum 
\begin{eqnarray*}
&&\ln \left(1+W_{{\bf r},{\bf r}'}
+W_{{\bf r},{\bf r}''}+W_{{\bf r}',{\bf r}''} \right)\\
&&\simeq_{m\to 0} 
\left(1+W_{{\bf r},{\bf r}'}
+W_{{\bf r},{\bf r}''}+W_{{\bf r}',{\bf r}''} \right)^{m}-1\\
&&=
\sum_{\genfrac{}{}{0pt}{}{(p,q,s)}{ 0< p+q+s \leq m} }
\sum_{p_{1} + p_{2} =p} 
\sum_{q_{1} +q_{2} =q} 
\sum_{s_{1} +s_{2} =s} 
w_{{\bf r},{\bf r}'}^{p_{1}}  w_{{\bf r}',{\bf r}}^{p_{2}}
w_{{\bf r},{\bf r}''}^{q_{1}} w_{{\bf r}'',{\bf r}}^{q_{2}}
w_{{\bf r}',{\bf r}''}^{s_{1}}w_{{\bf r}'',{\bf r}'}^{s_{2}}\\
&& ~~~~~~~~~~ ~~~~~~~~~~\times 
y^{2 (p+q+s)}
\left(\frac{|{\bf r}-{\bf r}'|}{a_{o}}\right)^{-2 \beta Jp}
\left(\frac{|{\bf r}-{\bf r}''|}{a_{o}}\right)^{-2 \beta Jq}
\left(\frac{|{\bf r}'-{\bf r}''|}{a_{o}}\right)^{-2 \beta J s}
\end{eqnarray*}

By averaging the second term over $V^{>}_{{\bf r}}$, and using 
$z_{\pm}^{{\bf r}}=ye^{\pm \beta v_{{\bf r}}}$, we get 
\begin{eqnarray*}
&&\overline{
w_{{\bf r},{\bf r}'}^{p_{1}}  w_{{\bf r}',{\bf r}}^{p_{2}}
w_{{\bf r},{\bf r}''}^{q_{2}} w_{{\bf r}'',{\bf r}}^{q_{2}}
w_{{\bf r}',{\bf r}''}^{s_{3}}w_{{\bf r}'',{\bf r}'}^{s_{2}}
}^{V^{>}}\\
&=&
(z_{+}^{{\bf r}}z_{-}^{{\bf r}'})^{p_{1}}
(z_{-}^{{\bf r}}z_{+}^{{\bf r}'})^{p_{2}}
(z_{+}^{{\bf r}}z_{-}^{{\bf r}''})^{q_{1}}
(z_{-}^{{\bf r}}z_{+}^{{\bf r}''})^{q_{2}}
(z_{+}^{{\bf r}'}z_{-}^{{\bf r}''})^{s_{1}}
(z_{-}^{{\bf r}'}z_{+}^{{\bf r}''})^{s_{2}}\\
&\times &
\left(\frac{|{\bf r}-{\bf r}'|}{a_{o}}
	\right)^{2 \sigma \beta^{2}J^{2} (p_{1}-p_{2})^{2}}
\left(\frac{|{\bf r}-{\bf r}''|}{a_{o}}
	\right)^{2 \sigma \beta^{2}J^{2} (q_{1}-q_{2})^{2}}
\left(\frac{|{\bf r}'-{\bf r}''|}{a_{o}}
	\right)^{2 \sigma \beta^{2} J^{2} (s_{1}-s_{2})^{2}}
\end{eqnarray*}
Inserting this result in the above expression
averaged over $V^{>}_{{\bf r}}$, the sum over the partitions
of the interval $[0,m]$ in 
$0 \leq p_1+p_2+q_1+q_2+s_1+s_2 \leq m$ can be
exactly rewritten as a sum over all {\it neutral} triplets 
${\bf n}, {\bf n'}, {\bf n''}$ of 
$m$ component vector charges with components
$0,\pm 1$ (see figure \ref{fig:secondorder}).
Thus one recovers the expression 
\[
\sum_{n+n'+n''=0} 
Y[n](r) Y[n'](r') Y[n''](r'')e^{-S_{n,n',n''}} -1
\]
We now note that the last three logarithmic terms in (\ref{logs}) 
just give additional restrictions on this sum  
(as they correspond respectively to ${\bf n}=0,{\bf n}'=0,{\bf n}''=0$). 
Thus, averaging over the local fugacities,
we end up with the expected second term 
(see \ref{replicah}) of the expansion 
of $\overline{Z^m}$ in power of the vector fugacities $Y[{\bf n}]$:

\begin{eqnarray}
&& \sum_{\genfrac{}{}{0pt}{}{n+n'+n''=0}{ {\bf n,n',n''} \neq 0}} 
Y[{\bf n}] Y[{\bf n'}] Y[{\bf n''}] e^{- \beta H_{{\bf n,n',n''}}}
\end{eqnarray}

 These simple combinatorics can be done on higher order terms : it gives the
equivalence term by term between the $m\to 0$ limit of the expansion 
in $Y[{\bf n}]$ and 
the expansion in the number of independent sites of the moments of the
free energy.

\section{Regularization of replicated Coulomb gases}\label{app-cutoff}

In this Appendix we discuss the consequences on the RG equations
of the choice of cutoff made in this paper for the replicated Coulomb gas.
We illustrate for simplicity only the case of zero disorder.
Although we will be mainly concerned with Coulomb gas,
most of this discussion can be applied to other general replica field 
theory.

Let us first recall the results for a single component CG ($m=1$).
We restrict to the most relevant charges $\pm 1$ for simplicity.
It is defined by the action 
\begin{equation}
- \beta H = K
\sum_{|{\bf r}-{\bf r}'|\geq a_{o}} 
n({\bf r})~ G(|{\bf r}-{\bf r}'|) ~n({\bf r}')
+ \ln (y) \sum_{{\bf r}} n^{2}({\bf r})
\end{equation}
The corresponding RG equation was derived by Kosterlitz:
\begin{equation}
\label{cutoff:RG1}
\partial_l y(l) = \left(2- K\right) y + \mathcal{O}(y^3)
\end{equation}

\begin{figure}[htbp]
\centerline{\fig{8cm}{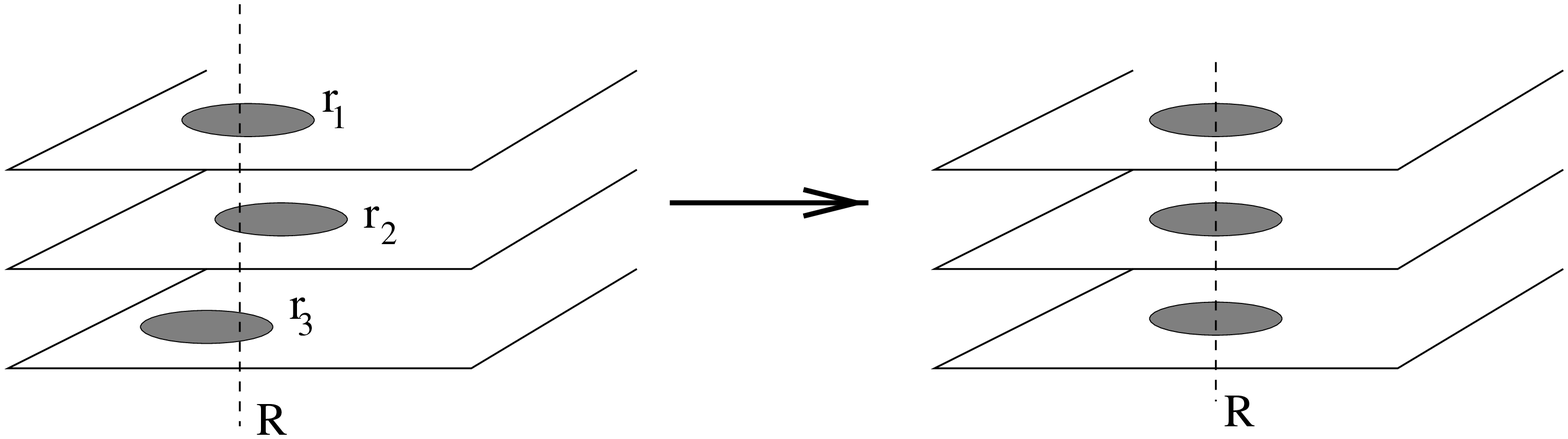}}
\caption{Schematic representation of the definition of the replicated
vector charges and their cutoff}
\label{fig:cutoff}
\end{figure}

We now consider $m$ copies of this model in the absence of
disorder, as illustrated in Fig. (\ref{fig:cutoff}).
They are a priori physically completely 
uncoupled. The most natural cutoff procedure in that case would be 
independent cutoffs (e.g. hard core for each) (left figure).
Another procedure, which becomes much more convenient
in the presence of disorder, is to reformulate the $m$ copies
as a single Coulomb gas of vector charges with $m$ components.
However in that case the cutoff is by definition columnar (right figure)
(e.g hard core vector charge are a hard columnar disk)
and in a sense the copies are coupled, via the cutoff.
We now check that in the pure case the ensuing vector CG RG 
equations are still perfectly compatible with (\ref{cutoff:RG1})
as they should. They read:
\begin{equation}
\partial_l Y[{\bf n}] = \left( 2 -  K {\bf n} \cdot {\bf n} \right) Y[{\bf n}] 
+ c_{2} \sum_{{\bf n}'\neq 0,{\bf n}} Y[{\bf n}'] Y[{\bf n}-{\bf n}'] 
+ \text{higher order terms} \nonumber
\end{equation}

Let us first illustrate the case of two copies $m=2$.
We can choose $Y[1,0]=Y[-1,0]=Y[0,1]=Y[0,-1] \equiv Y_1$
and $Y[1,1]=Y[-1,1]=Y[1,-1]=Y[-1,-1] \equiv Y_2$. Then 
considering all the possible fusions within this set 
the RG equations read:
\begin{eqnarray*}
&& \partial_l Y_1 = (2-K) Y_1 + 4 c_2 Y_1 Y_2 \\
&& \partial_l Y_2 = (2- 2 K) Y_2 + 2 c_2 Y_1^2
\end{eqnarray*}
A solution of these equations, to the order $o(y^3)$
at which we are working, is:
\begin{eqnarray*}
Y_l[1,0] = \lambda_1 y_l \quad Y_l[1,1] = \lambda_2 y_l^2
\end{eqnarray*}
with $\lambda_2 = c_2 \lambda_1^2$, where $y_l$ satisfies the
single copy equation (\ref{cutoff:RG1}). This can be generalized to
higher charges so that in general one can find solutions of the type:
$$
Y_{p} = Y_{[\underbrace{1,1,\dots,1}_{p},0,\dots 0]} = 
\lambda_{i} y_l^{i}
$$
with coefficients $\lambda_{p}$ which can be determined for a given
regularisation procedure. They depend on 
both the initial ($m=1$) and the replica regularisation (see fig \ref{fig:cutoff}). 

This example illustrates how a generic cutoff procedure will
produce non trivial coefficients $\lambda_i$.
Thus if one reinterprets formally $Y[{\bf n}] = < y^{n_+ + n_-} >_{\Phi(y)}$ in terms of 
a ``disorder'' $\Phi(y)$ as done in this paper
\footnote{with $z_+=z_-=y$}, one must keep
in mind that even the pure system corresponds to a non 
trivial ``bare disorder'' in the fugacities, which solely originates
from the (convenient) choice of a {\it columnar} 
cutoff\footnote{in the interpretation as a branching process, it does correspond
to the disorder in the tree structure. See the end of Section \ref{part:rappel-KPP}
for a particular choice which does not introduce additional disorder}.
It is thus clear that the definition
of the local fugacity distribution is strongly cutoff dependent.

\section{Explicit derivation of the expansion in the number of points}

\label{part:appexp}

In this appendix we derive formula (\ref{fk}) for the expansion of the free
energy in the number of points. The same method can be applied to
other physical observables.

To organize this expansion we start by introducing {\it fictitious site 
dependent fugacities} for
the charges : $\zeta_{{\bf r}_{i}}$. These are introduced only as a trick 
in this Appendix and should not
be confused with the real disordered fugacities of (\ref{decomposition}). 
Indeed,
to recover the original model, we will set them back at the end either to 
$\zeta_{{\bf r}_{i}} =1$
(for the lattice model (\ref{square-bis})) or $\zeta_{{\bf r}_{i}} =y$ for the
continuum model, where $y$ is the corresponding fugacity for the charge
in the pure case. These fugacities are introduced by writing a more general
form of the the partition function (\ref{square-bis}) as 
\begin{equation} \label{square-fict}
Z[V, \zeta]  = \sum_{p}
\sum'_{\{n_{1},\dots n_{p} \}}
\sum_{{\bf r}_{1}\neq \dots \neq {\bf r}_{p}}
\left(\prod_{i} \zeta_{{\bf r}_{i}}^{n_i^2} \right)
e^{\beta J \sum_{{\bf r}_{i}\neq {\bf r}_{j}}
n_{i}G_{{\bf r}_{i}-{\bf r}_{j}} n_{j} +
\beta \sum_{i} n_{i}V_{{\bf r}_{i}}}
\end{equation}
Here and below, as in (\ref{Zcont}), all formulaes can be extended to the 
continuum model
by replacing discrete sums over distincts sites 
${\bf r}_{1}\neq {\bf r}_{2}\dots \neq {\bf r}_{p}$ by 
integrals with, {\it e.g} hard core conditions
$|{\bf r}_{1}- {\bf r}_{2}|\geq a_{o}$. Note that in the above
expression (\ref{square-fict}) we do not make use of 
the decomposition (\ref{decomposition}) and $V$ denotes
the original disorder.

Let us consider for simplicity a system of $N$ distinct points
${\bf r}_1,...{\bf r}_N$. The free energy functional 
$F[V, \zeta] = - T \ln Z[V, \zeta ]$
is a function of the $N$ variables
$\zeta_{{\bf r}_1}$,.., $\zeta_{{\bf r}_N}$.
Let us write the conventional Taylor expansion of the free
energy around $\zeta=0$:

\begin{eqnarray}\label{taylor}
F[V, \zeta] &=&
\sum_{p_1,p_2,..p_N = 0}^{+\infty} 
\frac{1}{p_1 ! \dots p_N !}
\frac{\partial^{p_1+..p_N} F[V,\zeta]}
{\partial \zeta_{{\bf r}_1}^{p_1} \dots 
\partial \zeta_{{\bf r}_N}^{p_N}}|_{\zeta=0} 
\times \zeta_{{\bf r}_1}^{p_1}... 
\zeta_{{\bf r}_N}^{p_N}
\end{eqnarray}
We now separate in this sum the terms which have
only one non zero $p_i$, then only two non zero $p_i$, etc..
Thus we can rewrite:

\begin{eqnarray}\label{taylor-bis}
F[V, \zeta] &=&
\sum_{k=1}^{+\infty}
\sum_{\{ r_{i_1},..,r_{i_k} \}}
\sum_{q_1,..q_k = 1}^{+\infty}
\frac{1}{q_1 ! \dots q_k !}
\frac{\partial^{q_1+..q_k} F[V,\zeta]}
{\partial \zeta_{{\bf r}_{i_1}}^{q_1} \dots \partial 
\zeta_{{\bf r}_{i_k}}^{q_k}}|_{\zeta=0}
\times \zeta_{{\bf r}_{i_1}}^{q_1}... 
\zeta_{{\bf r}_{i_k}}^{q_k} \\
&=& 
\nonumber
\sum_{k=1}^{+\infty}
\sum_{\{ r_{i_1},..,r_{i_k} \}}
f^{(k)}_{r_{i_1},..,r_{i_k}}[V,\zeta]
\end{eqnarray}
where the sum is over all distincts {\it sets} 
${\{ r_{i_1},..,r_{i_k} \}}$ of $k$ distincts points
(among the $N$ points) and the sum over each $q_i$ goes
from $1$ to $+\infty$.

We have thus obtained an expansion of $F[V,\zeta]$ as a sum of terms
of the form $f^{(k)}_{{\bf r}_{1}\dots {\bf r}_{k}}$ which 
depends exactly and only on the variables $\zeta$ (and thus also only on the 
variables $V$) evaluated at the $k$ {\it distinct points} 
${{\bf r}_{i_1}\dots {\bf r}_{i_k}}$. For a neutral Coulomb gas it
starts with $k=2$ and reads:
\begin{eqnarray}
\label{free-zeta}
F[V,\zeta]&=& \sum_{\{ {\bf r}_{i_1} \neq {\bf r}_{i_2} \}}
f^{(2)}_{{\bf r}_{i_1},{\bf r}_{i_2}}[V,\zeta]  
+ \sum_{\{ {\bf r}_{i_1} \neq {\bf r}_{i_2} \neq {\bf r}_{i_3} \}}
f^{(3)}_{{\bf r}_{i_1},{\bf r}_{i_2},{\bf r}_{i_3}}[V,\zeta]
+\dots
\end{eqnarray}
with the definition 
\begin{eqnarray}\label{def-fk}
f^{(k)}_{{\bf r}_{i_1} \dots {\bf r}_{i_k}}[V,\zeta] &=&
\sum_{q_1,..q_k = 1}^{+\infty}
\frac{1}{q_1 ! \dots q_k !}
\frac{\partial^{q_1+..q_k} F[V,\zeta]}
{\partial \zeta_{{\bf r}_{i_1}}^{q_1} \dots \partial 
\zeta_{{\bf r}_{i_k}}^{q_k}}|_{\zeta=0}
\times \zeta_{{\bf r}_{i_1}}^{q_1}... 
\zeta_{{\bf r}_{i_k}}^{q_k}
\end{eqnarray}
Note that in this last expression we can drop out the dependence of 
$F[V,\zeta]$ on the fugacities $\zeta_{{\bf r}}$ and the potential 
$V_{{\bf r}}$ at points different from ${\bf r}_{i_1},\dots {\bf r}_{i_k}$.

A more explicit expression can be obtained
by summing over the $q_{i}$ in (\ref{def-fk}):
\begin{eqnarray}\label{fk-expl}
f^{(k)}_{{\bf r}_{1}\dots {\bf r}_{k}} [V,\zeta] &=&
\sum_{l=0}^{k} (-1)^{k-l}\sum_{i_{1},\dots i_{l}\in [1,\dots k]}
F_{{\bf r}_{i_{1}},\dots {\bf r}_{i_{l}}}[V,\zeta]
\end{eqnarray}
 where $F_{{\bf r}_{i_{1}},\dots {\bf r}_{i_{l}}}[V,\zeta]$ is the free energy
associated with the system of sites ${\bf r}_{i_{1}},\dots {\bf r}_{i_{l}}$
(instead of the full lattice). Equivalently, it does depend only on the
fugacities $\zeta_{{\bf r}}$ (or potential $V_{{\bf r}}$) at
 points ${\bf r}_{i_{1}},\dots {\bf r}_{i_{l}}$. 
 After setting $\zeta_{{\bf r}}=1$ (lattice model) 
or $\zeta_{{\bf r}}=y$ (continuum model) this gives the definition of the
$f^{(k)}[V]$ that we use throughout this paper, namely
the equation (\ref{fk}).
This last expression (\ref{fk-expl}) allows to explicitly compute the
coefficient $f^{(k)}[V]$ of the expansion (\ref{free-zeta}) for arbitrary order $k$. 
In the Section (\ref{part:exp}) we use it to
give explicit expressions
for the first few terms of the expansion of the free energy.

\section{Fusion corrections to the free energy expansion}
\label{app:kpoint}

As shown in the previous Section the free energy expansion
involves the partition function $Z^{(p)}_{ {\bf r}_1 .. {\bf r}_p }$
of a system of finite number of sites. We illustrate here how the 
rule of fusion of environments described in Section \ref{part:directRG}
works more generally. Let us consider a system of $p \ge 4$ sites, 
with charges restricted to $0, \pm 1$
and of energy and fugacities given by (\ref{Zcont}). Upon increase
of the cutoff one must take into account fusion of
the sites ${\bf r}_1$ and ${\bf r}_2$ into the site 
$\tilde{{\bf r}} = \frac{1}{2} ({\bf r}_1 + {\bf r}_2)$, one finds
that:
\begin{eqnarray}
Z^{(p)}_{ {\bf r}_1, {\bf r}_2, .. {\bf r}_p } \to \tilde{Z}^{(p-1)}_{\tilde{{\bf r}},  {\bf r}_3, ... {\bf r}_p }
\end{eqnarray}
with:
\begin{multline}\label{sump}
\tilde{Z}^{(p-1)}_{\tilde{{\bf r}}, {\bf r}_3 ... {\bf r}_p } = 
Z^{(p-2)}_{{\bf r}_3 .. {\bf r}_p } +
\tilde{W}_{12} Z^{(p-2)}_{{\bf r}_3 .. {\bf r}_p } \\
+ 
\sum_{q =2, q ~ \text{even}}^{q=p-1} 
\sum_{i_2,...i_q \in [3,..p] } ( W^{(q)}_{1,{\bf r}_{i_2},..{\bf r}_{i_q}} +
W^{(q)}_{2,{\bf r}_{i_2},..{\bf r}_{i_q}} )
\end{multline}
where $\tilde{W}_{12} = z^{+}_{{\bf r}_1} z^{-}_{{\bf r}_2} +
z^{+}_{{\bf r}_2} z^{-}_{{\bf r}_1}$ and 
\begin{eqnarray}
W^{(q)}_{1,{\bf r}_{i_2},..{\bf r}_{i_q} } = \sum_{n_1 = \pm 1} 
\sum_{\genfrac{}{}{0pt}{}{n_{r_i} = \pm 1 }{n_1 + \sum_{i} n_{r_i} =0}} 
e^{- \beta H[n,r] }
\end{eqnarray}
where $H$ has been defined in (\ref{Zcont}) and the charge $n_1$ is located
in $\tilde{{\bf r}}$. Note that $W^{(1)}_{1,{\bf r}_{3}} = W_{\tilde{{\bf r}}, {\bf r}_{3}}$ the dipole
weight defined in Section  \ref{part:directRG1}. In (\ref{sump}) the first term
corresponds to the total weight of configurations 
with no charge in ${\bf r}_1$ and ${\bf r}_2$. The
second term  correspond to configurations with a dipole in $( {\bf r}_1, {\bf r}_2)$. Its
original expression is complicated but simplifies when ${\bf r}_1$ and 
${\bf r}_2$ are fused into $\tilde{{\bf r}}$
( the interaction energy
of any given other charge $n_\alpha$ in ${\bf r}_\alpha$ with $n_{{\bf r}_1}$
becomes opposite to the one with $n_{{\bf r}_2}$ (up to higher order terms)
and $W_{{\bf r}_1, {\bf r}_2}$ simplifies into $\tilde{W}_{12}$
to lowest order in $dl$ as explained in Section \ref{part:directRG}). 
The last two terms correspond to all configurations with one charge either
in ${\bf r}_1$ or in ${\bf r}_2$. 

The important property with respect to the free energy expansion is that
one can factor the term $1+\tilde{W}_{12}$ and rewrite:

\begin{eqnarray}
\ln  \tilde{Z}^{(p-1)}_{\tilde{{\bf r}},  {\bf r}_3 ... {\bf r}_p }
=
\ln (1 + \tilde{W}_{12}) +
\ln ( Z^{(p-2)}_{{\bf r}_3 .. {\bf r}_p } + 
\sum_{q =2, q~ \text{even}}^{q=p-1} 
\sum_{i_2,...i_q \in [3,..p] }  \tilde{W}^{(q)}_{\tilde{{\bf r}},{\bf r}_{i_2},..{\bf r}_{i_q} }) \nonumber
\end{eqnarray}

where $\tilde{W}^{(q)}$ as the same definition as
$W^{(q)}$ except that the fugacity at $\tilde{{\bf r}}$ has been
modified according to the fusion rule for fugacities (\ref{rule2}). The total factor inside the
last logarithm is exactly the partition function 
$Z^{(p-1)}_{\tilde{{\bf r}}, {\bf r}_3 ... {\bf r}_p}$
of a system of $p-1$ sites with the new fugacity given by
(\ref{rule2}) on the site $\tilde{{\bf r}}$.

\section{Higher charges and extensions}
\label{highercharges}

In this Appendix we briefly indicate how higher charges can be included in the
same (fusion-diffusion) formalism and why including them 
does not affect any of our results.
Let us consider for instance the charges $\pm 2$ and define
$n_{++}$ and $n_{--}$ respectively as the number of component charges
$+2$ and $-2$ in the vector charge ${\bf n}$.
The parametrization
(\ref{eq:ydef}) can be readily extended to encode also for the charges $\pm 2$ simply by
writing the vector fugacity $Y[{\bf n}]$ as an average
over a function $\Phi(z_{+},z_{-},z_{++},z_{--})$ as
\begin{eqnarray}
Y[n_{+},n_{-},n_{++},n_{--}] = \int_{z_{+},z_{-},z_{++},z_{--}}
\Phi(z_{+},z_{-},z_{++},z_{--}) z_{+}^{n_{+}} z_{-}^{n_{-}} 
z_{++}^{n_{++}} z_{--}^{n_{--}} \nonumber
\end{eqnarray}
where the random variables $z_{++}$ and $z_{--}$ represent the
random local fugacities for the charges $\pm 2$. 
Similar manipulations as in Appendix (\ref{part:replimit}) lead to a RG 
equation for a normalized $P(z_{+},z_{-},z_{++},z_{--})$
which we will not write explicitly. It does contain a diffusion operator 
as well as a fusion term. For illustration we simply give the
diffusion operator, expressed using more convenient variables
$z_{\pm}=e^{\beta(u \pm v)}$ and $z_{++}=z_{+}^2 e^{\beta(u'+v')}$ 
and $z_{--}=z_{-}^2 e^{\beta (u'-v')}$. It reads, for the
corresponding probability distribution $\tilde{P}(u,u',v,v')$:
\begin{eqnarray}
2 - J(\partial_u + 2 \partial_{u'}) + \sigma J^2 \partial_v^2
\end{eqnarray}
The detailed study of the corresponding RG equation, together
with the fusion term will not 
be reported here. Instead let us indicate how the
irrelevance of the higher charges can be justified. Since one has:
\begin{eqnarray}
 2 - {\bf n}.K.{\bf n} =&& 2 - \beta J (n_+ + n_- + 4 (n_{++} + n_{--})) \\
&& + \sigma \beta^2 J^2 (n_+ - n_- + 2 (n_{++} - n_{--}))^2
\nonumber 
\end{eqnarray}
it is clear that the reduced probability
\begin{eqnarray}
Q(z_{++},z_{--})=\int_{z_+,z_-} P(z_{+},z_{-},z_{++},z_{--})
\end{eqnarray}
satisfies the same diffusion equation as $P(z_{+},z_{-})$
in (\ref{rgeqp}) but with the change $J \rightarrow 4 J$
and $\sigma \rightarrow  \sigma/4$. It becomes relevant by power counting 
only at $\sigma= 4 \sigma_c = 1/2$. Thus it is less relevant
than $P(z_{+},z_{-})$ in the region
of the phase diagram of interest and we can rightly neglect the
new fugacities $z_{++}$ and $z_{--}$. One can also check that the
fusion terms imply that $P(z_{++} \sim 1)$ is of order
$P(z_{+} \sim 1)^2$ when this parameter is small (since at
lowest order the RG equation contains a term proportional to
$\delta(z_{++} - z'_{+} z''_{+})$). The rare events which
involve the charges $\pm 2$ are thus subdominant.

\section{Evaluation of screening integrals at criticality}
\label{part:integrals}

To evaluate the integral (\ref{sct0.1}) in the screening equation for $J$
we consider separately three intervals
$u-X_l < b$, $X_l + b < u < - X_l - b$ and $u > - X_l - b$, where 
$b$ is a number such that $\tilde{p}_l(x)$ can be replaced by its asymptotic
expression for $x>b$ with good accuracy. Only in the middle interval can we
use the universal tail expressions for both factors $p_l$. Thus we have,
using symetries:

\begin{eqnarray}
 \int du p_l(u - X_l) p_l(- u - X_l) =&& 2 \int_{-\infty}^b dv p_l(v)
p_l(- v - 2 X_l)  
\nonumber \\
&& + \int_{X_l +b}^{- X_l -b} p_l(u - X_l) p_l(- u - X_l) 
\label{cutintwo}
\end{eqnarray}

In the first integral we can replace the second function by its
asymptotic form and get:

\begin{eqnarray}
\frac{2 A}{D} e^{2 X_l/\sqrt{D}} \int_{-\infty}^b dv p_l(v) (-2 X_l - v) 
e^{\frac{v}{\sqrt{D}}} e^{- \frac{(v + 2 X_l)^2}{8 D l}}
\sim C X_l e^{2 X_l}
\end{eqnarray}
since, remembering that the integral $\int_{-\infty}^{+\infty} dv p_l(v) =1$,
the above integral is at most a finite number. The second integral
in (\ref{cutintwo}) is estimated as:
\begin{eqnarray}
\frac{A^2}{D^2} e^{2 X_l/\sqrt{D}}  \int_{X_l +b}^{- X_l -b} du
(X_l^2 - u^2) e^{- \frac{u^2 + X_l^2}{4 D l}} 
\nonumber
\end{eqnarray}
and gives the leading contribution. For $X_l/l \to 0$ 
it behaves as $\sim X_l^3 e^{2 X_l/\sqrt{D}}$ and gives the
estimate in the text. It is encouraging to note that the 
leading behaviour precisely originates 
from the interval where $p_l(u)$ can be replaced everywhere by
its universal asymptotic form.

The integral (\ref{sct0.2}) in the screening equation for $\sigma$
can also be estimated by considering for each variable $u$, $u'$ two intervals 
$u<b$ and $u>b$. The end result is that we find to leading order:

\begin{eqnarray}
&& \partial_l J \sim \frac{1}{\sqrt{D}} \partial_l \sigma \sim
\frac{A}{D^2} (-X_l)^3 e^{2 X_l/\sqrt{D}}
\int_{-1}^{+1} dv (1-v^2) e^{- \frac{X_l^2}{ 4 D l} (1 + v^2)}
\end{eqnarray}


\end{document}